\documentclass[twocolumn,showkeys,showpacs,preprintnumbers,prd,superscriptaddress,nofootinbib]{revtex4-1}

\usepackage{graphicx}
\usepackage{epsf}
\usepackage{bm}
\usepackage{amsmath}
\usepackage{amsfonts}
\usepackage{amssymb}
\usepackage{epstopdf}
\usepackage{natbib}
\usepackage{hyperref}
\usepackage{color}
\usepackage{verbatim}
\usepackage{multirow}
\usepackage{bm}
\usepackage{hyperref}
\hypersetup{colorlinks   = true,
            urlcolor     = blue,
            citecolor    = blue,
            linkcolor    = blue,
            menucolor    = blue,
            anchorcolor  = blue,
            filecolor    = blue}
            
\date{\today}

\begin{document}

\title{Consistency tests of $\Lambda$CDM from the early integrated Sachs-Wolfe effect: Implications for early-time new physics and the Hubble tension}

\author{Sunny Vagnozzi}
\email{sunny.vagnozzi@ast.cam.ac.uk}
\affiliation{Kavli Institute for Cosmology (KICC) and Institute of Astronomy,\\University of Cambridge, Madingley Road, Cambridge CB3 0HA, United Kingdom}

\begin{abstract}
\noindent New physics increasing the expansion rate just prior to recombination is among the least unlikely solutions to the Hubble tension, and would be expected to leave an important signature in the early Integrated Sachs-Wolfe (eISW) effect, a source of Cosmic Microwave Background (CMB) anisotropies arising from the time-variation of gravitational potentials when the Universe was not completely matter dominated. Why, then, is there no clear evidence for new physics from the CMB alone, and why does the $\Lambda$CDM model fit CMB data so well? These questions and the vastness of the Hubble tension theory model space motivate general consistency tests of $\Lambda$CDM. I perform an eISW-based consistency test of $\Lambda$CDM introducing the parameter $A_{\rm eISW}$, which rescales the eISW contribution to the CMB power spectra. A fit to \textit{Planck} CMB data yields $A_{\rm eISW}=0.988 \pm 0.027$, in perfect agreement with the $\Lambda$CDM expectation $A_{\rm eISW}=1$, and posing an important challenge for early-time new physics, which I illustrate in a case study focused on early dark energy (EDE). I explicitly show that the increase in $\omega_c$ needed for EDE to preserve the fit to the CMB, which has been argued to worsen the fit to weak lensing and galaxy clustering measurements, is specifically required to lower the amplitude of the eISW effect, which would otherwise exceed $\Lambda$CDM's prediction by $\approx 20\%$: this is a generic problem beyond EDE and likely applying to most models enhancing the expansion rate around recombination. Early-time new physics models invoked to address the Hubble tension are therefore faced with the significant challenge of making a similar prediction to $\Lambda$CDM for the eISW effect, while not degrading the fit to other measurements in doing so.
\end{abstract}



\maketitle

\section{Introduction}
\label{sec:introduction}

The six-parameter $\Lambda$CDM model has proven to be extremely successful in explaining a wide range of cosmological and astrophysical observations, including observations of the Cosmic Microwave Background (CMB), the clustering of the large-scale structure (LSS), the magnitude-redshift relation of distant Type Ia Supernovae (SNeIa), and light element abundances~\cite{Riess:1998cb,Perlmutter:1998np,Troxel:2017xyo,Scolnic:2017caz,Aghanim:2018eyx,Bianchini:2019vxp,Aiola:2020azj,Alam:2020sor,Asgari:2020wuj,Mossa:2020gjc}. However, discrepancies between independent inferences of cosmological parameters under the assumption of $\Lambda$CDM might be a sign of the model's incompleteness, and pave the way towards new physics: this should not come as a surprise, given that several of the ingredients of $\Lambda$CDM, not least the nature of the dark sector, remain puzzling to date~\cite{Sahni:2004ai,Bertone:2004pz,Huterer:2017buf}. Among these discrepancies, the most significant one is the ``Hubble tension'', which refers to a persisting mismatch between several early- and late-time inferences of the Hubble constant $H_0$~\cite{Verde:2019ivm,DiValentino:2021izs,Perivolaropoulos:2021jda}.

The possibility that the Hubble tension calls for new physics is being given very serious consideration in the literature (see e.g. Refs.~\cite{Mortsell:2018mfj,Poulin:2018cxd,Agrawal:2019lmo,Alexander:2019rsc,Niedermann:2019olb,Sakstein:2019fmf,Ye:2020btb,Zumalacarregui:2020cjh,Hill:2020osr,Chudaykin:2020acu,Das:2020wfe,Braglia:2020bym,Niedermann:2020dwg,Ivanov:2020ril,DAmico:2020ods,Ye:2020oix,Niedermann:2020qbw,Murgia:2020ryi,Smith:2020rxx,Chudaykin:2020igl,CarrilloGonzalez:2020oac,Oikonomou:2020qah,Seto:2021xua,Tian:2021omz,Freese:2021rjq,Nojiri:2021dze,Vagnozzi:2018jhn,Nunes:2018xbm,Poulin:2018zxs,Yang:2018euj,Banihashemi:2018oxo,Guo:2018ans,Graef:2018fzu,Kreisch:2019yzn,Pandey:2019plg,Martinelli:2019dau,Vattis:2019efj,Lin:2019qug,Li:2019yem,DiValentino:2019exe,Yang:2019nhz,Pan:2019gop,Yang:2019nhz,Vagnozzi:2019ezj,Visinelli:2019qqu,Cai:2019bdh,Schoneberg:2019wmt,Pan:2019hac,DiValentino:2019ffd,Sola:2019jek,Escudero:2019gvw,DiValentino:2019jae,Liu:2019awo,Hart:2019dxi,Akarsu:2019hmw,Benisty:2019pxb,Yang:2020zuk,Choi:2020tqp,Lucca:2020zjb,Hogg:2020rdp,Benevento:2020fev,Barker:2020gcp,Gomez-Valent:2020mqn,Akarsu:2020yqa,Ballesteros:2020sik,Haridasu:2020xaa,Alestas:2020mvb,Jedamzik:2020krr,Ballardini:2020iws,Banerjee:2020xcn,Elizalde:2020mfs,Gonzalez:2020fdy,Capozziello:2020nyq,Das:2020xke,Banihashemi:2020wtb,Choudhury:2020tka,Brinckmann:2020bcn,Moshafi:2020rkq,Alestas:2021xes}). This new physics should be able to accommodate a higher value of $H_0$ from CMB data, while complying with constraints from other datasets: examples are Baryon Acoustic Oscillations (BAO)~\cite{Beutler:2011hx,Ross:2014qpa,Alam:2020sor} and Hubble flow SNeIa measurements~\cite{Scolnic:2017caz}, which severely restrict the possibility of solving the Hubble tension via global late-time new physics. These measurements single out the least unlikely scenarios to be those operating at early times, prior to and around recombination, and lowering the sound horizon at recombination by $\approx 7\%$~\cite{Bernal:2016gxb,Addison:2017fdm,Lemos:2018smw,Aylor:2018drw,Knox:2019rjx,Arendse:2019hev,Efstathiou:2021ocp}. However, it is fair to say that none of the many proposed new physics models have succeeded at the task of solving the Hubble tension, while at the same time not degrading the fit to other datasets or worsening other discrepancies.

The theory model space with regards to the Hubble tension is extremely vast and, besides the few data-driven indications discussed above, is short of a clear direction. In this sense, two types of analyses can be very useful to either point towards a definite direction towards which to move, or further restrict the set of viable directions: \textit{i)} model-agnostic tests of new physics, and \textit{ii)} consistency tests of the $\Lambda$CDM model.~\footnote{A particularly interesting consistency test of $\Lambda$CDM is via the use of so-called ``metaparameters'', see e.g. Refs.~\cite{Motloch:2020lhu,Chu:2004qx}.} The analysis I will pursue in this paper, while by construction more focused on \textit{ii)}, will contain a combination of these two features.

A non-negligible amount of early-time new physics needs to be present around the time of recombination for it to solve the Hubble tension~\cite{Knox:2019rjx}. One would therefore expect this new physics to already show up in \textit{CMB data alone}, before even looking at local $H_0$ measurements. The key question motivating my analysis is then the following: ``\textit{Why is there no clear evidence for new physics from CMB data alone?}'', or the closely related ``\textit{Why does the $\Lambda$CDM model fit CMB data so well?}'' While the two might at first glance sound like trivial questions, I believe that finding a clear answer to these can teach us valuable general lessons not only on why several early-time modifications to $\Lambda$CDM ultimately fail at solving the Hubble tension, but perhaps also on which direction(s) one should look towards in an attempt to construct successful early-time new physics models.

Where should a non-negligible amount of early-time new physics first show up in the CMB? The answer for many models, I argue, has to do with the early Integrated Sachs-Wolfe (eISW) effect. The eISW effect is a contribution to CMB anisotropies arising from time-varying gravitational potentials at early times, immediately after recombination, when the Universe was not entirely matter-dominated~\cite{Sachs:1967er,Rees:1968zza}. It is then not hard to understand why new physics altering the expansion rate around recombination would almost inevitably leave an important imprint on the eISW effect. In fact, one can generically expect that many types of early-time new physics models introduced to solve the Hubble tension will \textit{enhance} the amplitude of the eISW effect.

These considerations motivate a consistency test of the $\Lambda$CDM model, to test whether its predictions for the eISW effect fit CMB data well. I perform such a consistency test by introducing the parameter $A_{\rm eISW}$, which artificially rescales the eISW contribution to the CMB power spectra. This is closely reminiscent of the better known $A_{\rm lens}$ lensing amplitude, rescaling the amplitude of lensing in the CMB power spectra~\cite{Calabrese:2008rt}. Inferring a value of $A_{\rm eISW}$ strongly inconsistent with the standard value $A_{\rm eISW}=1$ could be a clear sign of early-time new physics in the CMB alone, whereas the converse could present an important challenge for these models.

From a fit to the \textit{Planck} 2018 CMB temperature and polarization anisotropy measurements, I find $A_{\rm eISW}=0.988 \pm 0.027$, showing that $\Lambda$CDM's prediction for the amplitude of the eISW effect is in perfect agreement with data. To illustrate the implications for early-time new physics, I provide a case study focused on the well-known early dark energy (EDE) model, one of the leading contenders to solve the Hubble tension~\cite{Mortsell:2018mfj,Poulin:2018cxd,Agrawal:2019lmo,Alexander:2019rsc,Niedermann:2019olb,Sakstein:2019fmf,Ye:2020btb,Zumalacarregui:2020cjh,Hill:2020osr,Chudaykin:2020acu,Das:2020wfe,Braglia:2020bym,Niedermann:2020dwg,Ivanov:2020ril,DAmico:2020ods,Ye:2020oix,Niedermann:2020qbw,Murgia:2020ryi,Smith:2020rxx,Chudaykin:2020igl,CarrilloGonzalez:2020oac,Oikonomou:2020qah,Seto:2021xua,Tian:2021omz,Freese:2021rjq,Nojiri:2021dze}. EDE's success hinges upon its ability to accommodate a higher value of $H_0$ while fitting CMB data as well as $\Lambda$CDM. This comes at the price of an increase in the cold dark matter (DM) density $\omega_c$, which has been argued to worsen the fit to weak lensing (WL) and galaxy clustering measurements, possibly leading to the demise of the EDE model as a solution to the Hubble tension~\cite{Hill:2020osr,Ivanov:2020ril,DAmico:2020ods} (see however a partial rebuttal of these results in Refs.~\cite{Murgia:2020ryi,Smith:2020rxx}). I explicitly show that the increase in $\omega_c$ is required to lower the amplitude of the eISW effect, which would otherwise be overpredicted by $\approx 20\%$, well beyond what is allowed by the data.

The problems faced by EDE in the context of the eISW effect, and the required increase in $\omega_c$, might actually more generally be a stumbling block for several other early-time new physics models proposed to solve the Hubble tension, particularly if these models enhance the expansion rate around recombination. It is in this sense that matching $\Lambda$CDM's prediction for the eISW effect presents an important challenge for early-time new physics. While not being a strict \textit{no-go theorem}, these results place further restrictions on the possibility of solving the Hubble tension with early-time new physics alone, and support recent findings along a related line~\cite{Krishnan:2020obg,Jedamzik:2020zmd,Lin:2021sfs,Dainotti:2021pqg,Krishnan:2021dyb,Vagnozzi:2021tjv,Krishnan:2021jmh}.

The rest of this paper is then organized as follows. In Sec.~\ref{sec:eisw} I review the physics of the early ISW effect, and discuss how $A_{\rm eISW}$ is introduced from a practical point of view, alongside the impact of this parameter on the CMB power spectrum. In Sec.~\ref{sec:data} I discuss the datasets and analysis methodology I make use of. The results of this analysis are presented in Sec.~\ref{sec:results}. The importance of these results in the context of early-time new physics is investigated through a case study focused on early dark energy in Sec.~\ref{sec:worked}. Finally, in Sec.~\ref{sec:conclusions} I provide concluding remarks. I recommend the very busy reader to skip to the key results presented in Tab.~\ref{tab:parametersaeiswvslcdm} and Fig.~\ref{fig:aeisw_lcdm_tri}, whereas the slightly less busy reader could also consult Tab.~\ref{tab:parametersaeiswvsextended} and Fig.~\ref{fig:aeisw_extended} if they are interested in the extended parameter space results, or Figs.~\ref{fig:plot_cmb_lcdm_ede} and~\ref{fig:plot_cmb_eisw_lcdm_ede} if they are interested in the early dark energy case study.

\section{Early ISW effect}
\label{sec:eisw}

I work in the Newtonian gauge on a spatially flat background, where the perturbed line element is characterized by the two scalar potentials $\Psi$ and $\Phi$: the former is the Newtonian gravitational potential, \textit{i.e.} the perturbation to the metric element $g_{00}=-1-2\Psi$, while the latter is the perturbation to the metric element $g_{ij}=a^2\delta_{ij}(1+2\Phi)$ (see Ref.~\cite{Ma:1995ey} for a pedagogical discussion). I denote by $\Theta \equiv \delta T_{\gamma}/\bar{T}_{\gamma}$ the relative photon temperature shift, quantifying the deviation of the CMB photon distribution from that of a perfect blackbody, and by $\Theta_{\ell}$ the $\ell$-th multipole moment of $\Theta$ (after moving to Fourier space).

Focusing on a given multipole $\ell$, $\Theta_{\ell}(k)$ receives contributions from several source terms, including gravitational redshift of CMB photons at the surface of last-scattering (Sachs-Wolfe effect), Doppler shifting, and CMB polarization. The effect I will be interested in is the integrated Sachs-Wolfe (ISW) effect, a contribution to the CMB anisotropies resulting from time-varying gravitational potentials. Since gravitational potentials are constant in a matter-dominated Universe, the ISW effect can only be active \textit{i)} at early times, when potentials decay in the presence of a non-negligible radiation component, and \textit{ii)} at late times, when their decay is due to the dark energy component responsible for cosmic acceleration. The former contribution corresponds to the early ISW (eISW) effect, while the latter corresponds to the late ISW (lISW) effect. In this paper, of the two I will only be interested in the eISW effect.

To linear order in temperature perturbations, the contribution of the eISW effect to $\Theta_{\ell}(k)$, which I denote by $\Theta_{\ell}^{\rm eISW}(k)$, is given by:
\begin{eqnarray}
\Theta_{\ell}^{\rm eISW}(k) = \int_{0}^{\eta_m}d\eta\,e^{-\tau(\eta)} \left [ \dot{\Psi}(k,\eta)-\dot{\Phi}(k,\eta) \right ]j_{\ell}(k\Delta\eta)\,, \nonumber \\
\label{eq:eisw}
\end{eqnarray}
with $\eta$ denoting conformal time, $\tau(\eta)$ the optical depth to a given conformal time, $j_{\ell}$ the spherical Bessel function of order $\ell$, and $\Delta \eta \equiv \eta-\eta_0$, where $\eta_0$ is the conformal time today. Finally, $\eta_m$ is the conformal time at an arbitrary point deep inside the matter-domination era (the exact value of $\eta_m$ is irrelevant, since gravitational potentials are constant in a matter-dominated Universe).

The eISW effect is expected to be dominant around recombination, when the contribution of radiation to the Universe's energy budget is still important. Because of this, Eq.~(\ref{eq:eisw}) can be approximated by setting the argument of $j_{\ell}$ to $k\Delta_{\rm rec} \equiv k(\eta_{\rm rec}-\eta_0)$, with $\eta_{\rm rec}$ the conformal time at recombination. If I further make the approximation $\Psi \approx -\Phi$, valid in the absence of anisotropic stress, Eq.~(\ref{eq:eisw}) is then well approximated by:
\begin{eqnarray}
\Theta_{\ell}^{\rm eISW}(k) \propto 2j_{\ell} \left ( k\Delta_{\rm rec} \right ) \left [ \Psi(k\,,\eta_m)-\Psi(k,\eta_{\rm rec}) \right ]\,.
\label{eq:eiswapprox}
\end{eqnarray}
From Eq.~(\ref{eq:eiswapprox}) one sees that the eISW effect mainly receives contributions from perturbations with wavenumber $k \sim 1/\eta_{\rm rec}$. Combined with the fact that the eISW effect adds in phase with the Sachs-Wolfe contribution to $\Theta_{\ell}$ (as the two are multiplied by the same Bessel function), this shows that the main consequence of the eISW effect is to boost the height of the first acoustic peaks, and the first one in particular.

%\begin{widetext}
%\begin{center}
\begin{figure*}[!t]
\includegraphics[width=0.7\linewidth]{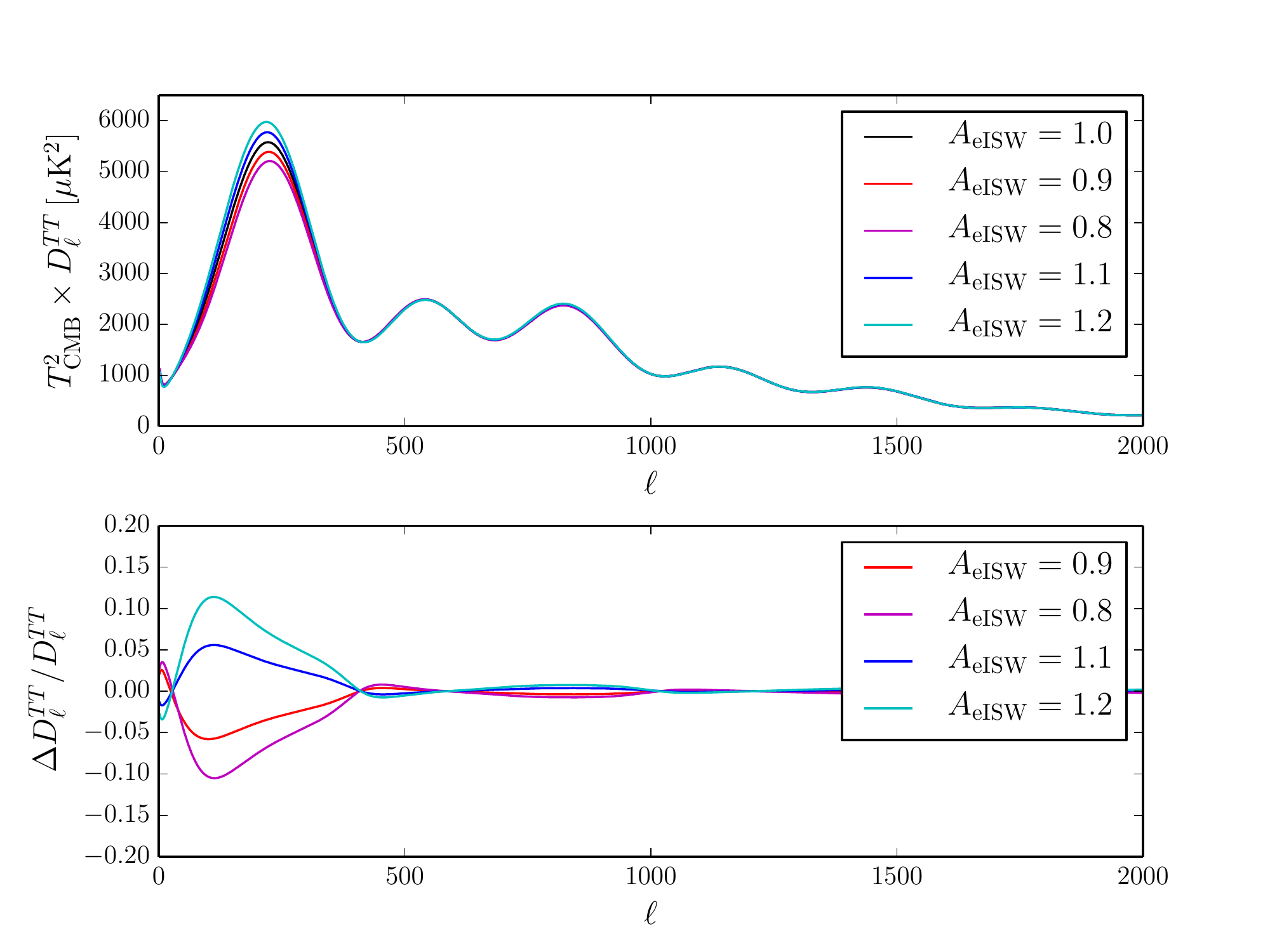}
\caption{Impact of varying $A_{\rm eISW}$ on the CMB temperature anisotropy power spectrum. \textit{Upper panel}: CMB temperature anisotropy power spectrum for different values of $A_{\rm eISW}$, specified by the color coding, with the black curve corresponding to the standard case $A_{\rm eISW}=1$. Notice that, as standard in the field, plotted on the $y$ axis is $T_{\rm CMB}^2D_{\ell}^{TT} \equiv T_{\rm CMB}^2\ell(\ell+1)C_{\ell}^{TT}$, with $T_{\rm CMB} \approx 2.725\,{\rm K}$ the CMB temperature today. \textit{Bottom panel}: relative differences between the power spectra shown in the upper panel, relative to the baseline $A_{\rm eISW}=1$ power spectrum, with the color coding as in the upper panel. It is clear that varying $A_{\rm eISW}$ mostly affects scales around the first acoustic peak, with $A_{\rm eISW}>0$ [$A_{\rm eISW}<0$] enhancing [suppressing] power.}
\label{fig:plot_aeisw_new_cmb}
\end{figure*}
%\end{center}
%\end{widetext}

My goal is now to isolate the eISW contribution to CMB power spectra, and to phenomenologically rescale this contribution by a factor whose value will then be determined by the data. Throughout this paper, I will refer to this factor as the ``eISW amplitude'', and will denote it by $A_{\rm eISW}$. To do so, I multiply the integrand of Eq.~(\ref{eq:eisw}) by a function $f(\eta)$ which takes the value $A_{\rm eISW}$ if $z(\eta)>z_t=30$, and $1$ otherwise. The choice of $z_t=30$ is purely phenomenological, and dictated by the fact that the minimum of the ISW source term occurs roughly for $z \sim 30$. I have explicitly checked that other reasonable choices of $z_t$ have no effect on my results, insofar as $z_t$ is chosen to be deep within the matter-dominated era.

Note that at least three earlier works previously introduced $A_{\rm eISW}$, which was constrained using data from WMAP7 and the South Pole Telescope (SPT)~\cite{Hou:2011ec}, from the \textit{Planck} 2015 data release~\cite{Cabass:2015xfa}, and finally from the \textit{Planck} 2018 temperature data alone~\cite{Kable:2020hcw}. It is of course worth revisiting constraints on $A_{\rm eISW}$ not only in light of the \textit{Planck} legacy data release (including polarization data, for reasons discussed towards the end of Sec.~\ref{sec:results}, and assessing the stability of the results against a different choice of high-$\ell$ \textit{Planck} likelihood~\cite{Efstathiou:2019mdh}), but especially with an eye towards early-time new physics models, given their potential to address the Hubble tension. Note, furthermore, that $A_{\rm eISW}$ bears close resemblance to the better known $A_{\rm lens}$, a phenomenological scaling parameter which rescales the amount of lensing in the CMB power spectra~\cite{Calabrese:2008rt}. \textit{Planck} primary CMB measurements exhibit a $\gtrsim 2\sigma$ preference for $A_{\rm lens}>1$~\cite{Aghanim:2018eyx}: this ``lensing anomaly'' is not supported by the amplitude of the \textit{Planck} CMB lensing power spectrum $C_{\ell}^{\phi\phi}$ reconstructed from the temperature four-point function~\cite{Aghanim:2018oex}, nor by the latest Atacama Cosmology Telescope results~\cite{Aiola:2020azj}, and might be the result of a statistical fluctuation~\cite{Efstathiou:2019mdh}. Much as $A_{\rm lens}$, $A_{\rm eISW}$ is also a purely phenomenological scaling parameter: rather than as being standard cosmological parameters, $A_{\rm eISW}$ and $A_{\rm lens}$ should rather be viewed as consistency test parameters.

In Fig.~\ref{fig:plot_aeisw_new_cmb}, I show the effect of varying $A_{\rm eISW}$ on the CMB temperature power spectrum, while keeping all the other cosmological parameters fixed. Relative to the standard case $A_{\rm eISW}=1$, from Fig.~\ref{fig:plot_aeisw_new_cmb} one sees that the main effect of varying $A_{\rm eISW}$ is to enhance [suppress] power for $A_{\rm eISW}>1$ [$A_{\rm eISW}<1$], with the change in power being most evident for the first peak, as I anticipated earlier from Eq.~(\ref{eq:eiswapprox}). This is clear from the bottom panel of Fig.~\ref{fig:plot_aeisw_new_cmb}, showing that the largest relative enhancement/suppression of power occurs at multipoles $\ell \sim 100$, \textit{i.e.} around the first peak. Varying $A_{\rm eISW}$ also has a subdominant effect on the higher acoustic peaks.

In the rest of this paper, rather than fixing $A_{\rm eISW}$ to the standard value $A_{\rm eISW}=1$, I will allow the eISW amplitude to vary and let the data constrain this parameter, first within a seven-parameter $\Lambda$CDM+$A_{\rm eISW}$ model, and then within extensions thereof. This will serve as an important consistency test of $\Lambda$CDM. Inferring a value of $A_{\rm eISW}$ consistent with $1$ would indicate that the $\Lambda$CDM prediction for the eISW effect fits the \textit{Planck} data remarkably well, which in turn could pose a challenge for early-time new physics. Conversely, any deviation from $A_{\rm eISW}=1$ could provide clues as to where and what kinds of early-time modifications are necessary or even allowed. This consistency test can therefore act as a signpost in the large theoretical parameter space of proposed early-time modifications to $\Lambda$CDM motivated by the Hubble tension.

\section{Datasets and methodology}
\label{sec:data}

In my baseline analysis, I only make use of measurements of CMB temperature anisotropy and polarization power spectra, as well as their cross-spectra, from the \textit{Planck} 2018 legacy data release. In particular, I combine the high-$\ell$ \texttt{Plik} likelihood for TT ($30 \leq \ell \lesssim 2500$) as well as TE and EE ($30 \leq \ell \lesssim 2000$), the low-$\ell$ TT-only ($2 \leq \ell < 29$) likelihood based on the \texttt{Commander} component-separation algorithm in pixel space, and the low-$\ell$ EE-only ($2 \leq \ell <29$) \texttt{SimAll} likelihood~\cite{Aghanim:2019ame}. This combination is typically referred to as \textit{Planck TTTEEE+lowE} by the Planck collaboration, while I will refer to it as \textit{\textbf{Planck}}. Where deemed necessary, I will explicitly remind the reader that it is the high-$\ell$ \texttt{Plik} likelihood which I am making use of, referring to the full CMB dataset as \textit{\textbf{Planck (\texttt{Plik})}}.

To assess the robustness of my results, I will also consider combinations of the \textit{Planck} dataset with the following external datasets:
\begin{itemize}
\item Big Bang Nucleosynthesis (BBN) prior on $100\omega_b = 2.233 \pm 0.036$, informed by the latest improved measurement of the rate of deuterium burning by LUNA~\cite{Mossa:2020gjc}. I refer to this dataset/prior as \textit{\textbf{BBN}}.
\item Baryon Acoustic Oscillation measurements from the 6dFGS~\cite{Beutler:2011hx}, SDSS-MGS~\cite{Ross:2014qpa}, and BOSS DR12~\cite{Alam:2016hwk} surveys. I refer to this dataset as \textit{\textbf{BAO}}.
\item Distance moduli measurements in the range $0.01<z<2.3$ from the \textit{Pantheon} SNeIa sample~\cite{Scolnic:2017caz}. I refer to this dataset as \textit{\textbf{Pantheon}}.
\end{itemize}
In particular, I will consider the constraints obtained within the \textit{Planck}+\textit{BBN} and \textit{Planck}+\textit{BAO}+\textit{Pantheon} dataset combinations, which I will then compare to the \textit{Planck}-only constraints.

Finally, to further assess the robustness of my results, I replace the \texttt{Plik} TTTEEE likelihood with the \texttt{CamSpec 12.5HMcl} likelihood, which has access to a larger sky fraction, and is hence statistically more powerful (see Ref.~\cite{Efstathiou:2019mdh} for a detailed discussion). When the \texttt{CamSpec 12.5HMcl} likelihood is combined with the \textit{Commander} low-$\ell$ TT and \textit{SimAll} low-$\ell$ EE likelihoods, I refer to the resulting combination as \textit{\textbf{Planck (\texttt{CamSpec})}}.

In terms of models, I begin by considering a seven-parameter model extending the six-parameter $\Lambda$CDM model by allowing the eISW amplitude $A_{\rm eISW}$ to vary. I refer to this model as $\Lambda$CDM+$A_{\rm eISW}$. I set an uniform prior in the range $A_{\rm eISW} \in [0;2]$.

To assess the robustness of my results against extensions of this minimal parameter space, I extend the $\Lambda$CDM+$A_{\rm eISW}$ model by varying one or two additional parameters which are otherwise fixed to standard values. In particular, the following parameters are varied in addition to the 7 standard ones:
\begin{itemize}
\item the effective number of relativistic degrees of freedom $N_{\rm eff}$, otherwise fixed to $N_{\rm eff}=3.046$;
\item the primordial Helium abundance $Y_P$, otherwise fixed to the value obtained from standard BBN given the values of $\omega_b$ and $N_{\rm eff}$ (in other words, when varying $Y_P$, I set \texttt{bbn\_consistency=F});
\item the lensing amplitude $A_{\rm lens}$, otherwise fixed to $A_{\rm lens}=1$;
\item the running of the scalar spectral index $\alpha_s \equiv dn_s/d\ln k$, otherwise fixed to $\alpha_s=0$;
\item the running of the running of the scalar spectral index $\beta_s \equiv d\alpha_s/d\ln k = d^2n_s/d(\ln k)^2$, otherwise fixed to $\beta_s=0$.
\end{itemize}
Therefore, I consider in total 5 extended models. Uniform priors are set on $N_{\rm eff} \in [0;10]$, $Y_P \in [0.1;0.5]$, $A_{\rm lens} \in [0;10]$, $\alpha_s \in [-1;1]$, and $\beta_s \in [-1;1]$, as done by the \textit{Planck} collaboration. Note that when $\beta_s$ is varied, I vary $\alpha_s$ as well.

Theoretical predictions for the CMB power spectra and the background expansion are obtained using the Boltzmann solver \texttt{CAMB}~\cite{Lewis:1999bs}. I use Monte Carlo Markov Chain (MCMC) methods to sample the posterior distributions for the parameters of the 6 cosmological models considered ($\Lambda$CDM+$A_{\rm eISW}$ as well as the 5 extensions thereof), using the cosmological sampler \texttt{CosmoMC}~\cite{Lewis:2002ah} to generate the MCMC chains. I assess the convergence of the MCMC chains by using the Gelman-Rubin parameter $R-1$~\cite{Gelman:1992zz}, and set the requirement $R-1<0.03$ for the MCMC chains to be considered converged.

\section{Results}
\label{sec:results}

I begin by discussing the results obtained within the baseline seven-parameter $\Lambda$CDM+$A_{\rm eISW}$ model from the \textit{Planck} (\texttt{Plik}) dataset alone. In this case, I find $A_{\rm eISW}=0.988 \pm 0.027$ at 68\% confidence level (C.L.), perfectly consistent with the standard $\Lambda$CDM expectation of $A_{\rm eISW}=1$. The inferred values of the other 6 parameters are reported in Tab.~\ref{tab:parametersaeiswvslcdm} (right column), and compared to their values inferred within the $\Lambda$CDM model (where $A_{\rm eISW}=1$ is fixed).

\begin{table*}[!ht]
\centering
\scalebox{1.2}{
\begin{tabular}{|c||cc|}       
\hline\hline
Parameter & \multicolumn{2}{c|}{\textit{Planck}} \\
 & $\Lambda$CDM & $\Lambda$CDM+$A_{\rm eISW}$ \\ \hline
100$\omega_b$ & $2.235 \pm 0.015$ & $2.241 \pm 0.020$ \\
$\omega_c$ & $0.1202 \pm 0.0013$ & $0.1203 \pm 0.0014$ \\
$\theta_s$ & $1.0409 \pm 0.0003$ & $1.0409 \pm 0.0003$ \\
$\tau$ & $0.0544 \pm 0.0078$ & $0.0541 \pm 0.0078$ \\
$\ln(10^{10}A_s)$ & $3.045 \pm 0.016$ & $3.046 \pm 0.016$ \\
$n_s$ & $0.965 \pm 0.004$ & $0.963 \pm 0.005$ \\
$A_{\rm eISW}$ & $1.0$ & $0.988 \pm 0.027$ \\
\hline
$H_0\,[{\rm km}/{\rm s}/{\rm Mpc}]$ & $67.26 \pm 0.57$ & $67.28 \pm 0.62$ \\
$\Omega_m$ & $0.317 \pm 0.008$ & $0.317 \pm 0.009$ \\
\hline \hline                                                  
\end{tabular}}
\caption{68\%~C.L. constraints on the cosmological parameters of the six-parameter $\Lambda$CDM (left column) and seven-parameter $\Lambda$CDM+$A_{\rm eISW}$ (right column) models, obtained from the \textit{Planck} (\texttt{Plik}) dataset alone. The eISW amplitude $A_{\rm eISW}$ is fixed to the standard value $A_{\rm eISW}=1$ within the $\Lambda$CDM model. The final two rows are separated from the previous rows to highlight the fact that the two associated parameters ($H_0$ and $\Omega_m$) are derived parameters.}
\label{tab:parametersaeiswvslcdm}                                              
\end{table*}

From Tab.~\ref{tab:parametersaeiswvslcdm}, it is clear that the inferred values of all 6 $\Lambda$CDM parameters are extremely stable against the extension where $A_{\rm eISW}$ is allowed to vary, with their uncertainties in most cases barely increasing. This result indicates a simple but very important fact: $\Lambda$CDM's prediction for the amplitude of the eISW effect agrees perfectly with the \textit{Planck} data. Conversely, as far as the eISW effect is concerned, there is no obvious sign or need for new physics beyond $\Lambda$CDM. As I discussed earlier, and will discuss again later in the context of EDE, this simple observation has important consequences for early-time modifications to $\Lambda$CDM, include modifications motivated by attempts to solve the Hubble tension.

The largest parameters shifts are observed for the two parameters most strongly correlated with $A_{\rm eISW}$, \textit{i.e.} $\omega_b$ and $n_s$, which however shift by no more than $0.3\sigma$ when allowing $A_{\rm eISW}$ to vary. In particular, I find $A_{\rm eISW}$ to be negatively correlated with $\omega_b$ and positively correlated with $n_s$. Overall, the most noticeable effect of allowing $A_{\rm eISW}$ to vary is a $\approx 30\%$ broadening of the uncertainty on $\omega_b$. The reason why $A_{\rm eISW}$ is most strongly correlated with $\omega_b$ and $n_s$ can be understood by recalling the effect of these two parameters on the CMB temperature power spectrum (see e.g. the top right and bottom left panels of Fig.~4.5 of Ref.~\cite{Vagnozzi:2019utt}). Increasing $\omega_b$ increases the relative odd/even peak height, with the effect being most noticeable especially for the first peak: the effect is therefore similar to that of increasing $A_{\rm eISW}$, which explains the negative correlation between these two parameters. Similarly, increasing $n_s$ reduces power on large angular scales (while increasing it on small angular scales), an effect which for scales around the first peak is similar to that of decreasing $A_{\rm eISW}$, explaining the positive correlation between these two parameters. Moreover, the mild positive correlation existing between $\omega_b$ and $n_s$ within $\Lambda$CDM is considerably reduced, to the extent of there being almost no correlation, when varying $A_{\rm eISW}$.

%\begin{figure}[!ht]
%\includegraphics[width=1.0\linewidth]{aeisw.pdf}
%\caption{Results}
%\label{fig:aeisw}
%\end{figure}

%\begin{figure}[!ht]
%\includegraphics[width=1.0\linewidth]{aeisw_tri.pdf}
%\caption{Results}
%\label{fig:aeisw_tri}
%\end{figure}

\begin{figure}[!ht]
\includegraphics[width=1.0\linewidth]{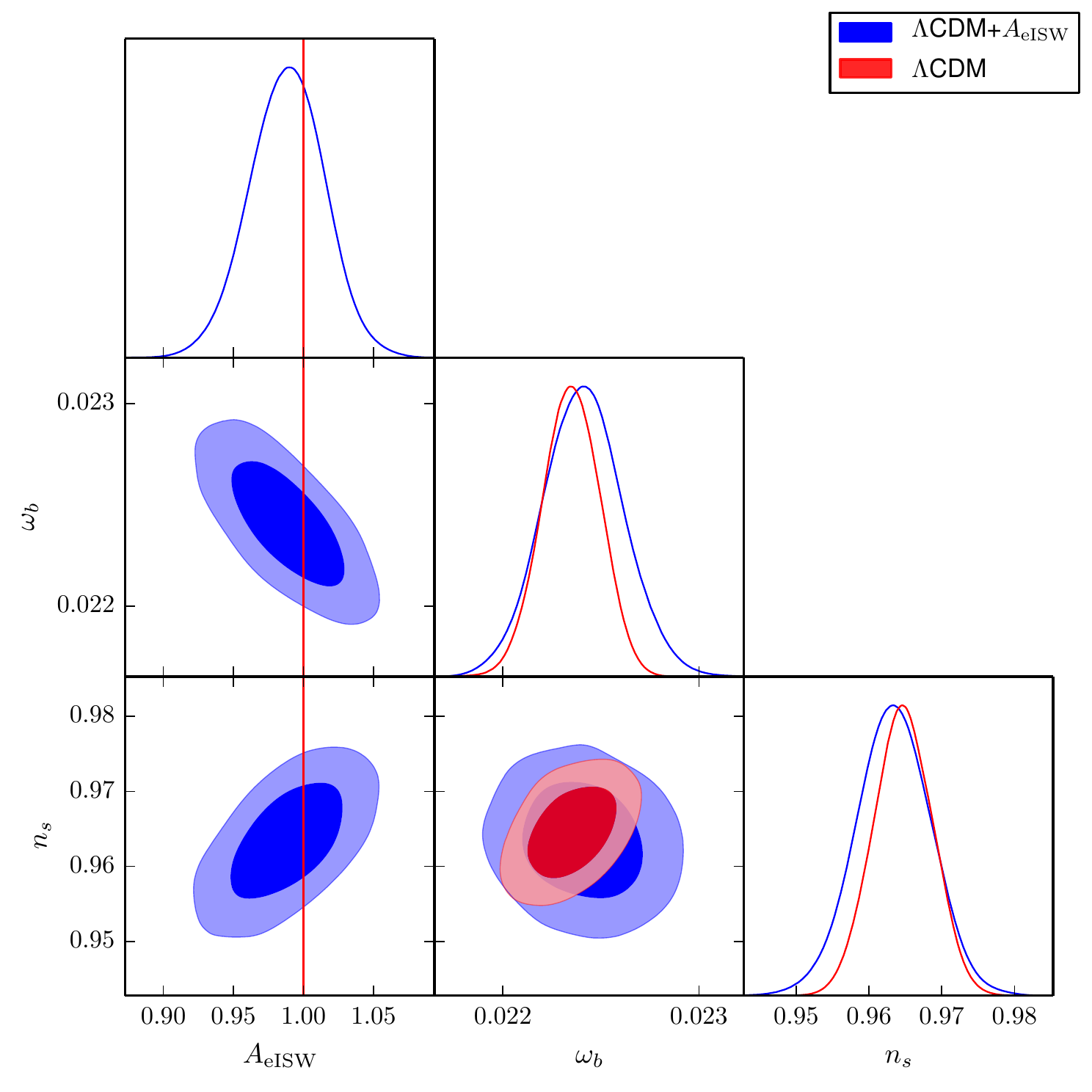}
\caption{Triangular plot showing 2D joint and 1D marginalized posterior probability distributions for the eISW amplitude $A_{\rm eISW}$, the physical baryon density $\omega_b$, and the scalar spectral index $n_s$, obtained from a fit to the \textit{Planck} dataset. The blue contours are obtained within the seven-parameter $\Lambda$CDM+$A_{\rm eISW}$ model, while the red contours are obtained within the six-parameter $\Lambda$CDM model, where $A_{\rm eISW}$ is fixed to $A_{\rm eISW}=1$, as indicated by the vertical red line.}
\label{fig:aeisw_lcdm_tri}
\end{figure}

I now assess the robustness of my results against \textit{i)} the inclusion of additional external datasets, or \textit{ii)} the use of the \texttt{CamSpec} likelihood in place of \texttt{Plik} in analyzing the high-$\ell$ \textit{Planck} measurements. The results are summarized in Tab.~\ref{tab:parametersplanckvsexternal}. When adding the \textit{BBN} prior on $\omega_b$ to the baseline \textit{Planck} dataset, the largest improvement is as expected on $\omega_b$, whose uncertainty decreases by $\approx 10\%$, with the \textit{BBN} prior cutting out part of the parameter space at higher $\omega_b$. Given the direction of the $A_{\rm eISW}$-$\omega_b$ degeneracy discussed earlier, this results in part of the parameter space at lower $A_{\rm eISW}$ being cut, as well as the corresponding parameter uncertainty slightly decreasing. From the \textit{Planck}+\textit{BBN} dataset combination, I infer $A_{\rm eISW}=0.993 \pm 0.025$, which further improves the consistency with the standard value $A_{\rm eISW}=1$.

\begin{table*}[!t]
\centering
\scalebox{1.2}{
\begin{tabular}{|c||cccc|}       
\hline\hline
Parameter & \multicolumn{4}{c|}{$\Lambda$CDM+$A_{\rm eISW}$} \\
 & \textit{Planck} (\texttt{Plik}) & \textit{Planck} (\texttt{CamSpec}) & \textit{Planck}+\textit{BBN} & \textit{Planck}+\textit{BAO}+\textit{Pantheon} \\ \hline
100$\omega_b$ & $2.241 \pm 0.020$ & $2.219 \pm 0.020$ & $2.239 \pm 0.018$ & $2.251 \pm 0.020$ \\
$\omega_c$ & $0.1203 \pm 0.0014$ & $0.1197 \pm 0.0013$ & $0.1203 \pm 0.0013$ & $0.1192 \pm 0.0010$ \\
$\theta_s$ & $1.0409 \pm 0.0003$ & $1.0411 \pm 0.0003$ & $1.0409 \pm 0.0003$ & $1.0410 \pm 0.0003$ \\
$\tau$ & $0.0541 \pm 0.0078$ & $0.0527 \pm 0.0083$ & $0.0548 \pm 0.0076$ & $0.0557 \pm 0.0081$ \\
$\ln(10^{10}A_s)$ & $3.046 \pm 0.016$ & $3.038 \pm 0.017$ & $3.047 \pm 0.016$ & $3.047 \pm 0.016$ \\
$n_s$ & $0.963 \pm 0.005$ & $0.969 \pm 0.006$ & $0.964 \pm 0.005$ & $0.966 \pm 0.005$ \\
$A_{\rm eISW}$ & $0.988 \pm 0.027$ & $1.016 \pm 0.027$ & $0.993 \pm 0.025$ & $0.986 \pm 0.027$ \\
\hline
$H_0\,[{\rm km}/{\rm s}/{\rm Mpc}]$ & $67.28 \pm 0.62$ & $67.37 \pm 0.57$ & $67.26 \pm 0.56$ & $67.80 \pm 0.47$ \\
$\Omega_m$ & $0.317 \pm 0.009$ & $0.314 \pm 0.008$ & $0.317 \pm 0.008$ & $0.310 \pm 0.006$ \\
\hline \hline                                                  
\end{tabular}}
\caption{68\%~C.L. constraints on the cosmological parameters of the seven-parameter $\Lambda$CDM+$A_{\rm eISW}$ model, obtained from various datasets/dataset combinations as indicated in the upper section of the Table.}
\label{tab:parametersplanckvsexternal}                                              
\end{table*}

Qualitatively similar results are obtained when considering the \textit{Planck}+\textit{BAO}+\textit{Pantheon} dataset combination. While larger parameter shifts are observed compared to the previous case, these remain overall small, and still well below the $1\sigma$ level. In particular, I infer $A_{\rm eISW}=0.986 \pm 0.027$, still perfectly consistent with the standard value $A_{\rm eISW}=1$.

Finally, I consider again \textit{Planck} data alone, but this time adopting the \texttt{CamSpec} likelihood in place of the \texttt{Plik} high-$\ell$ (\textit{TTTEEE}) one. In this case, I find larger parameter shifts, although again these remain all relatively small, always below the $1\sigma$ level. The magnitude of these shifts are consistent with those reported by the \textit{Planck} collaboration~\cite{Aghanim:2018eyx} and investigated in Ref.~\cite{Efstathiou:2019mdh}. The largest observed shift is a $\approx 0.8\sigma$ shift towards lower values of $\omega_b$. Again, given the previously discussed mutual degeneracies between $\omega_b$, $n_s$, and $A_{\rm eISW}$, this shift in $\omega_b$ is unsurprisingly accompanied by correlated shifts in $n_s$ and $A_{\rm eISW}$ towards larger values. Overall, the inferred value of $A_{\rm eISW}=1.016 \pm 0.027$ remains in excellent agreement with the standard value $A_{\rm eISW}=1$.

\begin{figure}[!ht]
\includegraphics[width=1.0\linewidth]{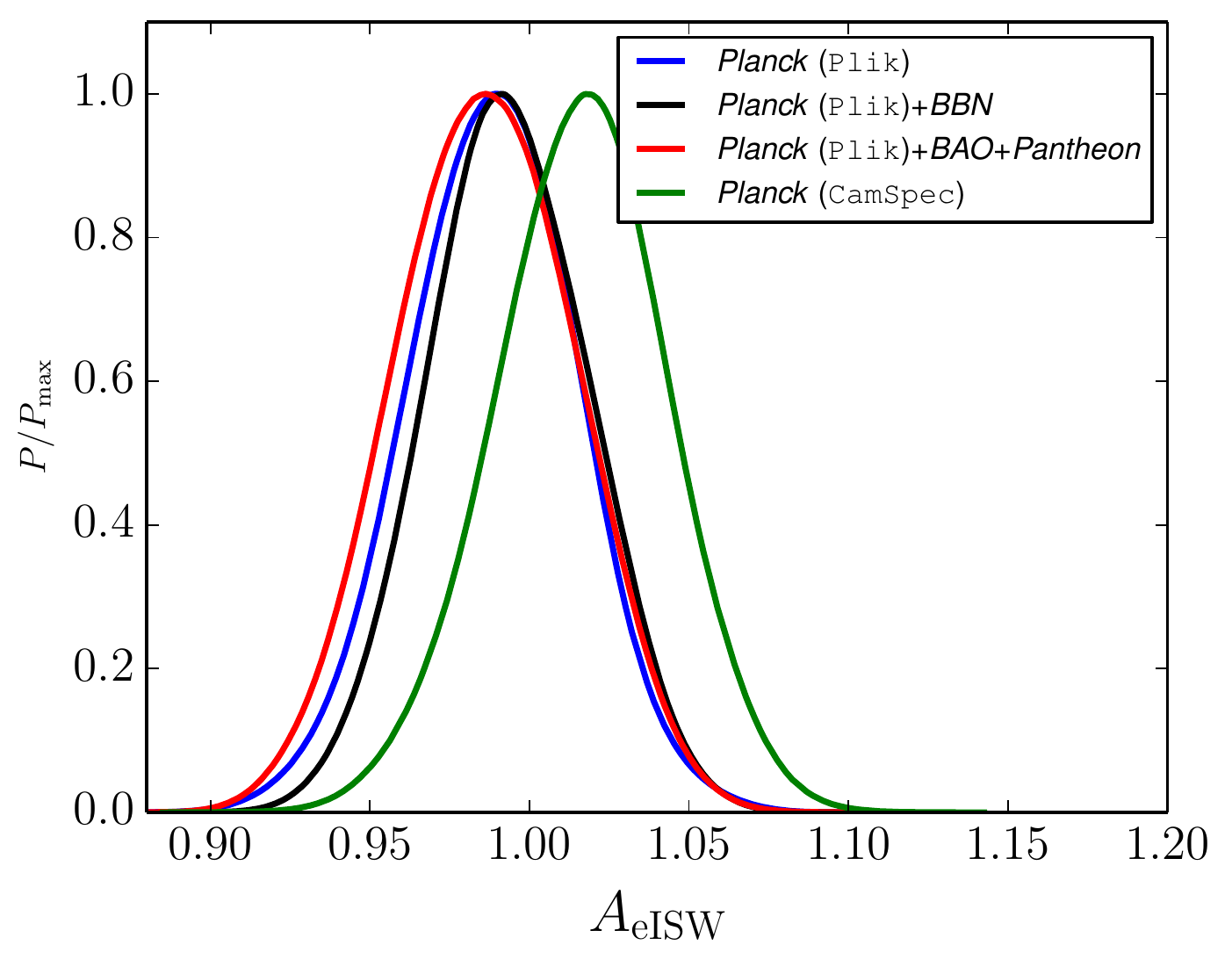}
\caption{1D marginalized normalized posterior distribution for $A_{\rm eISW}$, obtained within the baseline seven-parameter $\Lambda$CDM+$A_{\rm eISW}$ model, from four different datasets/dataset combinations: the baseline \textit{Planck} data using the high-$\ell$ \texttt{Plik} likelihood (blue curve); a combination of \textit{Planck} and a BBN prior on $\omega_b$ (black curve); a combination of \textit{Planck} with late-time BAO and \textit{Pantheon} Hubble flow SNeIa measurements (red curve); and the baseline \textit{Planck} data, replacing the high-$\ell$ \texttt{Plik} likelihood with the \texttt{CamSpec} one (green curve). It is clear that the inferred value of $A_{\rm eISW}$ is overall very stable against these choices of datasets/dataset combinations.}
\label{fig:aeisw_robustness}
\end{figure}

The main message is therefore that the \textit{Planck}-only result, which sees remarkable consistency between the inferred value of $A_{\rm eISW}=0.988 \pm 0.027$ and the standard value $A_{\rm eISW}=1$, is very stable against \textit{i)} the addition of external datasets (\textit{BBN} and \textit{BAO}+\textit{Pantheon}), and \textit{ii)} the choice of high-$\ell$ \textit{Planck} likelihood (\texttt{Plik} or \texttt{CamSpec}). A visual representation of these results is given in Fig.~\ref{fig:aeisw_robustness}, where I plot the 1D marginalized $A_{\rm eISW}$ posterior distributions, within the $\Lambda$CDM+$A_{\rm eISW}$ model, given all the dataset combinations/choices discussed.

I now assess the robustness of the previous results against extended parameter spaces, considering the one- and two-parameter extensions discussed in Sec.~\ref{sec:data}. The results are reported in Tab.~\ref{tab:parametersaeiswvsextended}, focusing only on $\omega_b$, $n_s$, $A_{\rm eISW}$, and of course the additional parameters themselves. I have chosen to vary $N_{\rm eff}$, $Y_P$, $A_{\rm lens}$, $\alpha_s$, and $\beta_s$, as these are the parameters I expect to be most strongly correlated with $A_{\rm eISW}$ (for reasons similar to those previously discussed for $\omega_b$ and $n_s$).

\begin{table*}[!t]
\centering
\scalebox{1.0}{
\begin{tabular}{|c||cccccc|}       
\hline\hline
Parameter & \multicolumn{6}{c|}{\textit{Planck}} \\
 & $\Lambda$CDM+$A_{\rm eISW}$ & +$N_{\rm eff}$ & +$Y_P$ & +$A_{\rm lens}$ & +$\alpha_s$ & +$\alpha_s$+$\beta_s$ \\ \hline
100$\omega_b$ & $2.241 \pm 0.020$ & $2.225 \pm 0.028$ & $2.234 \pm 0.025$ & $2.275 \pm 0.024$ & $2.243 \pm 0.020$ & $2.243 \pm 0.021$ \\
$n_s$ & $0.963 \pm 0.005$ & $0.958 \pm 0.009$ & $0.961 \pm 0.008$ & $0.969 \pm 0.005$ & $0.963 \pm 0.006$ & $0.958 \pm 0.007$ \\
$A_{\rm eISW}$ & $0.988 \pm 0.027$ & $0.994 \pm 0.029$ & $0.991 \pm 0.029$ & $0.975 \pm 0.027$ & $0.992 \pm 0.027$ & $0.978 \pm 0.030$ \\
$X$ & -- & $2.89 \pm 0.19$ & $0.239 \pm 0.013$ & $1.192 \pm 0.065$ & $-0.006 \pm 0.007$ & \small{$0.005 \pm 0.011\,,0.017 \pm 0.014$} \\
$X$ ($A_{\rm eISW}=1$) & -- & $2.92 \pm 0.19$ & $0.240 \pm 0.013$ & $1.180 \pm 0.065$ & $-0.006 \pm 0.007$ & \small{$0.001 \pm 0.010\,,0.012 \pm 0.013$} \\
\hline \hline                                                  
\end{tabular}}
\caption{As in Tab.~\ref{tab:parametersplanckvsexternal}, but focused on one- and two-parameter extensions of the baseline seven-parameter $\Lambda$CDM+$A_{\rm eISW}$ model, with the extra parameters being: the effective number of relativistic species $N_{\rm eff}$; the primordial Helium fraction $Y_P$; the lensing amplitude $A_{\rm lens}$; the running of the scalar spectral index $\alpha_s$; and the running of the running of the scalar spectral index $\beta_s$. Note that whenever $\beta_s$ is varied, $\alpha_s$ is varied as well. These extra parameters are indicated by $X$ in the parameters column. The row labeled ``$X$ ($A_{\rm eISW}=1$)'' reports constraints on the extra parameter(s) $X$ within the seven- or eight-parameter $\Lambda$CDM+$X$ model, where $A_{\rm eISW}$ is fixed to the standard value $A_{\rm eISW}=1$.}
\label{tab:parametersaeiswvsextended}                                              
\end{table*}

Overall, I find that the inferred value of $A_{\rm eISW}$ is very stable against these extensions. The largest shifts are observed when allowing $A_{\rm lens}$ to vary. This is hardly surprising, given that the \textit{Planck} data appear to show a moderate preference for $A_{\rm lens}>1$: this ``lensing anomaly'' (and the closely related apparent preference for a spatially closed Universe from \textit{Planck} primary CMB data) is an issue which is well-known and well-documented in the literature (see e.g. Refs.~\cite{Park:2017xbl,Aghanim:2019ame,Handley:2019tkm,Efstathiou:2019mdh,DiValentino:2019qzk,Efstathiou:2020wem,DiValentino:2020hov,DiValentino:2020evt,DiValentino:2020srs,Vagnozzi:2020rcz,DiValentino:2020kpf,Vagnozzi:2020dfn,Cao:2021ldv,Dhawan:2021mel,Gonzalez:2021ojp}). However, even within the $\Lambda$CDM+$A_{\rm eISW}$+$A_{\rm lens}$ model, the associated parameter shifts relative to the $\Lambda$CDM+$A_{\rm eISW}$ model (with $A_{\rm lens}=1$ fixed) remain small. The largest observed shift is for $\omega_b$, which increases by $\approx 1\sigma$. On the other hand, $A_{\rm eISW}$ decreases by $\approx 0.3\sigma$ to $A_{\rm eISW}=0.975 \pm 0.027$, but still remains perfectly consistent with the standard value $A_{\rm eISW}=1$ within better than $1\sigma$.

Shifts of comparable magnitude are observed in $n_s$ and $A_{\rm eISW}$ when allowing $\alpha_s$ and $\beta_s$ to vary. In particular, within the $\Lambda$CDM+$A_{\rm eISW}$+$\alpha$+$\beta_s$ model, I infer $A_{\rm eISW}=0.978 \pm 0.030$, which also remains perfectly consistent with the standard value $A_{\rm eISW}=1$ within better than $1\sigma$, accompanied by $\approx 1\sigma$ hints for non-zero $\beta_s$. As discussed in Ref.~\cite{Cabass:2016ldu}, the $1\sigma$ hint for non-zero $\beta_s$, already reported by the \textit{Planck} collaboration~\cite{Aghanim:2018eyx}, is due to the fact that a positive $\beta_s$ provides a slightly better fit to the low-$\ell$ part of the CMB power spectrum. While remaining consistent with $\alpha_s=0$ within $1\sigma$, the slight preference for negative $\alpha_s$ when this parameter is varied and $\beta_s$ is fixed is instead related to the mild tensions between high-$\ell$ and low-$\ell$ multipoles in the CMB temperature power spectrum, discussed in Refs.~\cite{Addison:2015wyg,Aghanim:2018eyx}.

\begin{figure}[!ht]
\includegraphics[width=1.0\linewidth]{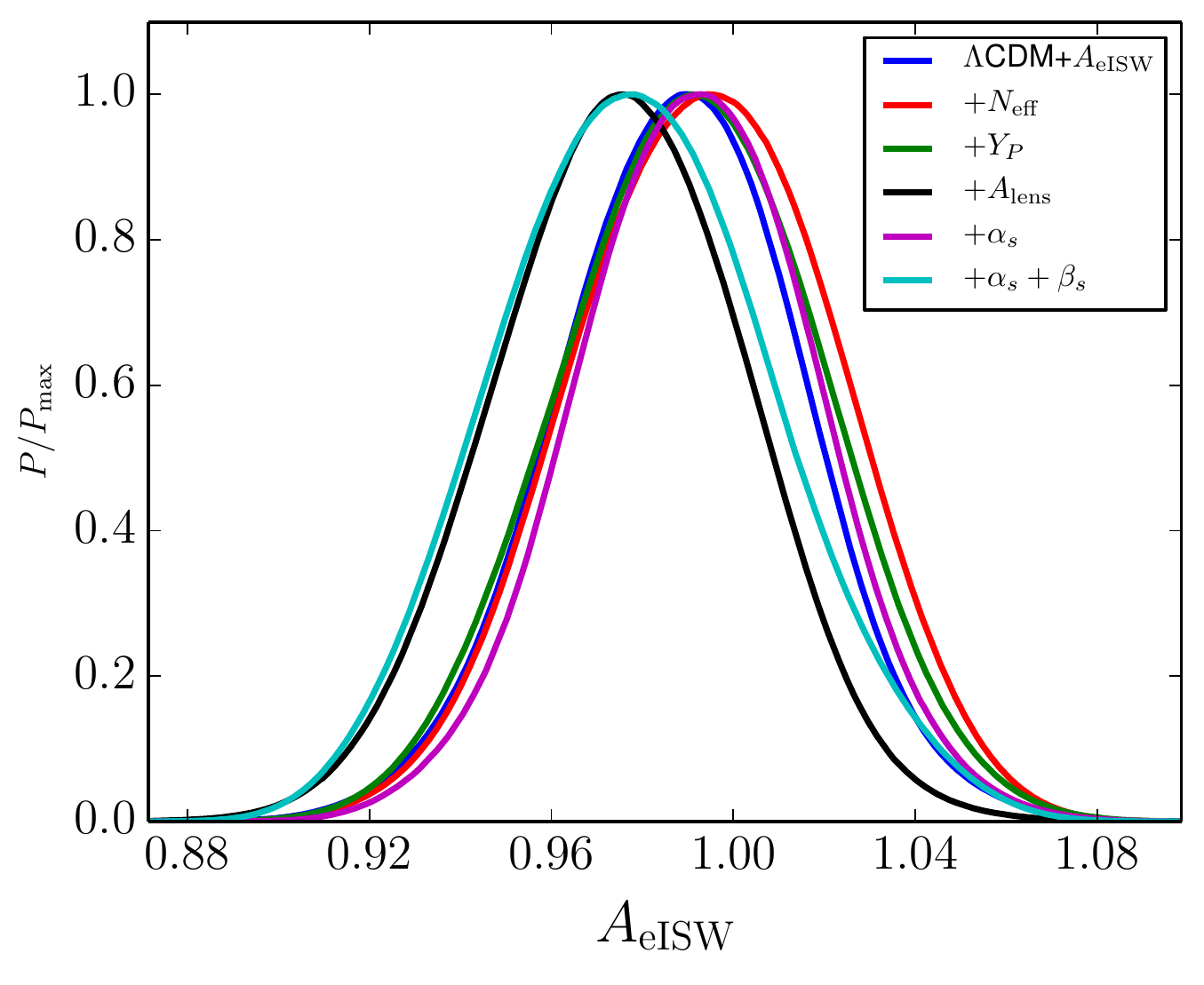}
\caption{1D marginalized normalized posterior distribution for $A_{\rm eISW}$, obtained from the \textit{Planck} data, within the baseline seven-parameter $\Lambda$CDM+$A_{\rm eISW}$ model (blue curve), as well as extensions where the following parameters are allowed to vary as well: the effective number of relativistic species $N_{\rm eff}$ (red curve), the Helium fraction $Y_P$ (green curve), the lensing amplitude $A_{\rm lens}$ (black curve), the running of the scalar spectral index $\alpha_s$ (magenta curve), and both the running $\alpha_s$ and the running of the running of the scalar spectral index $\beta_s$ (cyan curve). It is clear that the inferred value of $A_{\rm eISW}$ is overall very stable against these parameter space extensions.}
\label{fig:aeisw_extended}
\end{figure}

Overall, the main message here is that the inferred value of $A_{\rm eISW}$ being highly consistent with the standard value $A_{\rm eISW}=1$, obtained within the baseline $\Lambda$CDM+$A_{\rm eISW}$ model, is very stable against the minimal one- and two-parameter extensions I have studied. A visual representation of these results is given in Fig.~\ref{fig:aeisw_extended}, where I plot the 1D marginalized $A_{\rm eISW}$ posterior distributions, obtained from the \textit{Planck} dataset, given the baseline and extended models considered.

These results place important restrictions on early-time new physics models, for instance those invoked to address the Hubble tension. Clearly, in order to fit the CMB data well, these models will need to make a prediction for the eISW effect which is similar, or at least not too distant, from that of $\Lambda$CDM. Since a non-negligible amount of early-time new physics is required to significantly alleviate the Hubble tension, the resulting impact on the eISW effect is expected to be equally non-negligible, thereby posing a significant challenge to these models. One possible solution is that of shifting the values of some of the standard $\Lambda$CDM parameters (such as $\omega_c$) in order to re-adjust the eISW amplitude, a shift which however may come at the price of degrading the fit to other datasets. In the following Section, I will provide a case study of early dark energy, with the goal of illustrating the challenges it faces in relation to the eISW effect, as already anticipated in Section~\ref{sec:introduction}.

Note that at least three earlier works previously introduced $A_{\rm eISW}$, constrained to $A_{\rm eISW}=0.979 \pm 0.055$ from WMAP7+SPT~\cite{Hou:2011ec}, to $A_{\rm eISW}=1.06 \pm 0.04$ from the \textit{Planck} 2015 temperature and large-scale polarization data~\cite{Cabass:2015xfa}, and to $A_{\rm eISW}=1.064 \pm 0.042$ from the \textit{Planck} 2018 temperature data alone~\cite{Kable:2020hcw}. It is of course worth revisiting these constraints \textit{i)} in light of the full \textit{Planck} 2018 legacy release data (including polarization data and assessing the stability of the results against a different choice of high-$\ell$ \textit{Planck} likelihood~\cite{Efstathiou:2019mdh}), and \textit{ii)} in view of the Hubble tension and the possible implications of the results for early-time new physics. Concerning \textit{i)}, particularly crucial are the inclusion of small-scale \textit{TTTEEE} polarization data, and the significant improvements which led to the \texttt{SimAll} large-scale \textit{EE} likelihood. As discussed in detail in Ref.~\cite{Galli:2014kla}, the TE cross-spectrum is remarkably good at constraining $\omega_b$, while the EE power spectrum is extremely good at constraining $\omega_b$, $\omega_c$, and $n_s$, and in particular at breaking the $\omega_b$-$n_s$ degeneracy (due to the fact that the amplitude of the EE power spectrum is sensitive to $c_s^2$, the square of the photon-baryon sound speed). This is particularly true when the TE and EE spectra are combined with measurements of the TT power spectrum, as this combination is very efficient at breaking degeneracies between the main cosmological parameters. However, EE alone is already able to constrain $\omega_b$, $\omega_c$, and $n_s$ as well as or better than TT, even at a lower signal-to-noise level~\cite{Galli:2014kla}.

For the reasons discussed above, the use of the latest large- and small-scale polarization data, included in this analysis, is crucial to further improve constraints on $A_{\rm eISW}$. This results in an uncertainty which is reduced by up to a factor of $2$ compared to the results of the earlier Refs.~\cite{Hou:2011ec,Cabass:2015xfa,Kable:2020hcw}, while the central value of $A_{\rm eISW}$ is also in better agreement with the standard value $A_{\rm eISW}=1$.

Future Stage III and Stage IV CMB experiments (mostly ground-based) will significantly improve CMB power spectrum measurements compared to \textit{Planck}, especially insofar as the E-mode power spectrum is concerned, significantly improving constraints on the base $\Lambda$CDM parameters. As far as $\omega_b$ and $n_s$ are concerned, their determinations will improve by up to a factor of 2 for Stage III experiments such as \textit{Simons Observatory} (see Tab.~3 of Ref.~\cite{Ade:2018sbj}) and by up to a factor of 3 or more for planned Stage IV experiments such as CMB-S4 (see Tab.~8-1 of Ref.~\cite{Abazajian:2016yjj}). While a fully-fledged forecast is beyond the scope of this paper, I expect comparable improvements in the achievable $A_{\rm eISW}$ sensitivity. This means that $A_{\rm eISW}$ might potentially be inferred to $\lesssim 1\%$ precision from Stage IV experiments: this would set even more important limits on the extent to which early-time new physics can operate.

\section{Early ISW effect and new physics: case study with early dark energy}
\label{sec:worked}

\begin{table*}[!ht]
\centering
\scalebox{1.2}{
\begin{tabular}{|c||ccc|}       
\hline\hline
Parameter & $\Lambda$CDM & EDE (high $\omega_c$) & EDE (low $\omega_c$) \\ \hline
100$\omega_b$ & $2.253$ & $2.253$ & $2.253$ \\
$\omega_c$ & $0.1177$ & $0.1322$ & $0.1177$ \\
$H_0\,[{\rm km}/{\rm s}/{\rm Mpc}]$ & $68.21$ & $72.19$ & $72.19$ \\
$\tau$ & $0.085$ & $0.072$ & $0.072$ \\
$\ln(10^{10}A_s)$ & $3.0983$ & $3.0978$ & $3.0978$ \\
$n_s$ & $0.9686$ & $0.9889$ & $0.9889$ \\
$f_{\rm EDE}$ & -- & 0.122 & 0.122 \\
$\log_{10}z_c$ & -- & 3.562 & 3.562 \\
$\theta_i$ & -- & 2.83 & 2.83 \\
$n$ & -- & 3  & 3\\
\hline \hline                                                  
\end{tabular}}
\caption{Values of the cosmological parameters used in the numerical example comparing $\Lambda$CDM against two EDE models in Sec.~\ref{sec:worked} (see Figs.~\ref{fig:plot_cmb_lcdm_ede} and~\ref{fig:plot_cmb_eisw_lcdm_ede}). The difference between the two EDE models is in the value of the physical cold DM density $\omega_c$, which is fixed to $\omega_c=0.1320$ in the ``high'' case (middle column), and to $\omega_c=0.1177$ (as in the $\Lambda$CDM model it is being compared against) in the ``low'' case (right column).}
\label{tab:parametersede}

\end{table*}

In this section, I will discuss how the previous results affect early-time new physics in practice. For convenience, I will present a case study focused on the eISW effect in early dark energy (EDE) models (see e.g. Refs.~\cite{Mortsell:2018mfj,Poulin:2018cxd,Agrawal:2019lmo,Alexander:2019rsc,Niedermann:2019olb,Sakstein:2019fmf,Ye:2020btb,Zumalacarregui:2020cjh,Hill:2020osr,Chudaykin:2020acu,Das:2020wfe,Braglia:2020bym,Niedermann:2020dwg,Ivanov:2020ril,DAmico:2020ods,Ye:2020oix,Niedermann:2020qbw,Murgia:2020ryi,Smith:2020rxx,Chudaykin:2020igl,CarrilloGonzalez:2020oac,Oikonomou:2020qah,Seto:2021xua,Tian:2021omz,Freese:2021rjq,Nojiri:2021dze} for examples of EDE models). The arguments I will present in this Section, however, are likely to apply more broadly than just to EDE, but more generally to most model whose effect is to enhance the expansion rate around recombination.

Early dark energy is among the most promising (or ``least unlikely'' in the words of Ref.~\cite{Knox:2019rjx}) proposed solutions to the Hubble tension. EDE falls within a class of models which increase the expansion rate just prior to recombination. This reduces the sound horizon at last-scattering, allowing for a higher $H_0$ from CMB data without running afoul of late-time constraints from BAO and Hubble flow SNeIa. In the implementation I will consider here, the role of EDE is played by an ultra-light scalar field, with mass of order ${\cal O}(10^{-27})\,{\rm eV}$. At early times, the field is displaced from the minimum of its potential, while being held in place by Hubble friction, therefore behaving as an effective dark energy component. Once the expansion rate drops below the mass of the field, Hubble friction is no longer important and the field is free to roll down the potential and oscillate around the minimum. If the potential around the minimum is sufficiently steep (in particular steeper than quartic), EDE then redshifts faster than radiation, and rapidly becomes a subdominant component of the Universe's energy budget.

Consider an EDE axion-like field $\phi$, \textit{i.e.} a pseudo-scalar enjoying a global $U(1)$ shift symmetry, broken by non-perturbative effects generating a potential $V(\phi)$, which in turn dictates the dynamics of the EDE field.~\footnote{This type of EDE field could arise from the so-called ``string axiverse''~\cite{Svrcek:2006yi,Arvanitaki:2009fg,Kamionkowski:2014zda,Karwal:2016vyq,Visinelli:2018utg}, featuring multiple axion-like particles spanning various decades in mass.} Here I will consider a potential of the form:
\begin{eqnarray}
V(\phi) = m^2f^2\left ( 1-\cos \left ( \frac{\phi}{f} \right ) \right )^n\,,
\label{eq:potential}
\end{eqnarray}
where $m$ is a mass scale and $f$ is the EDE decay constant, at which the global $U(1)$ symmetry is broken. A potential of this form requires a careful fine-tuning of the hierarchy of instanton actions~\cite{Rudelius:2015xta,Montero:2015ofa}, but for integer values of $n$ the fine-tuning is only restricted to the first $n$ terms. I will fix $n=3$, as this is the minimum integer value which allows EDE to redshift faster than radiation once the field reaches the minimum of the potential. At its minimum, the potential is locally $V \propto \phi^6$, so that the subsequent effective equation of state of EDE is $w_{\rm EDE}=1/2$. The dynamics of the EDE field in an expanding Universe are governed by the Klein-Gordon equation:
\begin{eqnarray}
\ddot{\phi}+3H\dot{\phi}+\frac{dV(\phi)}{d\phi}=0\,,
\label{eq:kleingordon}
\end{eqnarray}
with the dot denoting a derivative with respect to time.

The cosmological dynamics of this EDE model are more simply described in terms of three parameters: the initial field displacement (or misalignment angle) $\theta_i = \phi_i/f$, with $\phi_i$ being the initial value at which $\phi$ is held by Hubble friction, as well as the ``critical redshift'' $z_c$ and the ``EDE fraction'' $f_{\rm EDE}$. More specifically, at the critical redshift $z_c$, approximately corresponding to the moment just before the field starts to oscillate around its minimum, EDE provides its maximum fractional contribution to the energy budget of the Universe $f_{\rm EDE} \equiv \rho_{\rm EDE}(z)/(3M_{\rm Pl}^2H^2(z))\vert_{z=z_c}$, with $\rho_{\rm EDE}(z)$ and $M_{\rm Pl}$ being the EDE energy density and reduced Planck mass respectively. Alleviating the Hubble tension requires an EDE fraction of order $f_{\rm EDE} \simeq 10\%$.

To illustrate the issues EDE faces in relation to the eISW effect, I will consider a numerical example closely following that of Ref.~\cite{Hill:2020osr}, focusing on $n=3$ EDE. I will compare EDE to $\Lambda$CDM, with the parameters of both models in this numerical example being summarized in Tab.~\ref{tab:parametersede}. The EDE parameters (right column) are set to their best-fit values as determined from a fit to \textit{Planck}, BAO, \textit{Pantheon}, and redshift-space distortions measurements, alongside a SH0ES prior on $H_0$, as reported in Ref.~\cite{Poulin:2018cxd}. The key to the success of the EDE proposal is its ability to accommodate a higher value of $H_0$ while fitting the CMB power spectra as well as $\Lambda$CDM (with a lower $H_0$). In the context of this numerical example, the $\Lambda$CDM parameters (left column) are chosen so that the resulting CMB power spectra are essentially indistinguishable from those within the EDE model.

\begin{figure*}[!ht]
\includegraphics[width=0.495\linewidth]{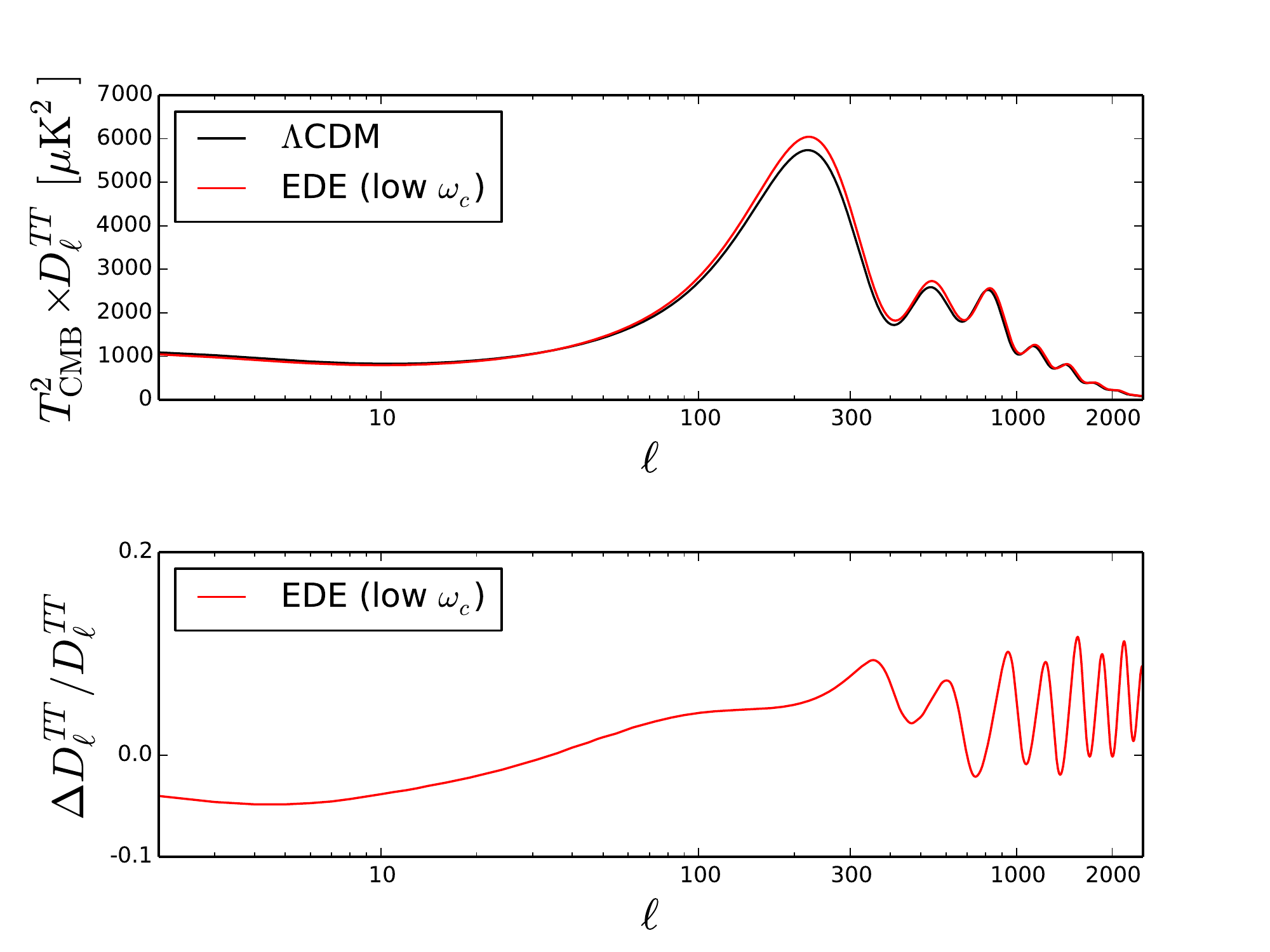}
\includegraphics[width=0.495\linewidth]{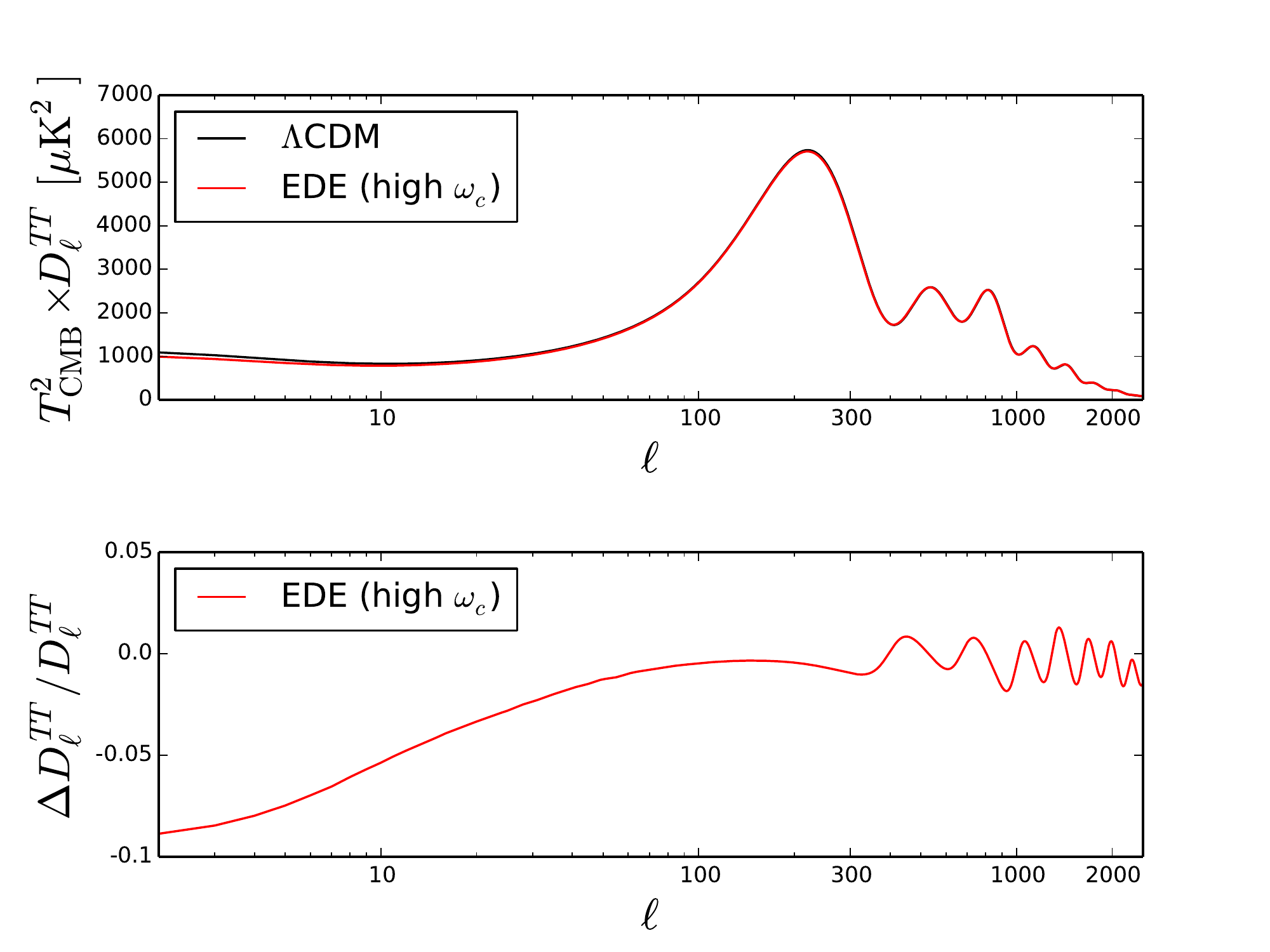}
\caption{CMB temperature anisotropy power spectra for $\Lambda$CDM and early dark energy (EDE). \textit{Left panel}: the upper panel shows the CMB temperature anisotropy power spectrum for a $\Lambda$CDM model (black curve) with parameters given by the left column of Tab.~\ref{tab:parametersede}, and for an EDE model (red curve) with parameters given by the middle column of Tab.~\ref{tab:parametersede}, where in particular the physical cold DM density is fixed to the ``low'' value $\omega_c=0.1177$. The lower panel shows the relative differences between the EDE and $\Lambda$CDM power spectra shown in the upper panel, with the former clearly predicting an excess of power compared to the latter, particularly around the scale of the first peak. \textit{Right panel}: as for the left panel, but with the EDE parameters given by the right column of Tab.~\ref{tab:parametersede}, where in particular the physical cold DM density is fixed to the ``high'' value $\omega_c=0.1320$. As the lower panel shows, this increase in $\omega_c$ makes the two power spectra nearly indistinguishable.}
\label{fig:plot_cmb_lcdm_ede}
\end{figure*}

As can clearly be seen from Tab.~\ref{tab:parametersede}, and already noted elsewhere (see e.g. Refs.~\cite{Poulin:2018cxd,Hill:2020osr}), the fact that EDE can accommodate a higher $H_0$ while preserving the fit to CMB data comes at the cost of important shifts in some of the standard $\Lambda$CDM parameters. In particular, both $\omega_c$ and $n_s$ need to increase, rather substantially in the case of $\omega_c$. To show the importance of the increase in $\omega_c$, in Fig.~\ref{fig:plot_cmb_lcdm_ede} I show the resulting $\Lambda$CDM and EDE temperature power spectra, with parameters as given in Tab.~\ref{tab:parametersede}, with the EDE model with ``low''/``high'' $\omega_c$ given in the left/right panel. The corresponding bottom panels show the relative differences in the EDE power spectrum relative to $\Lambda$CDM. Fig.~\ref{fig:plot_cmb_lcdm_ede} has been produced using \texttt{CLASS\_EDE}~\cite{Hill:2020osr},~\footnote{Available at \href{https://github.com/mwt5345/class\_ede}{github.com/mwt5345/class\_ede}.} a modified version of the Boltzmann solver \texttt{CLASS}~\cite{Blas:2011rf}.

From Fig.~\ref{fig:plot_cmb_lcdm_ede} we see that an EDE model with low $\omega_c$ predicts excess power particularly around the first acoustic peak, and more generally for all multipoles $\ell \gtrsim 100$. The power spectrum of the EDE model with high $\omega_c$, however, is essentially indistinguishable from the $\Lambda$CDM one. In both models, the slight decrease of power at large scales ($\ell \lesssim 30$) is swamped by cosmic variance and hence virtually impossible to observe. Analogous effects are observed in the CMB polarization power spectra and temperature-polarization cross-spectra, which for conciseness I do not show here.

The physical origin of the shift in $\omega_c$ can be traced back to the effect of EDE during the time when its contribution to the Universe's energy budget is non-negligible. During this time, the increase in the expansion rate brought upon by EDE suppresses the growth of perturbations: this is analogous to how cosmic acceleration at late times also suppresses the growth of structure. In order to preserve the fit to the CMB power spectra, this needs to be accompanied by an increase in $\omega_c$ which compensates for the decreased efficiency in the growth of structure.~\footnote{A smaller increase in $n_s$ is instead also needed to compensate for the scale-dependent suppression of growth due to the fact that EDE is dynamically relevant only for a short amount of time.} The increase in $\omega_c$, however, directly increases the late-time amplitude of matter fluctuations $\sigma_8$, exacerbating the discrepancy between CMB and weak lensing (WL) probes, and degrading the fit to large-scale structure (LSS) clustering data. This degraded fit to WL and LSS data, ultimately, limits the success of the EDE proposal in solving the Hubble tension~\cite{Hill:2020osr,Ivanov:2020ril,DAmico:2020ods} (see however a partial rebuttal of these results in Refs.~\cite{Murgia:2020ryi,Smith:2020rxx}).

Here, I will show that these shifts in $\omega_c$ are essentially required to preserve the amplitude of the eISW effect predicted by $\Lambda$CDM, which perfectly fits \textit{Planck} data, as per my earlier results discussed in Sec.~\ref{sec:results}, inferring $A_{\rm eISW}$ to be remarkably consistent with the standard value $A_{\rm eISW}=1$. Let us focus on the EDE model with low $\omega_c$, which I argue increases the amplitude of the eISW effect compared to $\Lambda$CDM (with the same value of $\omega_c$). This comes about for two reasons, working at the perturbation and background levels respectively:
\begin{itemize}
\item At the perturbation level, the EDE-induced suppression of the growth of perturbations is accompanied by a related enhanced decay of the gravitational potentials $\Phi$ and $\Psi$, which decay more quickly than if the Universe were filled with radiation alone. This leads to a scale-dependent enhancement of the eISW effect. The increase in $\omega_c$ helps counteracting this enhanced decay of the gravitational potentials.
\item At the background level, the presence of an extra component delays the onset of matter domination. This increases the time over which gravitational potentials vary with time, leading to a global enhancement of the eISW effect. The increase in $\omega_c$ helps counteracting this effect by anticipating the onset of matter domination.
\end{itemize}
Note that both effects are similar to how increasing the energy density of dark energy at late times enhances the decay of gravitational potentials and anticipates the onset of dark energy domination, leading to an enhanced late ISW effect, visible in the low-$\ell$ part of CMB temperature power spectrum, or more clearly in cross-correlations between CMB and LSS probes~\cite{Ho:2008bz,Giannantonio:2013kqa,Renk:2016olm,Giusarma:2018jei,Vagnozzi:2019kvw}. The first of the two points above, \textit{i.e.} the excess decay of the Weyl potential $\Phi-\Psi$, was first pointed out in the context of an EDE-type model in Ref.~\cite{Lin:2019qug}, see also Sec.~IIIE and Fig.~8 of Ref.~\cite{Niedermann:2020dwg} in the context of the related new EDE (NEDE) model (the different sign convention in the perturbed line element means that the Weyl potential in Ref.~\cite{Niedermann:2020dwg} is given by $\Phi+\Psi$ rather than $\Phi-\Psi$).

Besides these two physical effects enhancing the eISW amplitude, there is potentially a third effect which comes into play purely at the perturbation level. As first pointed out in Refs.~\cite{Lin:2019qug,Niedermann:2020dwg}, once EDE starts decaying, the decaying fluid supports its own acoustic oscillations, which in turn source the gravitational potential. The subsequent Jeans stabilization of these acoustic oscillations then leads to an enhanced decay of the Weyl potential, which in turn enhances the eISW effect. This effect can be clearly seen in Fig.~8 of Ref.~\cite{Niedermann:2020dwg} to the right of the vertical dotted line (whereas the previous two effects show up to the left thereof). It is worth noting that, while the previous two physical effects apply to generic EDE implementations, this third effect is significantly more model-dependent, and depends on specific details of the EDE model, such as sound speed, perturbation mode, trigger dynamics, and viscosity parameter.

\begin{figure*}[!ht]
\includegraphics[width=0.495\linewidth]{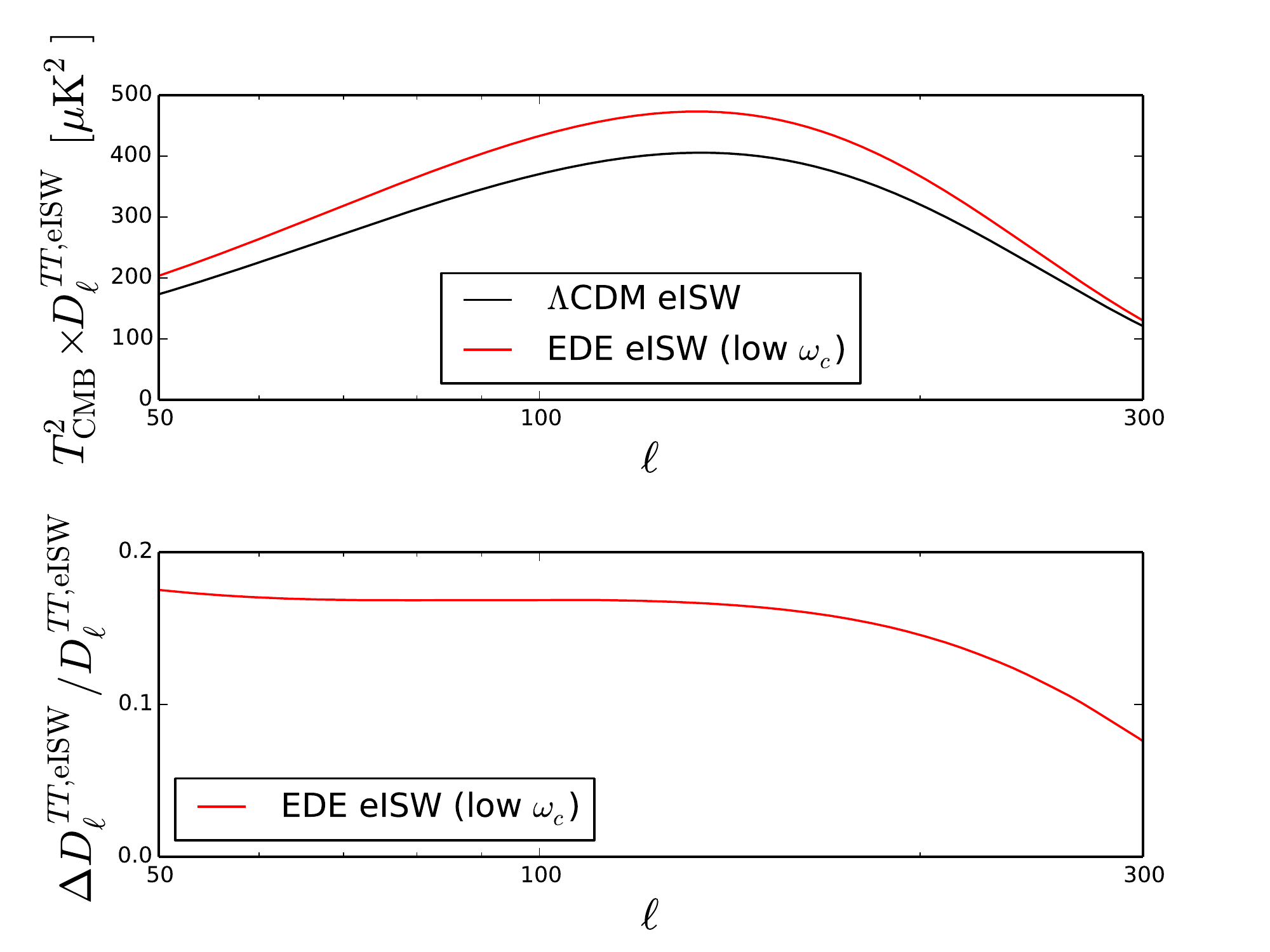}
\includegraphics[width=0.495\linewidth]{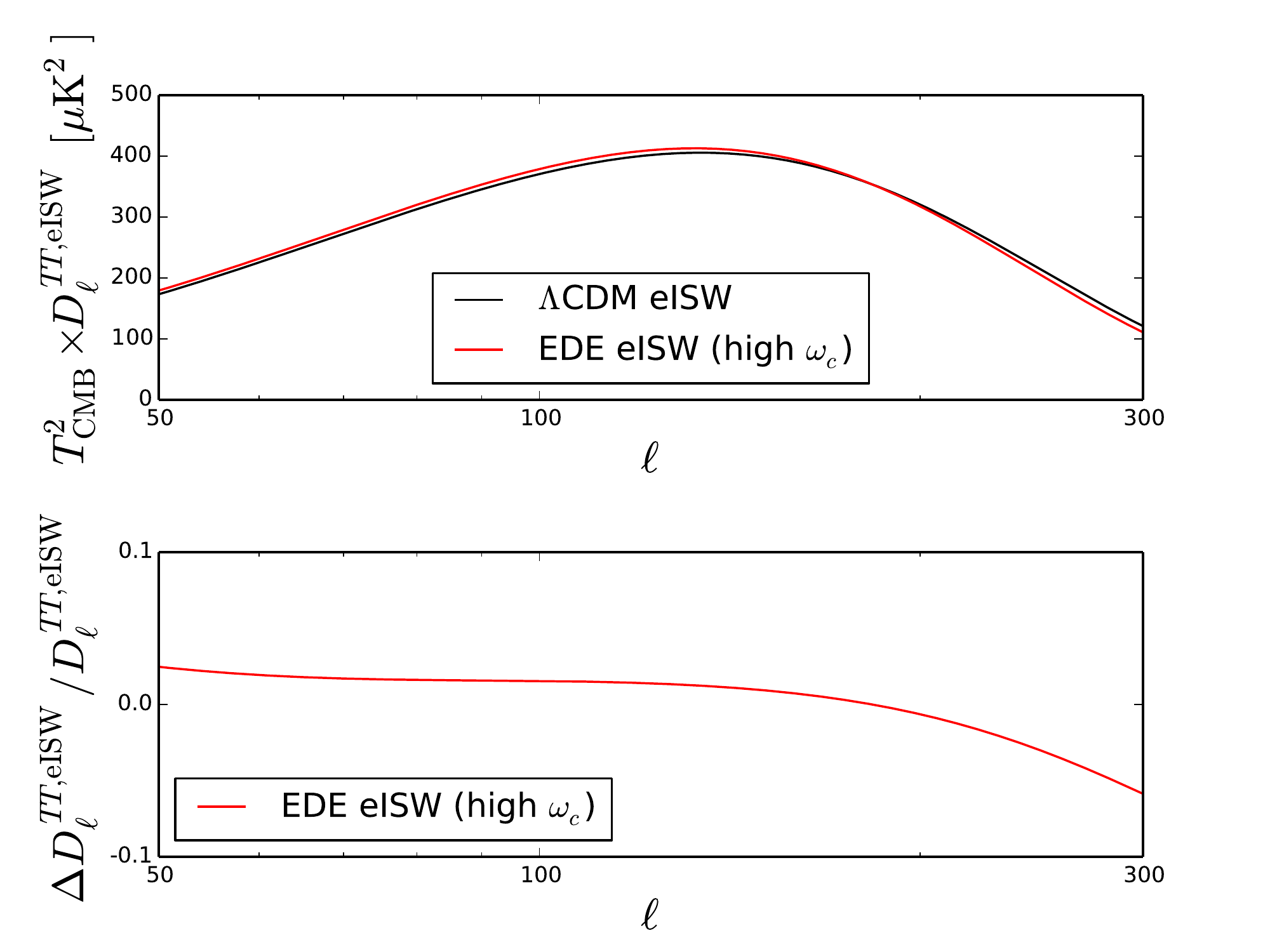}
\caption{As in Fig.~\ref{fig:plot_cmb_lcdm_ede}, but focusing on the early ISW (eISW) contribution to the CMB temperature anisotropy power spectrum. It is clear that the increase in $\omega_c$ significantly decreases the excess eISW contribution on most scales of interest. Note that the $x$ axis only covers multipoles in the range $50<\ell<300$, where the eISW effect is most prominent, as shown in Fig.~\ref{fig:plot_aeisw_new_cmb}.}
\label{fig:plot_cmb_eisw_lcdm_ede}
\end{figure*}

To confirm that the observed shifts in $\omega_c$ are indeed required to reduce the amplitude of the eISW effect, I produce plots comparing $\Lambda$CDM and EDE analogous to those in Fig.~\ref{fig:plot_cmb_lcdm_ede}, this time isolating the eISW contribution to the CMB temperature power spectrum, $C_{\ell}^{TT,{\rm eISW}}$. The result is shown in Fig.~\ref{fig:plot_cmb_eisw_lcdm_ede}, from which one clearly sees that an EDE model with low $\omega_c$ (left panel) predicts an enhanced eISW effect on all scales. In particular, I find a nearly $20\%$ excess in the eISW power on multipoles $\ell \lesssim 500$, particularly evident around the scale of the first peak (see the left panel of Fig.~\ref{fig:plot_cmb_lcdm_ede}). Raising $\omega_c$ (right panel) considerably suppresses this effect, leading to an excess in eISW power at the ${\cal O}(3\%)$ level, which in turn significantly improves the fit to the CMB temperature power spectrum, leading to predictions which are essentially indistinguishable from those of $\Lambda$CDM (see the right panel of Fig.~\ref{fig:plot_cmb_lcdm_ede}).

As my earlier results in Sec.~\ref{sec:results} have shown, a ${\cal O}(20\%)$ increase in the eISW effect is excluded at $>5\sigma$ by the \textit{Planck} measurements, and therefore needs to be counteracted by means of shifts in some of the standard $\Lambda$CDM parameters, in this case $\omega_c$. My results in this section and in particular Fig.~\ref{fig:plot_cmb_eisw_lcdm_ede} therefore show that, if the increase in $\omega_c$ (and corresponding worsened fit to WL and LSS data) leads to the demise of the EDE scenario in terms of solving the Hubble tension (modulo the rebuttal arguments discussed in Refs.~\cite{Murgia:2020ryi,Smith:2020rxx}), it is really the eISW effect which ultimately is to blame. More precisely, what is to ``blame'' is the fact that $\Lambda$CDM's prediction for the eISW effect is in excellent agreement with the \textit{Planck} measurements, which show no obvious evidence for new physics at early times.

Are the EDE-specific arguments presented here more general, \textit{i.e.} do they apply more generally to other early-time new physics scenarios? I believe that the answer is yes, at least insofar as this early-time new physics increases the expansion rate around recombination. In fact, from very general considerations one can expect that this type of early-time new physics will \textit{i)} lead to a decay of the gravitational potentials (much as dark energy at late times), and \textit{ii)} slow down the growth of perturbations. To confirm this, one should solve the equation governing the time-evolution of gravitational potentials, which is generically given by (see e.g. Ref.~\cite{Dodelson:2003ft}):
\begin{eqnarray}
k^2\Phi+3{\cal H} \left ( \dot{\phi}-3{\cal H}\Psi \right ) = 4\pi Ga^2 \sum_i\rho_i\delta_i\,,
\label{eq:phi}
\end{eqnarray}
with ${\cal H}$ being the conformal Hubble rate, $G$ Newton's constant, $a$ the scale factor, the dot denoting a derivative with respect to conformal time $\eta$, and the sum on the right-hand side running over all components contributing to the Universe's energy budget, with energy densities and density contrasts given by $\rho_i$ and $\delta_i$ respectively. Approximate solutions to Eq.~(\ref{eq:phi}) are known only for idealistic situations where the Universe is dominated by a single component, \textit{e.g.} deep in the matter era ($\Phi(\eta)={\rm const}$) or deep in the radiation era ($\Phi(\eta) \propto [\sin(\eta)-\eta\cos(\eta)]/\eta^3$). In all other situations, including the simultaneous presence of radiation, matter, and a new component at early times (such as EDE), Eq.~(\ref{eq:phi}) needs to be solved numerically. However, given the form of Eq.~(\ref{eq:phi}), it is reasonable to expect that any component whose effect is to speed up (even if slightly) the expansion of the Universe and hence raise ${\cal H}$, will contribute to the decay of $\Phi$, which in itself is already decaying during the radiation-dominated era, thus enhancing the eISW effect.

Similarly, the equation for the growth of matter perturbations $\delta$ does not take a simple closed form in a generic Universe where radiation, matter, and a new component at early times are simultaneously present. However, if the new component is subdominant, the equation governing $\delta$ should approximately reduce to the form:
\begin{eqnarray}
\ddot{\delta}+\frac{\dot{\delta}}{\eta} = {\cal S}(k,\eta)\,,
\label{eq:growthofmatter}
\end{eqnarray}
with source term ${\cal S}(k,\eta)$ given by:
\begin{eqnarray}
{\cal S}(k,\eta) = -3\ddot{\Phi}+k^2\Phi-\frac{3\dot{\Phi}}{\eta}\,.
\label{eq:source}
\end{eqnarray}
Given the previous considerations on the behaviour of $\Phi$ in the presence of a new component speeding up the expansion of the Universe, and the form of Eqs.~(\ref{eq:growthofmatter},\ref{eq:source}), one can generically expect $\delta$ to grow more slowly in the presence of such a component.

I wish to clarify that the arguments presented above concerning the enhanced decay of potentials and reduced growth of perturbations are not a rigorous \textit{no-go theorem}, nor is it my intention to elevate them to such. While I expect these arguments to hold generically for models which decrease the sound horizon by introducing a new dark energy-like component, or in any case through a speed-up of the expansion of the Universe around recombination, I stress that the validity of these conclusions should be checked on a case-by-case basis for each given model. Note that a similar argument (at a similar level of rigor) was also presented in Ref.~\cite{Hill:2020osr}. In any case if, as I expect, these arguments apply more generically to models beyond EDE, I would also expect that these models would have to compensate for the enhanced eISW effect. The most direct, but by no means only way of doing so, would be that of raising $\omega_c$, which however would exacerbate the $S_8$ discrepancy,~\footnote{See however Ref.~\cite{Nunes:2021ipq} for a recent work arguing that the $S_8$ discrepancy may be compatible with a statistical fluctuation.} worsening the fit to WL and LSS clustering measurements, and possibly leading to the demise of these models.

However, model-dependent aspects make it important to assess the details of early-type models on a case-by-case basis to judge their prospects. Specific early-type models may be able to introduce ingredients which can at least partially counteract the enhanced eISW effect. The NEDE model is an example in this sense (see the discussion in Ref.~\cite{Niedermann:2020qbw}), as it does not significantly worsen the $S_8$ discrepancy already present within $\Lambda$CDM. Another example is that of the majoron model, which can damp neutrino free-streaming and inject additional energy density in the neutrino sector prior to recombination, thereby enhancing the pre-recombination expansion rate~\cite{Escudero:2021rfi}, and requiring an increase in $\omega_c$. However, these neutrino-majoron interactions leave peculiar imprints in the low-$\ell$ part of the CMB power spectrum, where the eISW effect is relevant (see Fig.~5 of Ref.~\cite{Escudero:2021rfi}): these imprints might be partially responsible for the model's good performance in solving the Hubble tension when confronted against CMB and LSS data, despite the increase in $\omega_c$. The NEDE and majoron cases are just two examples of models introducing specific model-dependent ingredients which can partially counteract the two effects I previously identified as being responsible for the enhanced eISW effect: other ingredients which also do so are of course possible. However, it is likely that these ingredients will not be able to fully resolve the discrepancies between CMB and LSS/WL data already present within $\Lambda$CDM, and that further extensions will be required to do so.~\footnote{In the context of NEDE, these ingredients could include interactions between the scalar field and the visible sector and/or a significant amount of oscillations around the new vacuum after the phase transition, all of which affect the way the Weyl potential decays, and therefore the way the eISW effect is enhanced (see e.g. the discussion towards the end of Ref.~\cite{Niedermann:2020qbw}).}

Other types of early-time new physics which accommodate a higher $H_0$ without speeding up the expansion of the Universe around recombination should be immune to these conclusions, although they are of course subject to stringent constraints from the full CMB power spectra. One example is the strongly-interacting neutrino model~\cite{Kreisch:2019yzn}, whose worsened fit to CMB polarization measurements limits its success in solving the Hubble tension~\cite{Das:2020xke,Choudhury:2020tka,Brinckmann:2020bcn}, although it is worth noting that the preference for a higher value of $N_{\rm eff}$ within this model indirectly leads to an enhancement of the pre-recombination expansion rate.

My overall findings are similar in spirit to those of recent works along a related line, which question the ability of early-time new physics, or more precisely early-time new physics lowering the sound horizon alone, to fully resolve the Hubble tension~\cite{Krishnan:2020obg,Jedamzik:2020zmd,Lin:2021sfs,Dainotti:2021pqg,Krishnan:2021dyb,Vagnozzi:2021tjv,Krishnan:2021jmh}. There is, of course, no question around the fact that late-time new physics \textit{alone} falls short of fully solving the Hubble tension, due to constraints imposed by BAO and SNeIa measurements~\cite{Bernal:2016gxb,Addison:2017fdm,Lemos:2018smw,Aylor:2018drw,Knox:2019rjx,Arendse:2019hev,Efstathiou:2021ocp} (with the possible exception of models invoking local effects such as those considered in Refs.~\cite{Lombriser:2019ahl,Desmond:2019ygn,Ding:2019mmw,Desmond:2020wep,Alestas:2020zol,Cai:2021wgv,Marra:2021fvf}). My results, alongside those of Refs.~\cite{Krishnan:2020obg,Jedamzik:2020zmd,Lin:2021sfs,Dainotti:2021pqg,Krishnan:2021dyb,Vagnozzi:2021tjv,Krishnan:2021jmh} (though perhaps less model-independent than the latter), show that early-time new physics models face equally severe (if not more severe) stumbling blocks as late-time new physics ones.

\section{Conclusions}
\label{sec:conclusions}

\textit{Why does $\Lambda$CDM fit CMB measurements so well?} \textit{Why is there no evidence for new physics from CMB data alone?} The answers to these questions, and to other general model-agnostic questions testing the consistency of $\Lambda$CDM, can provide a compass for navigating the \textit{mare magnum} which is the theory space of proposed solutions to the Hubble tension. Early-time new physics, particularly models which raise the expansion rate around recombination to lower the sound horizon, should leave an important imprint on the early ISW (eISW) effect, which should typically be enhanced. Motivated by this, I have performed an eISW-based consistency test of $\Lambda$CDM by introducing the phenomenological scaling parameter $A_{\rm eISW}$, artificially rescaling amplitude of the eISW contribution to the CMB power spectra.

From a fit to the \textit{Planck} 2018 legacy data release temperature and polarization data within a seven-parameter $\Lambda$CDM+$A_{\rm eISW}$ model, I infer $A_{\rm eISW}=0.988 \pm 0.027$, in perfect agreement with the standard value $A_{\rm eISW}=1$. This result shows that $\Lambda$CDM's prediction for the eISW effect is in perfect agreement with data, and there is room for no more than a $\approx 3\%$ enhancement/suppression of the eISW effect relative to this prediction. More importantly, this result poses important restrictions on early-time new physics models, which will need to (approximately) match this prediction.

I have illustrated the implications of these results for new physics focusing on the well-known early dark energy (EDE) model. For EDE to fit CMB data as well as $\Lambda$CDM while accommodating a higher $H_0$ requires an increase in $\omega_c$ -- this has been argued to worsen the fit to WL and galaxy clustering measurements, leading to the conclusion that EDE fails to restore cosmic concordance. I have explicitly shown that the increase in $\omega_c$ is required to bring the amplitude of the eISW effect (which would otherwise be overpredicted by $\approx 20\%$) in better agreement with $\Lambda$CDM's prediction. I have argued that this problem should go beyond EDE and be a general feature of early-time new physics, at least for models which raise the expansion rate around recombination.

My findings join further existing restrictions concerning the ability of early-time new physics lowering the sound horizon alone to fully resolve the Hubble tension~\cite{Krishnan:2020obg,Jedamzik:2020zmd,Lin:2021sfs,Dainotti:2021pqg,Krishnan:2021dyb,Vagnozzi:2021tjv,Krishnan:2021jmh}. On the other hand, the late-time expansion history is too well constrained by BAO and uncalibrated SNeIa measurements for late-time new physics to be able to fully resolve the Hubble tension on their own~\cite{Bernal:2016gxb,Addison:2017fdm,Lemos:2018smw,Aylor:2018drw,Knox:2019rjx,Arendse:2019hev,Efstathiou:2021ocp}. Overall, this suggests that a definitive resolution of the Hubble tension might require one or more among: \textit{i)} a combination of modifications to $\Lambda$CDM, both at early and late times, \textit{ii)} highly non-trivial early-time modifications to $\Lambda$CDM which are able to match $\Lambda$CDM's prediction for the eISW effect while not degrading the fit to other datasets, \textit{iii)} local new physics (in agreement with the findings of Ref.~\cite{Lin:2021sfs}, and including models of the types proposed by Refs.~\cite{Desmond:2019ygn,Desmond:2020wep,Alestas:2020zol,Marra:2021fvf,Cai:2021wgv}); or \textit{iv)} convincingly identifying systematic errors in one or more of the involved datasets (see e.g. Refs.~\cite{Efstathiou:2020wxn,Mortsell:2021nzg,Mortsell:2021tcx}).

It is of course entirely possible that clear signs of new physics will appear in future CMB data alone~\cite{Abazajian:2016yjj,Ade:2018sbj,Abitbol:2019nhf}. In this case, whatever emerges will undoubtedly teach us something fundamental about Nature. Until then, however, general consistency tests of $\Lambda$CDM will keep providing important taffrails while steering through the vast sea that is the theory space of new physics.

\begin{acknowledgments}
\noindent I am grateful to George Efstathiou for many useful discussions which led to the development of this work, and for encouraging me to think about general consistency tests of $\Lambda$CDM. I thank Alex Reeves, Janina Renk, and Blake Sherwin for very useful discussions around the subject of this work, and for help with the codes, Josh Kable for bringing a very relevant work to my attention, and Florian Niedermann and Martin Sloth for valuable comments on a previous draft. I am supported by the Isaac Newton Trust and the Kavli Foundation through a Newton-Kavli Fellowship, and by a grant from the Foundation Blanceflor Boncompagni Ludovisi, n\'{e}e Bildt. I acknowledge a College Research Associateship at Homerton College, University of Cambridge. This work was performed using resources provided by the Cambridge Service for Data Driven Discovery (CSD3) operated by the University of Cambridge Research Computing Service (\href{https://www.hpc.cam.ac.uk/}{www.hpc.cam.ac.uk}), provided by Dell EMC and Intel using Tier-2 funding from the Engineering and Physical Sciences Research Council (capital grant EP/P020259/1), and DiRAC funding from the Science and Technology Facilities Council (\href{https://www.dirac.ac.uk/}{www.dirac.ac.uk}).
\end{acknowledgments}

\bibliography{eISW}

\begin{thebibliography}{172}%
\makeatletter
\providecommand \@ifxundefined [1]{%
 \@ifx{#1\undefined}
}%
\providecommand \@ifnum [1]{%
 \ifnum #1\expandafter \@firstoftwo
 \else \expandafter \@secondoftwo
 \fi
}%
\providecommand \@ifx [1]{%
 \ifx #1\expandafter \@firstoftwo
 \else \expandafter \@secondoftwo
 \fi
}%
\providecommand \natexlab [1]{#1}%
\providecommand \enquote  [1]{``#1''}%
\providecommand \bibnamefont  [1]{#1}%
\providecommand \bibfnamefont [1]{#1}%
\providecommand \citenamefont [1]{#1}%
\providecommand \href@noop [0]{\@secondoftwo}%
\providecommand \href [0]{\begingroup \@sanitize@url \@href}%
\providecommand \@href[1]{\@@startlink{#1}\@@href}%
\providecommand \@@href[1]{\endgroup#1\@@endlink}%
\providecommand \@sanitize@url [0]{\catcode `\\12\catcode `\$12\catcode
  `\&12\catcode `\#12\catcode `\^12\catcode `\_12\catcode `\%12\relax}%
\providecommand \@@startlink[1]{}%
\providecommand \@@endlink[0]{}%
\providecommand \url  [0]{\begingroup\@sanitize@url \@url }%
\providecommand \@url [1]{\endgroup\@href {#1}{\urlprefix }}%
\providecommand \urlprefix  [0]{URL }%
\providecommand \Eprint [0]{\href }%
\providecommand \doibase [0]{http://dx.doi.org/}%
\providecommand \selectlanguage [0]{\@gobble}%
\providecommand \bibinfo  [0]{\@secondoftwo}%
\providecommand \bibfield  [0]{\@secondoftwo}%
\providecommand \translation [1]{[#1]}%
\providecommand \BibitemOpen [0]{}%
\providecommand \bibitemStop [0]{}%
\providecommand \bibitemNoStop [0]{.\EOS\space}%
\providecommand \EOS [0]{\spacefactor3000\relax}%
\providecommand \BibitemShut  [1]{\csname bibitem#1\endcsname}%
\let\auto@bib@innerbib\@empty
%</preamble>
\bibitem [{\citenamefont {Riess}\ \emph {et~al.}(1998)\citenamefont {Riess}
  \emph {et~al.}}]{Riess:1998cb}%
  \BibitemOpen
  \bibfield  {author} {\bibinfo {author} {\bibfnamefont {A.~G.}\ \bibnamefont
  {Riess}} \emph {et~al.} (\bibinfo {collaboration} {Supernova Search Team}),\
  }\href {\doibase 10.1086/300499} {\bibfield  {journal} {\bibinfo  {journal}
  {Astron. J.}\ }\textbf {\bibinfo {volume} {116}},\ \bibinfo {pages} {1009}
  (\bibinfo {year} {1998})},\ \Eprint {http://arxiv.org/abs/astro-ph/9805201}
  {arXiv:astro-ph/9805201} \BibitemShut {NoStop}%
\bibitem [{\citenamefont {Perlmutter}\ \emph {et~al.}(1999)\citenamefont
  {Perlmutter} \emph {et~al.}}]{Perlmutter:1998np}%
  \BibitemOpen
  \bibfield  {author} {\bibinfo {author} {\bibfnamefont {S.}~\bibnamefont
  {Perlmutter}} \emph {et~al.} (\bibinfo {collaboration} {Supernova Cosmology
  Project}),\ }\href {\doibase 10.1086/307221} {\bibfield  {journal} {\bibinfo
  {journal} {Astrophys. J.}\ }\textbf {\bibinfo {volume} {517}},\ \bibinfo
  {pages} {565} (\bibinfo {year} {1999})},\ \Eprint
  {http://arxiv.org/abs/astro-ph/9812133} {arXiv:astro-ph/9812133} \BibitemShut
  {NoStop}%
\bibitem [{\citenamefont {Troxel}\ \emph {et~al.}(2018)\citenamefont {Troxel}
  \emph {et~al.}}]{Troxel:2017xyo}%
  \BibitemOpen
  \bibfield  {author} {\bibinfo {author} {\bibfnamefont {M.~A.}\ \bibnamefont
  {Troxel}} \emph {et~al.} (\bibinfo {collaboration} {DES}),\ }\href {\doibase
  10.1103/PhysRevD.98.043528} {\bibfield  {journal} {\bibinfo  {journal} {Phys.
  Rev. D}\ }\textbf {\bibinfo {volume} {98}},\ \bibinfo {pages} {043528}
  (\bibinfo {year} {2018})},\ \Eprint {http://arxiv.org/abs/1708.01538}
  {arXiv:1708.01538 [astro-ph.CO]} \BibitemShut {NoStop}%
\bibitem [{\citenamefont {Scolnic}\ \emph {et~al.}(2018)\citenamefont {Scolnic}
  \emph {et~al.}}]{Scolnic:2017caz}%
  \BibitemOpen
  \bibfield  {author} {\bibinfo {author} {\bibfnamefont {D.~M.}\ \bibnamefont
  {Scolnic}} \emph {et~al.},\ }\href {\doibase 10.3847/1538-4357/aab9bb}
  {\bibfield  {journal} {\bibinfo  {journal} {Astrophys. J.}\ }\textbf
  {\bibinfo {volume} {859}},\ \bibinfo {pages} {101} (\bibinfo {year}
  {2018})},\ \Eprint {http://arxiv.org/abs/1710.00845} {arXiv:1710.00845
  [astro-ph.CO]} \BibitemShut {NoStop}%
\bibitem [{\citenamefont {Aghanim}\ \emph
  {et~al.}(2020{\natexlab{a}})\citenamefont {Aghanim} \emph
  {et~al.}}]{Aghanim:2018eyx}%
  \BibitemOpen
  \bibfield  {author} {\bibinfo {author} {\bibfnamefont {N.}~\bibnamefont
  {Aghanim}} \emph {et~al.} (\bibinfo {collaboration} {Planck}),\ }\href
  {\doibase 10.1051/0004-6361/201833910} {\bibfield  {journal} {\bibinfo
  {journal} {Astron. Astrophys.}\ }\textbf {\bibinfo {volume} {641}},\ \bibinfo
  {pages} {A6} (\bibinfo {year} {2020}{\natexlab{a}})},\ \Eprint
  {http://arxiv.org/abs/1807.06209} {arXiv:1807.06209 [astro-ph.CO]}
  \BibitemShut {NoStop}%
\bibitem [{\citenamefont {Bianchini}\ \emph {et~al.}(2020)\citenamefont
  {Bianchini} \emph {et~al.}}]{Bianchini:2019vxp}%
  \BibitemOpen
  \bibfield  {author} {\bibinfo {author} {\bibfnamefont {F.}~\bibnamefont
  {Bianchini}} \emph {et~al.} (\bibinfo {collaboration} {SPT}),\ }\href
  {\doibase 10.3847/1538-4357/ab6082} {\bibfield  {journal} {\bibinfo
  {journal} {Astrophys. J.}\ }\textbf {\bibinfo {volume} {888}},\ \bibinfo
  {pages} {119} (\bibinfo {year} {2020})},\ \Eprint
  {http://arxiv.org/abs/1910.07157} {arXiv:1910.07157 [astro-ph.CO]}
  \BibitemShut {NoStop}%
\bibitem [{\citenamefont {Aiola}\ \emph {et~al.}(2020)\citenamefont {Aiola}
  \emph {et~al.}}]{Aiola:2020azj}%
  \BibitemOpen
  \bibfield  {author} {\bibinfo {author} {\bibfnamefont {S.}~\bibnamefont
  {Aiola}} \emph {et~al.} (\bibinfo {collaboration} {ACT}),\ }\href {\doibase
  10.1088/1475-7516/2020/12/047} {\bibfield  {journal} {\bibinfo  {journal}
  {JCAP}\ }\textbf {\bibinfo {volume} {12}},\ \bibinfo {pages} {047} (\bibinfo
  {year} {2020})},\ \Eprint {http://arxiv.org/abs/2007.07288} {arXiv:2007.07288
  [astro-ph.CO]} \BibitemShut {NoStop}%
\bibitem [{\citenamefont {Alam}\ \emph {et~al.}(2021)\citenamefont {Alam} \emph
  {et~al.}}]{Alam:2020sor}%
  \BibitemOpen
  \bibfield  {author} {\bibinfo {author} {\bibfnamefont {S.}~\bibnamefont
  {Alam}} \emph {et~al.} (\bibinfo {collaboration} {eBOSS}),\ }\href {\doibase
  10.1103/PhysRevD.103.083533} {\bibfield  {journal} {\bibinfo  {journal}
  {Phys. Rev. D}\ }\textbf {\bibinfo {volume} {103}},\ \bibinfo {pages}
  {083533} (\bibinfo {year} {2021})},\ \Eprint
  {http://arxiv.org/abs/2007.08991} {arXiv:2007.08991 [astro-ph.CO]}
  \BibitemShut {NoStop}%
\bibitem [{\citenamefont {Asgari}\ \emph {et~al.}(2021)\citenamefont {Asgari}
  \emph {et~al.}}]{Asgari:2020wuj}%
  \BibitemOpen
  \bibfield  {author} {\bibinfo {author} {\bibfnamefont {M.}~\bibnamefont
  {Asgari}} \emph {et~al.} (\bibinfo {collaboration} {KiDS}),\ }\href {\doibase
  10.1051/0004-6361/202039070} {\bibfield  {journal} {\bibinfo  {journal}
  {Astron. Astrophys.}\ }\textbf {\bibinfo {volume} {645}},\ \bibinfo {pages}
  {A104} (\bibinfo {year} {2021})},\ \Eprint {http://arxiv.org/abs/2007.15633}
  {arXiv:2007.15633 [astro-ph.CO]} \BibitemShut {NoStop}%
\bibitem [{\citenamefont {Mossa}\ \emph {et~al.}(2020)\citenamefont {Mossa}
  \emph {et~al.}}]{Mossa:2020gjc}%
  \BibitemOpen
  \bibfield  {author} {\bibinfo {author} {\bibfnamefont {V.}~\bibnamefont
  {Mossa}} \emph {et~al.},\ }\href {\doibase 10.1038/s41586-020-2878-4}
  {\bibfield  {journal} {\bibinfo  {journal} {Nature}\ }\textbf {\bibinfo
  {volume} {587}},\ \bibinfo {pages} {210} (\bibinfo {year}
  {2020})}\BibitemShut {NoStop}%
\bibitem [{\citenamefont {Sahni}(2004)}]{Sahni:2004ai}%
  \BibitemOpen
  \bibfield  {author} {\bibinfo {author} {\bibfnamefont {V.}~\bibnamefont
  {Sahni}},\ }\href {\doibase 10.1007/b99562} {\bibfield  {journal} {\bibinfo
  {journal} {Lect. Notes Phys.}\ }\textbf {\bibinfo {volume} {653}},\ \bibinfo
  {pages} {141} (\bibinfo {year} {2004})},\ \Eprint
  {http://arxiv.org/abs/astro-ph/0403324} {arXiv:astro-ph/0403324} \BibitemShut
  {NoStop}%
\bibitem [{\citenamefont {Bertone}\ \emph {et~al.}(2005)\citenamefont
  {Bertone}, \citenamefont {Hooper},\ and\ \citenamefont
  {Silk}}]{Bertone:2004pz}%
  \BibitemOpen
  \bibfield  {author} {\bibinfo {author} {\bibfnamefont {G.}~\bibnamefont
  {Bertone}}, \bibinfo {author} {\bibfnamefont {D.}~\bibnamefont {Hooper}}, \
  and\ \bibinfo {author} {\bibfnamefont {J.}~\bibnamefont {Silk}},\ }\href
  {\doibase 10.1016/j.physrep.2004.08.031} {\bibfield  {journal} {\bibinfo
  {journal} {Phys. Rept.}\ }\textbf {\bibinfo {volume} {405}},\ \bibinfo
  {pages} {279} (\bibinfo {year} {2005})},\ \Eprint
  {http://arxiv.org/abs/hep-ph/0404175} {arXiv:hep-ph/0404175} \BibitemShut
  {NoStop}%
\bibitem [{\citenamefont {Huterer}\ and\ \citenamefont
  {Shafer}(2018)}]{Huterer:2017buf}%
  \BibitemOpen
  \bibfield  {author} {\bibinfo {author} {\bibfnamefont {D.}~\bibnamefont
  {Huterer}}\ and\ \bibinfo {author} {\bibfnamefont {D.~L.}\ \bibnamefont
  {Shafer}},\ }\href {\doibase 10.1088/1361-6633/aa997e} {\bibfield  {journal}
  {\bibinfo  {journal} {Rept. Prog. Phys.}\ }\textbf {\bibinfo {volume} {81}},\
  \bibinfo {pages} {016901} (\bibinfo {year} {2018})},\ \Eprint
  {http://arxiv.org/abs/1709.01091} {arXiv:1709.01091 [astro-ph.CO]}
  \BibitemShut {NoStop}%
\bibitem [{\citenamefont {Verde}\ \emph {et~al.}(2019)\citenamefont {Verde},
  \citenamefont {Treu},\ and\ \citenamefont {Riess}}]{Verde:2019ivm}%
  \BibitemOpen
  \bibfield  {author} {\bibinfo {author} {\bibfnamefont {L.}~\bibnamefont
  {Verde}}, \bibinfo {author} {\bibfnamefont {T.}~\bibnamefont {Treu}}, \ and\
  \bibinfo {author} {\bibfnamefont {A.~G.}\ \bibnamefont {Riess}},\ }\href
  {\doibase 10.1038/s41550-019-0902-0} {\bibfield  {journal} {\bibinfo
  {journal} {Nature Astron.}\ }\textbf {\bibinfo {volume} {3}},\ \bibinfo
  {pages} {891} (\bibinfo {year} {2019})},\ \Eprint
  {http://arxiv.org/abs/1907.10625} {arXiv:1907.10625 [astro-ph.CO]}
  \BibitemShut {NoStop}%
\bibitem [{\citenamefont {Di~Valentino}\ \emph
  {et~al.}(2021{\natexlab{a}})\citenamefont {Di~Valentino}, \citenamefont
  {Mena}, \citenamefont {Pan}, \citenamefont {Visinelli}, \citenamefont {Yang},
  \citenamefont {Melchiorri}, \citenamefont {Mota}, \citenamefont {Riess},\
  and\ \citenamefont {Silk}}]{DiValentino:2021izs}%
  \BibitemOpen
  \bibfield  {author} {\bibinfo {author} {\bibfnamefont {E.}~\bibnamefont
  {Di~Valentino}}, \bibinfo {author} {\bibfnamefont {O.}~\bibnamefont {Mena}},
  \bibinfo {author} {\bibfnamefont {S.}~\bibnamefont {Pan}}, \bibinfo {author}
  {\bibfnamefont {L.}~\bibnamefont {Visinelli}}, \bibinfo {author}
  {\bibfnamefont {W.}~\bibnamefont {Yang}}, \bibinfo {author} {\bibfnamefont
  {A.}~\bibnamefont {Melchiorri}}, \bibinfo {author} {\bibfnamefont {D.~F.}\
  \bibnamefont {Mota}}, \bibinfo {author} {\bibfnamefont {A.~G.}\ \bibnamefont
  {Riess}}, \ and\ \bibinfo {author} {\bibfnamefont {J.}~\bibnamefont {Silk}},\
  }\href@noop {} {\  (\bibinfo {year} {2021}{\natexlab{a}})},\ \Eprint
  {http://arxiv.org/abs/2103.01183} {arXiv:2103.01183 [astro-ph.CO]}
  \BibitemShut {NoStop}%
\bibitem [{\citenamefont {Perivolaropoulos}\ and\ \citenamefont
  {Skara}(2021)}]{Perivolaropoulos:2021jda}%
  \BibitemOpen
  \bibfield  {author} {\bibinfo {author} {\bibfnamefont {L.}~\bibnamefont
  {Perivolaropoulos}}\ and\ \bibinfo {author} {\bibfnamefont {F.}~\bibnamefont
  {Skara}},\ }\href@noop {} {\  (\bibinfo {year} {2021})},\ \Eprint
  {http://arxiv.org/abs/2105.05208} {arXiv:2105.05208 [astro-ph.CO]}
  \BibitemShut {NoStop}%
\bibitem [{\citenamefont {M\"ortsell}\ and\ \citenamefont
  {Dhawan}(2018)}]{Mortsell:2018mfj}%
  \BibitemOpen
  \bibfield  {author} {\bibinfo {author} {\bibfnamefont {E.}~\bibnamefont
  {M\"ortsell}}\ and\ \bibinfo {author} {\bibfnamefont {S.}~\bibnamefont
  {Dhawan}},\ }\href {\doibase 10.1088/1475-7516/2018/09/025} {\bibfield
  {journal} {\bibinfo  {journal} {JCAP}\ }\textbf {\bibinfo {volume} {09}},\
  \bibinfo {pages} {025} (\bibinfo {year} {2018})},\ \Eprint
  {http://arxiv.org/abs/1801.07260} {arXiv:1801.07260 [astro-ph.CO]}
  \BibitemShut {NoStop}%
\bibitem [{\citenamefont {Poulin}\ \emph {et~al.}(2019)\citenamefont {Poulin},
  \citenamefont {Smith}, \citenamefont {Karwal},\ and\ \citenamefont
  {Kamionkowski}}]{Poulin:2018cxd}%
  \BibitemOpen
  \bibfield  {author} {\bibinfo {author} {\bibfnamefont {V.}~\bibnamefont
  {Poulin}}, \bibinfo {author} {\bibfnamefont {T.~L.}\ \bibnamefont {Smith}},
  \bibinfo {author} {\bibfnamefont {T.}~\bibnamefont {Karwal}}, \ and\ \bibinfo
  {author} {\bibfnamefont {M.}~\bibnamefont {Kamionkowski}},\ }\href {\doibase
  10.1103/PhysRevLett.122.221301} {\bibfield  {journal} {\bibinfo  {journal}
  {Phys. Rev. Lett.}\ }\textbf {\bibinfo {volume} {122}},\ \bibinfo {pages}
  {221301} (\bibinfo {year} {2019})},\ \Eprint
  {http://arxiv.org/abs/1811.04083} {arXiv:1811.04083 [astro-ph.CO]}
  \BibitemShut {NoStop}%
\bibitem [{\citenamefont {Agrawal}\ \emph {et~al.}(2019)\citenamefont
  {Agrawal}, \citenamefont {Cyr-Racine}, \citenamefont {Pinner},\ and\
  \citenamefont {Randall}}]{Agrawal:2019lmo}%
  \BibitemOpen
  \bibfield  {author} {\bibinfo {author} {\bibfnamefont {P.}~\bibnamefont
  {Agrawal}}, \bibinfo {author} {\bibfnamefont {F.-Y.}\ \bibnamefont
  {Cyr-Racine}}, \bibinfo {author} {\bibfnamefont {D.}~\bibnamefont {Pinner}},
  \ and\ \bibinfo {author} {\bibfnamefont {L.}~\bibnamefont {Randall}},\
  }\href@noop {} {\  (\bibinfo {year} {2019})},\ \Eprint
  {http://arxiv.org/abs/1904.01016} {arXiv:1904.01016 [astro-ph.CO]}
  \BibitemShut {NoStop}%
\bibitem [{\citenamefont {Alexander}\ and\ \citenamefont
  {McDonough}(2019)}]{Alexander:2019rsc}%
  \BibitemOpen
  \bibfield  {author} {\bibinfo {author} {\bibfnamefont {S.}~\bibnamefont
  {Alexander}}\ and\ \bibinfo {author} {\bibfnamefont {E.}~\bibnamefont
  {McDonough}},\ }\href {\doibase 10.1016/j.physletb.2019.134830} {\bibfield
  {journal} {\bibinfo  {journal} {Phys. Lett. B}\ }\textbf {\bibinfo {volume}
  {797}},\ \bibinfo {pages} {134830} (\bibinfo {year} {2019})},\ \Eprint
  {http://arxiv.org/abs/1904.08912} {arXiv:1904.08912 [astro-ph.CO]}
  \BibitemShut {NoStop}%
\bibitem [{\citenamefont {Niedermann}\ and\ \citenamefont
  {Sloth}(2021{\natexlab{a}})}]{Niedermann:2019olb}%
  \BibitemOpen
  \bibfield  {author} {\bibinfo {author} {\bibfnamefont {F.}~\bibnamefont
  {Niedermann}}\ and\ \bibinfo {author} {\bibfnamefont {M.~S.}\ \bibnamefont
  {Sloth}},\ }\href {\doibase 10.1103/PhysRevD.103.L041303} {\bibfield
  {journal} {\bibinfo  {journal} {Phys. Rev. D}\ }\textbf {\bibinfo {volume}
  {103}},\ \bibinfo {pages} {L041303} (\bibinfo {year} {2021}{\natexlab{a}})},\
  \Eprint {http://arxiv.org/abs/1910.10739} {arXiv:1910.10739 [astro-ph.CO]}
  \BibitemShut {NoStop}%
\bibitem [{\citenamefont {Sakstein}\ and\ \citenamefont
  {Trodden}(2020)}]{Sakstein:2019fmf}%
  \BibitemOpen
  \bibfield  {author} {\bibinfo {author} {\bibfnamefont {J.}~\bibnamefont
  {Sakstein}}\ and\ \bibinfo {author} {\bibfnamefont {M.}~\bibnamefont
  {Trodden}},\ }\href {\doibase 10.1103/PhysRevLett.124.161301} {\bibfield
  {journal} {\bibinfo  {journal} {Phys. Rev. Lett.}\ }\textbf {\bibinfo
  {volume} {124}},\ \bibinfo {pages} {161301} (\bibinfo {year} {2020})},\
  \Eprint {http://arxiv.org/abs/1911.11760} {arXiv:1911.11760 [astro-ph.CO]}
  \BibitemShut {NoStop}%
\bibitem [{\citenamefont {Ye}\ and\ \citenamefont
  {Piao}(2020{\natexlab{a}})}]{Ye:2020btb}%
  \BibitemOpen
  \bibfield  {author} {\bibinfo {author} {\bibfnamefont {G.}~\bibnamefont
  {Ye}}\ and\ \bibinfo {author} {\bibfnamefont {Y.-S.}\ \bibnamefont {Piao}},\
  }\href {\doibase 10.1103/PhysRevD.101.083507} {\bibfield  {journal} {\bibinfo
   {journal} {Phys. Rev. D}\ }\textbf {\bibinfo {volume} {101}},\ \bibinfo
  {pages} {083507} (\bibinfo {year} {2020}{\natexlab{a}})},\ \Eprint
  {http://arxiv.org/abs/2001.02451} {arXiv:2001.02451 [astro-ph.CO]}
  \BibitemShut {NoStop}%
\bibitem [{\citenamefont {Zumalacarregui}(2020)}]{Zumalacarregui:2020cjh}%
  \BibitemOpen
  \bibfield  {author} {\bibinfo {author} {\bibfnamefont {M.}~\bibnamefont
  {Zumalacarregui}},\ }\href {\doibase 10.1103/PhysRevD.102.023523} {\bibfield
  {journal} {\bibinfo  {journal} {Phys. Rev. D}\ }\textbf {\bibinfo {volume}
  {102}},\ \bibinfo {pages} {023523} (\bibinfo {year} {2020})},\ \Eprint
  {http://arxiv.org/abs/2003.06396} {arXiv:2003.06396 [astro-ph.CO]}
  \BibitemShut {NoStop}%
\bibitem [{\citenamefont {Hill}\ \emph {et~al.}(2020)\citenamefont {Hill},
  \citenamefont {McDonough}, \citenamefont {Toomey},\ and\ \citenamefont
  {Alexander}}]{Hill:2020osr}%
  \BibitemOpen
  \bibfield  {author} {\bibinfo {author} {\bibfnamefont {J.~C.}\ \bibnamefont
  {Hill}}, \bibinfo {author} {\bibfnamefont {E.}~\bibnamefont {McDonough}},
  \bibinfo {author} {\bibfnamefont {M.~W.}\ \bibnamefont {Toomey}}, \ and\
  \bibinfo {author} {\bibfnamefont {S.}~\bibnamefont {Alexander}},\ }\href
  {\doibase 10.1103/PhysRevD.102.043507} {\bibfield  {journal} {\bibinfo
  {journal} {Phys. Rev. D}\ }\textbf {\bibinfo {volume} {102}},\ \bibinfo
  {pages} {043507} (\bibinfo {year} {2020})},\ \Eprint
  {http://arxiv.org/abs/2003.07355} {arXiv:2003.07355 [astro-ph.CO]}
  \BibitemShut {NoStop}%
\bibitem [{\citenamefont {Chudaykin}\ \emph {et~al.}(2020)\citenamefont
  {Chudaykin}, \citenamefont {Gorbunov},\ and\ \citenamefont
  {Nedelko}}]{Chudaykin:2020acu}%
  \BibitemOpen
  \bibfield  {author} {\bibinfo {author} {\bibfnamefont {A.}~\bibnamefont
  {Chudaykin}}, \bibinfo {author} {\bibfnamefont {D.}~\bibnamefont {Gorbunov}},
  \ and\ \bibinfo {author} {\bibfnamefont {N.}~\bibnamefont {Nedelko}},\ }\href
  {\doibase 10.1088/1475-7516/2020/08/013} {\bibfield  {journal} {\bibinfo
  {journal} {JCAP}\ }\textbf {\bibinfo {volume} {08}},\ \bibinfo {pages} {013}
  (\bibinfo {year} {2020})},\ \Eprint {http://arxiv.org/abs/2004.13046}
  {arXiv:2004.13046 [astro-ph.CO]} \BibitemShut {NoStop}%
\bibitem [{\citenamefont {Gogoi}\ \emph {et~al.}(2020)\citenamefont {Gogoi},
  \citenamefont {Sharma}, \citenamefont {Chanda},\ and\ \citenamefont
  {Das}}]{Das:2020wfe}%
  \BibitemOpen
  \bibfield  {author} {\bibinfo {author} {\bibfnamefont {A.}~\bibnamefont
  {Gogoi}}, \bibinfo {author} {\bibfnamefont {R.~K.}\ \bibnamefont {Sharma}},
  \bibinfo {author} {\bibfnamefont {P.}~\bibnamefont {Chanda}}, \ and\ \bibinfo
  {author} {\bibfnamefont {S.}~\bibnamefont {Das}},\ }\href@noop {} {\
  (\bibinfo {year} {2020})},\ \Eprint {http://arxiv.org/abs/2005.11889}
  {arXiv:2005.11889 [astro-ph.CO]} \BibitemShut {NoStop}%
\bibitem [{\citenamefont {Braglia}\ \emph {et~al.}(2020)\citenamefont
  {Braglia}, \citenamefont {Emond}, \citenamefont {Finelli}, \citenamefont
  {Gumrukcuoglu},\ and\ \citenamefont {Koyama}}]{Braglia:2020bym}%
  \BibitemOpen
  \bibfield  {author} {\bibinfo {author} {\bibfnamefont {M.}~\bibnamefont
  {Braglia}}, \bibinfo {author} {\bibfnamefont {W.~T.}\ \bibnamefont {Emond}},
  \bibinfo {author} {\bibfnamefont {F.}~\bibnamefont {Finelli}}, \bibinfo
  {author} {\bibfnamefont {A.~E.}\ \bibnamefont {Gumrukcuoglu}}, \ and\
  \bibinfo {author} {\bibfnamefont {K.}~\bibnamefont {Koyama}},\ }\href
  {\doibase 10.1103/PhysRevD.102.083513} {\bibfield  {journal} {\bibinfo
  {journal} {Phys. Rev. D}\ }\textbf {\bibinfo {volume} {102}},\ \bibinfo
  {pages} {083513} (\bibinfo {year} {2020})},\ \Eprint
  {http://arxiv.org/abs/2005.14053} {arXiv:2005.14053 [astro-ph.CO]}
  \BibitemShut {NoStop}%
\bibitem [{\citenamefont {Niedermann}\ and\ \citenamefont
  {Sloth}(2020)}]{Niedermann:2020dwg}%
  \BibitemOpen
  \bibfield  {author} {\bibinfo {author} {\bibfnamefont {F.}~\bibnamefont
  {Niedermann}}\ and\ \bibinfo {author} {\bibfnamefont {M.~S.}\ \bibnamefont
  {Sloth}},\ }\href {\doibase 10.1103/PhysRevD.102.063527} {\bibfield
  {journal} {\bibinfo  {journal} {Phys. Rev. D}\ }\textbf {\bibinfo {volume}
  {102}},\ \bibinfo {pages} {063527} (\bibinfo {year} {2020})},\ \Eprint
  {http://arxiv.org/abs/2006.06686} {arXiv:2006.06686 [astro-ph.CO]}
  \BibitemShut {NoStop}%
\bibitem [{\citenamefont {Ivanov}\ \emph {et~al.}(2020)\citenamefont {Ivanov},
  \citenamefont {McDonough}, \citenamefont {Hill}, \citenamefont {Simonovi\'c},
  \citenamefont {Toomey}, \citenamefont {Alexander},\ and\ \citenamefont
  {Zaldarriaga}}]{Ivanov:2020ril}%
  \BibitemOpen
  \bibfield  {author} {\bibinfo {author} {\bibfnamefont {M.~M.}\ \bibnamefont
  {Ivanov}}, \bibinfo {author} {\bibfnamefont {E.}~\bibnamefont {McDonough}},
  \bibinfo {author} {\bibfnamefont {J.~C.}\ \bibnamefont {Hill}}, \bibinfo
  {author} {\bibfnamefont {M.}~\bibnamefont {Simonovi\'c}}, \bibinfo {author}
  {\bibfnamefont {M.~W.}\ \bibnamefont {Toomey}}, \bibinfo {author}
  {\bibfnamefont {S.}~\bibnamefont {Alexander}}, \ and\ \bibinfo {author}
  {\bibfnamefont {M.}~\bibnamefont {Zaldarriaga}},\ }\href {\doibase
  10.1103/PhysRevD.102.103502} {\bibfield  {journal} {\bibinfo  {journal}
  {Phys. Rev. D}\ }\textbf {\bibinfo {volume} {102}},\ \bibinfo {pages}
  {103502} (\bibinfo {year} {2020})},\ \Eprint
  {http://arxiv.org/abs/2006.11235} {arXiv:2006.11235 [astro-ph.CO]}
  \BibitemShut {NoStop}%
\bibitem [{\citenamefont {D'Amico}\ \emph {et~al.}(2021)\citenamefont
  {D'Amico}, \citenamefont {Senatore}, \citenamefont {Zhang},\ and\
  \citenamefont {Zheng}}]{DAmico:2020ods}%
  \BibitemOpen
  \bibfield  {author} {\bibinfo {author} {\bibfnamefont {G.}~\bibnamefont
  {D'Amico}}, \bibinfo {author} {\bibfnamefont {L.}~\bibnamefont {Senatore}},
  \bibinfo {author} {\bibfnamefont {P.}~\bibnamefont {Zhang}}, \ and\ \bibinfo
  {author} {\bibfnamefont {H.}~\bibnamefont {Zheng}},\ }\href {\doibase
  10.1088/1475-7516/2021/05/072} {\bibfield  {journal} {\bibinfo  {journal}
  {JCAP}\ }\textbf {\bibinfo {volume} {05}},\ \bibinfo {pages} {072} (\bibinfo
  {year} {2021})},\ \Eprint {http://arxiv.org/abs/2006.12420} {arXiv:2006.12420
  [astro-ph.CO]} \BibitemShut {NoStop}%
\bibitem [{\citenamefont {Ye}\ and\ \citenamefont
  {Piao}(2020{\natexlab{b}})}]{Ye:2020oix}%
  \BibitemOpen
  \bibfield  {author} {\bibinfo {author} {\bibfnamefont {G.}~\bibnamefont
  {Ye}}\ and\ \bibinfo {author} {\bibfnamefont {Y.-S.}\ \bibnamefont {Piao}},\
  }\href {\doibase 10.1103/PhysRevD.102.083523} {\bibfield  {journal} {\bibinfo
   {journal} {Phys. Rev. D}\ }\textbf {\bibinfo {volume} {102}},\ \bibinfo
  {pages} {083523} (\bibinfo {year} {2020}{\natexlab{b}})},\ \Eprint
  {http://arxiv.org/abs/2008.10832} {arXiv:2008.10832 [astro-ph.CO]}
  \BibitemShut {NoStop}%
\bibitem [{\citenamefont {Niedermann}\ and\ \citenamefont
  {Sloth}(2021{\natexlab{b}})}]{Niedermann:2020qbw}%
  \BibitemOpen
  \bibfield  {author} {\bibinfo {author} {\bibfnamefont {F.}~\bibnamefont
  {Niedermann}}\ and\ \bibinfo {author} {\bibfnamefont {M.~S.}\ \bibnamefont
  {Sloth}},\ }\href {\doibase 10.1103/PhysRevD.103.103537} {\bibfield
  {journal} {\bibinfo  {journal} {Phys. Rev. D}\ }\textbf {\bibinfo {volume}
  {103}},\ \bibinfo {pages} {103537} (\bibinfo {year} {2021}{\natexlab{b}})},\
  \Eprint {http://arxiv.org/abs/2009.00006} {arXiv:2009.00006 [astro-ph.CO]}
  \BibitemShut {NoStop}%
\bibitem [{\citenamefont {Murgia}\ \emph {et~al.}(2021)\citenamefont {Murgia},
  \citenamefont {Abell\'an},\ and\ \citenamefont {Poulin}}]{Murgia:2020ryi}%
  \BibitemOpen
  \bibfield  {author} {\bibinfo {author} {\bibfnamefont {R.}~\bibnamefont
  {Murgia}}, \bibinfo {author} {\bibfnamefont {G.~F.}\ \bibnamefont
  {Abell\'an}}, \ and\ \bibinfo {author} {\bibfnamefont {V.}~\bibnamefont
  {Poulin}},\ }\href {\doibase 10.1103/PhysRevD.103.063502} {\bibfield
  {journal} {\bibinfo  {journal} {Phys. Rev. D}\ }\textbf {\bibinfo {volume}
  {103}},\ \bibinfo {pages} {063502} (\bibinfo {year} {2021})},\ \Eprint
  {http://arxiv.org/abs/2009.10733} {arXiv:2009.10733 [astro-ph.CO]}
  \BibitemShut {NoStop}%
\bibitem [{\citenamefont {Smith}\ \emph {et~al.}(2020)\citenamefont {Smith},
  \citenamefont {Poulin}, \citenamefont {Bernal}, \citenamefont {Boddy},
  \citenamefont {Kamionkowski},\ and\ \citenamefont {Murgia}}]{Smith:2020rxx}%
  \BibitemOpen
  \bibfield  {author} {\bibinfo {author} {\bibfnamefont {T.~L.}\ \bibnamefont
  {Smith}}, \bibinfo {author} {\bibfnamefont {V.}~\bibnamefont {Poulin}},
  \bibinfo {author} {\bibfnamefont {J.~L.}\ \bibnamefont {Bernal}}, \bibinfo
  {author} {\bibfnamefont {K.~K.}\ \bibnamefont {Boddy}}, \bibinfo {author}
  {\bibfnamefont {M.}~\bibnamefont {Kamionkowski}}, \ and\ \bibinfo {author}
  {\bibfnamefont {R.}~\bibnamefont {Murgia}},\ }\href@noop {} {\  (\bibinfo
  {year} {2020})},\ \Eprint {http://arxiv.org/abs/2009.10740} {arXiv:2009.10740
  [astro-ph.CO]} \BibitemShut {NoStop}%
\bibitem [{\citenamefont {Chudaykin}\ \emph {et~al.}(2021)\citenamefont
  {Chudaykin}, \citenamefont {Gorbunov},\ and\ \citenamefont
  {Nedelko}}]{Chudaykin:2020igl}%
  \BibitemOpen
  \bibfield  {author} {\bibinfo {author} {\bibfnamefont {A.}~\bibnamefont
  {Chudaykin}}, \bibinfo {author} {\bibfnamefont {D.}~\bibnamefont {Gorbunov}},
  \ and\ \bibinfo {author} {\bibfnamefont {N.}~\bibnamefont {Nedelko}},\ }\href
  {\doibase 10.1103/PhysRevD.103.043529} {\bibfield  {journal} {\bibinfo
  {journal} {Phys. Rev. D}\ }\textbf {\bibinfo {volume} {103}},\ \bibinfo
  {pages} {043529} (\bibinfo {year} {2021})},\ \Eprint
  {http://arxiv.org/abs/2011.04682} {arXiv:2011.04682 [astro-ph.CO]}
  \BibitemShut {NoStop}%
\bibitem [{\citenamefont {Carrillo~Gonz\'alez}\ \emph
  {et~al.}(2021)\citenamefont {Carrillo~Gonz\'alez}, \citenamefont {Liang},
  \citenamefont {Sakstein},\ and\ \citenamefont
  {Trodden}}]{CarrilloGonzalez:2020oac}%
  \BibitemOpen
  \bibfield  {author} {\bibinfo {author} {\bibfnamefont {M.}~\bibnamefont
  {Carrillo~Gonz\'alez}}, \bibinfo {author} {\bibfnamefont {Q.}~\bibnamefont
  {Liang}}, \bibinfo {author} {\bibfnamefont {J.}~\bibnamefont {Sakstein}}, \
  and\ \bibinfo {author} {\bibfnamefont {M.}~\bibnamefont {Trodden}},\ }\href
  {\doibase 10.1088/1475-7516/2021/04/063} {\bibfield  {journal} {\bibinfo
  {journal} {JCAP}\ }\textbf {\bibinfo {volume} {04}},\ \bibinfo {pages} {063}
  (\bibinfo {year} {2021})},\ \Eprint {http://arxiv.org/abs/2011.09895}
  {arXiv:2011.09895 [astro-ph.CO]} \BibitemShut {NoStop}%
\bibitem [{\citenamefont {Oikonomou}(2021)}]{Oikonomou:2020qah}%
  \BibitemOpen
  \bibfield  {author} {\bibinfo {author} {\bibfnamefont {V.~K.}\ \bibnamefont
  {Oikonomou}},\ }\href {\doibase 10.1103/PhysRevD.103.044036} {\bibfield
  {journal} {\bibinfo  {journal} {Phys. Rev. D}\ }\textbf {\bibinfo {volume}
  {103}},\ \bibinfo {pages} {044036} (\bibinfo {year} {2021})},\ \Eprint
  {http://arxiv.org/abs/2012.00586} {arXiv:2012.00586 [astro-ph.CO]}
  \BibitemShut {NoStop}%
\bibitem [{\citenamefont {Seto}\ and\ \citenamefont
  {Toda}(2021)}]{Seto:2021xua}%
  \BibitemOpen
  \bibfield  {author} {\bibinfo {author} {\bibfnamefont {O.}~\bibnamefont
  {Seto}}\ and\ \bibinfo {author} {\bibfnamefont {Y.}~\bibnamefont {Toda}},\
  }\href {\doibase 10.1103/PhysRevD.103.123501} {\bibfield  {journal} {\bibinfo
   {journal} {Phys. Rev. D}\ }\textbf {\bibinfo {volume} {103}},\ \bibinfo
  {pages} {123501} (\bibinfo {year} {2021})},\ \Eprint
  {http://arxiv.org/abs/2101.03740} {arXiv:2101.03740 [astro-ph.CO]}
  \BibitemShut {NoStop}%
\bibitem [{\citenamefont {Tian}\ and\ \citenamefont
  {Zhu}(2021)}]{Tian:2021omz}%
  \BibitemOpen
  \bibfield  {author} {\bibinfo {author} {\bibfnamefont {S.~X.}\ \bibnamefont
  {Tian}}\ and\ \bibinfo {author} {\bibfnamefont {Z.-H.}\ \bibnamefont {Zhu}},\
  }\href {\doibase 10.1103/PhysRevD.103.043518} {\bibfield  {journal} {\bibinfo
   {journal} {Phys. Rev. D}\ }\textbf {\bibinfo {volume} {103}},\ \bibinfo
  {pages} {043518} (\bibinfo {year} {2021})},\ \Eprint
  {http://arxiv.org/abs/2102.06399} {arXiv:2102.06399 [gr-qc]} \BibitemShut
  {NoStop}%
\bibitem [{\citenamefont {Freese}\ and\ \citenamefont
  {Winkler}(2021)}]{Freese:2021rjq}%
  \BibitemOpen
  \bibfield  {author} {\bibinfo {author} {\bibfnamefont {K.}~\bibnamefont
  {Freese}}\ and\ \bibinfo {author} {\bibfnamefont {M.~W.}\ \bibnamefont
  {Winkler}},\ }\href@noop {} {\  (\bibinfo {year} {2021})},\ \Eprint
  {http://arxiv.org/abs/2102.13655} {arXiv:2102.13655 [astro-ph.CO]}
  \BibitemShut {NoStop}%
\bibitem [{\citenamefont {Nojiri}\ \emph {et~al.}(2021)\citenamefont {Nojiri},
  \citenamefont {Odintsov}, \citenamefont {Saez-Chillon~Gomez},\ and\
  \citenamefont {Sharov}}]{Nojiri:2021dze}%
  \BibitemOpen
  \bibfield  {author} {\bibinfo {author} {\bibfnamefont {S.}~\bibnamefont
  {Nojiri}}, \bibinfo {author} {\bibfnamefont {S.~D.}\ \bibnamefont
  {Odintsov}}, \bibinfo {author} {\bibfnamefont {D.}~\bibnamefont
  {Saez-Chillon~Gomez}}, \ and\ \bibinfo {author} {\bibfnamefont {G.~S.}\
  \bibnamefont {Sharov}},\ }\href {\doibase 10.1016/j.dark.2021.100837}
  {\bibfield  {journal} {\bibinfo  {journal} {Phys. Dark Univ.}\ }\textbf
  {\bibinfo {volume} {32}},\ \bibinfo {pages} {100837} (\bibinfo {year}
  {2021})},\ \Eprint {http://arxiv.org/abs/2103.05304} {arXiv:2103.05304
  [gr-qc]} \BibitemShut {NoStop}%
\bibitem [{\citenamefont {Vagnozzi}\ \emph {et~al.}(2018)\citenamefont
  {Vagnozzi}, \citenamefont {Dhawan}, \citenamefont {Gerbino}, \citenamefont
  {Freese}, \citenamefont {Goobar},\ and\ \citenamefont
  {Mena}}]{Vagnozzi:2018jhn}%
  \BibitemOpen
  \bibfield  {author} {\bibinfo {author} {\bibfnamefont {S.}~\bibnamefont
  {Vagnozzi}}, \bibinfo {author} {\bibfnamefont {S.}~\bibnamefont {Dhawan}},
  \bibinfo {author} {\bibfnamefont {M.}~\bibnamefont {Gerbino}}, \bibinfo
  {author} {\bibfnamefont {K.}~\bibnamefont {Freese}}, \bibinfo {author}
  {\bibfnamefont {A.}~\bibnamefont {Goobar}}, \ and\ \bibinfo {author}
  {\bibfnamefont {O.}~\bibnamefont {Mena}},\ }\href {\doibase
  10.1103/PhysRevD.98.083501} {\bibfield  {journal} {\bibinfo  {journal} {Phys.
  Rev. D}\ }\textbf {\bibinfo {volume} {98}},\ \bibinfo {pages} {083501}
  (\bibinfo {year} {2018})},\ \Eprint {http://arxiv.org/abs/1801.08553}
  {arXiv:1801.08553 [astro-ph.CO]} \BibitemShut {NoStop}%
\bibitem [{\citenamefont {Nunes}(2018)}]{Nunes:2018xbm}%
  \BibitemOpen
  \bibfield  {author} {\bibinfo {author} {\bibfnamefont {R.~C.}\ \bibnamefont
  {Nunes}},\ }\href {\doibase 10.1088/1475-7516/2018/05/052} {\bibfield
  {journal} {\bibinfo  {journal} {JCAP}\ }\textbf {\bibinfo {volume} {05}},\
  \bibinfo {pages} {052} (\bibinfo {year} {2018})},\ \Eprint
  {http://arxiv.org/abs/1802.02281} {arXiv:1802.02281 [gr-qc]} \BibitemShut
  {NoStop}%
\bibitem [{\citenamefont {Poulin}\ \emph {et~al.}(2018)\citenamefont {Poulin},
  \citenamefont {Boddy}, \citenamefont {Bird},\ and\ \citenamefont
  {Kamionkowski}}]{Poulin:2018zxs}%
  \BibitemOpen
  \bibfield  {author} {\bibinfo {author} {\bibfnamefont {V.}~\bibnamefont
  {Poulin}}, \bibinfo {author} {\bibfnamefont {K.~K.}\ \bibnamefont {Boddy}},
  \bibinfo {author} {\bibfnamefont {S.}~\bibnamefont {Bird}}, \ and\ \bibinfo
  {author} {\bibfnamefont {M.}~\bibnamefont {Kamionkowski}},\ }\href {\doibase
  10.1103/PhysRevD.97.123504} {\bibfield  {journal} {\bibinfo  {journal} {Phys.
  Rev. D}\ }\textbf {\bibinfo {volume} {97}},\ \bibinfo {pages} {123504}
  (\bibinfo {year} {2018})},\ \Eprint {http://arxiv.org/abs/1803.02474}
  {arXiv:1803.02474 [astro-ph.CO]} \BibitemShut {NoStop}%
\bibitem [{\citenamefont {Yang}\ \emph {et~al.}(2018)\citenamefont {Yang},
  \citenamefont {Pan}, \citenamefont {Di~Valentino}, \citenamefont {Nunes},
  \citenamefont {Vagnozzi},\ and\ \citenamefont {Mota}}]{Yang:2018euj}%
  \BibitemOpen
  \bibfield  {author} {\bibinfo {author} {\bibfnamefont {W.}~\bibnamefont
  {Yang}}, \bibinfo {author} {\bibfnamefont {S.}~\bibnamefont {Pan}}, \bibinfo
  {author} {\bibfnamefont {E.}~\bibnamefont {Di~Valentino}}, \bibinfo {author}
  {\bibfnamefont {R.~C.}\ \bibnamefont {Nunes}}, \bibinfo {author}
  {\bibfnamefont {S.}~\bibnamefont {Vagnozzi}}, \ and\ \bibinfo {author}
  {\bibfnamefont {D.~F.}\ \bibnamefont {Mota}},\ }\href {\doibase
  10.1088/1475-7516/2018/09/019} {\bibfield  {journal} {\bibinfo  {journal}
  {JCAP}\ }\textbf {\bibinfo {volume} {09}},\ \bibinfo {pages} {019} (\bibinfo
  {year} {2018})},\ \Eprint {http://arxiv.org/abs/1805.08252} {arXiv:1805.08252
  [astro-ph.CO]} \BibitemShut {NoStop}%
\bibitem [{\citenamefont {Banihashemi}\ \emph {et~al.}(2020)\citenamefont
  {Banihashemi}, \citenamefont {Khosravi},\ and\ \citenamefont
  {Shirazi}}]{Banihashemi:2018oxo}%
  \BibitemOpen
  \bibfield  {author} {\bibinfo {author} {\bibfnamefont {A.}~\bibnamefont
  {Banihashemi}}, \bibinfo {author} {\bibfnamefont {N.}~\bibnamefont
  {Khosravi}}, \ and\ \bibinfo {author} {\bibfnamefont {A.~H.}\ \bibnamefont
  {Shirazi}},\ }\href {\doibase 10.1103/PhysRevD.101.123521} {\bibfield
  {journal} {\bibinfo  {journal} {Phys. Rev. D}\ }\textbf {\bibinfo {volume}
  {101}},\ \bibinfo {pages} {123521} (\bibinfo {year} {2020})},\ \Eprint
  {http://arxiv.org/abs/1808.02472} {arXiv:1808.02472 [astro-ph.CO]}
  \BibitemShut {NoStop}%
\bibitem [{\citenamefont {Guo}\ \emph {et~al.}(2019)\citenamefont {Guo},
  \citenamefont {Zhang},\ and\ \citenamefont {Zhang}}]{Guo:2018ans}%
  \BibitemOpen
  \bibfield  {author} {\bibinfo {author} {\bibfnamefont {R.-Y.}\ \bibnamefont
  {Guo}}, \bibinfo {author} {\bibfnamefont {J.-F.}\ \bibnamefont {Zhang}}, \
  and\ \bibinfo {author} {\bibfnamefont {X.}~\bibnamefont {Zhang}},\ }\href
  {\doibase 10.1088/1475-7516/2019/02/054} {\bibfield  {journal} {\bibinfo
  {journal} {JCAP}\ }\textbf {\bibinfo {volume} {02}},\ \bibinfo {pages} {054}
  (\bibinfo {year} {2019})},\ \Eprint {http://arxiv.org/abs/1809.02340}
  {arXiv:1809.02340 [astro-ph.CO]} \BibitemShut {NoStop}%
\bibitem [{\citenamefont {Graef}\ \emph {et~al.}(2019)\citenamefont {Graef},
  \citenamefont {Benetti},\ and\ \citenamefont {Alcaniz}}]{Graef:2018fzu}%
  \BibitemOpen
  \bibfield  {author} {\bibinfo {author} {\bibfnamefont {L.~L.}\ \bibnamefont
  {Graef}}, \bibinfo {author} {\bibfnamefont {M.}~\bibnamefont {Benetti}}, \
  and\ \bibinfo {author} {\bibfnamefont {J.~S.}\ \bibnamefont {Alcaniz}},\
  }\href {\doibase 10.1103/PhysRevD.99.043519} {\bibfield  {journal} {\bibinfo
  {journal} {Phys. Rev. D}\ }\textbf {\bibinfo {volume} {99}},\ \bibinfo
  {pages} {043519} (\bibinfo {year} {2019})},\ \Eprint
  {http://arxiv.org/abs/1809.04501} {arXiv:1809.04501 [astro-ph.CO]}
  \BibitemShut {NoStop}%
\bibitem [{\citenamefont {Kreisch}\ \emph {et~al.}(2020)\citenamefont
  {Kreisch}, \citenamefont {Cyr-Racine},\ and\ \citenamefont
  {Dor\'e}}]{Kreisch:2019yzn}%
  \BibitemOpen
  \bibfield  {author} {\bibinfo {author} {\bibfnamefont {C.~D.}\ \bibnamefont
  {Kreisch}}, \bibinfo {author} {\bibfnamefont {F.-Y.}\ \bibnamefont
  {Cyr-Racine}}, \ and\ \bibinfo {author} {\bibfnamefont {O.}~\bibnamefont
  {Dor\'e}},\ }\href {\doibase 10.1103/PhysRevD.101.123505} {\bibfield
  {journal} {\bibinfo  {journal} {Phys. Rev. D}\ }\textbf {\bibinfo {volume}
  {101}},\ \bibinfo {pages} {123505} (\bibinfo {year} {2020})},\ \Eprint
  {http://arxiv.org/abs/1902.00534} {arXiv:1902.00534 [astro-ph.CO]}
  \BibitemShut {NoStop}%
\bibitem [{\citenamefont {Pandey}\ \emph {et~al.}(2020)\citenamefont {Pandey},
  \citenamefont {Karwal},\ and\ \citenamefont {Das}}]{Pandey:2019plg}%
  \BibitemOpen
  \bibfield  {author} {\bibinfo {author} {\bibfnamefont {K.~L.}\ \bibnamefont
  {Pandey}}, \bibinfo {author} {\bibfnamefont {T.}~\bibnamefont {Karwal}}, \
  and\ \bibinfo {author} {\bibfnamefont {S.}~\bibnamefont {Das}},\ }\href
  {\doibase 10.1088/1475-7516/2020/07/026} {\bibfield  {journal} {\bibinfo
  {journal} {JCAP}\ }\textbf {\bibinfo {volume} {07}},\ \bibinfo {pages} {026}
  (\bibinfo {year} {2020})},\ \Eprint {http://arxiv.org/abs/1902.10636}
  {arXiv:1902.10636 [astro-ph.CO]} \BibitemShut {NoStop}%
\bibitem [{\citenamefont {Martinelli}\ \emph {et~al.}(2019)\citenamefont
  {Martinelli}, \citenamefont {Hogg}, \citenamefont {Peirone}, \citenamefont
  {Bruni},\ and\ \citenamefont {Wands}}]{Martinelli:2019dau}%
  \BibitemOpen
  \bibfield  {author} {\bibinfo {author} {\bibfnamefont {M.}~\bibnamefont
  {Martinelli}}, \bibinfo {author} {\bibfnamefont {N.~B.}\ \bibnamefont
  {Hogg}}, \bibinfo {author} {\bibfnamefont {S.}~\bibnamefont {Peirone}},
  \bibinfo {author} {\bibfnamefont {M.}~\bibnamefont {Bruni}}, \ and\ \bibinfo
  {author} {\bibfnamefont {D.}~\bibnamefont {Wands}},\ }\href {\doibase
  10.1093/mnras/stz1915} {\bibfield  {journal} {\bibinfo  {journal} {Mon. Not.
  Roy. Astron. Soc.}\ }\textbf {\bibinfo {volume} {488}},\ \bibinfo {pages}
  {3423} (\bibinfo {year} {2019})},\ \Eprint {http://arxiv.org/abs/1902.10694}
  {arXiv:1902.10694 [astro-ph.CO]} \BibitemShut {NoStop}%
\bibitem [{\citenamefont {Vattis}\ \emph {et~al.}(2019)\citenamefont {Vattis},
  \citenamefont {Koushiappas},\ and\ \citenamefont {Loeb}}]{Vattis:2019efj}%
  \BibitemOpen
  \bibfield  {author} {\bibinfo {author} {\bibfnamefont {K.}~\bibnamefont
  {Vattis}}, \bibinfo {author} {\bibfnamefont {S.~M.}\ \bibnamefont
  {Koushiappas}}, \ and\ \bibinfo {author} {\bibfnamefont {A.}~\bibnamefont
  {Loeb}},\ }\href {\doibase 10.1103/PhysRevD.99.121302} {\bibfield  {journal}
  {\bibinfo  {journal} {Phys. Rev. D}\ }\textbf {\bibinfo {volume} {99}},\
  \bibinfo {pages} {121302} (\bibinfo {year} {2019})},\ \Eprint
  {http://arxiv.org/abs/1903.06220} {arXiv:1903.06220 [astro-ph.CO]}
  \BibitemShut {NoStop}%
\bibitem [{\citenamefont {Lin}\ \emph {et~al.}(2019)\citenamefont {Lin},
  \citenamefont {Benevento}, \citenamefont {Hu},\ and\ \citenamefont
  {Raveri}}]{Lin:2019qug}%
  \BibitemOpen
  \bibfield  {author} {\bibinfo {author} {\bibfnamefont {M.-X.}\ \bibnamefont
  {Lin}}, \bibinfo {author} {\bibfnamefont {G.}~\bibnamefont {Benevento}},
  \bibinfo {author} {\bibfnamefont {W.}~\bibnamefont {Hu}}, \ and\ \bibinfo
  {author} {\bibfnamefont {M.}~\bibnamefont {Raveri}},\ }\href {\doibase
  10.1103/PhysRevD.100.063542} {\bibfield  {journal} {\bibinfo  {journal}
  {Phys. Rev. D}\ }\textbf {\bibinfo {volume} {100}},\ \bibinfo {pages}
  {063542} (\bibinfo {year} {2019})},\ \Eprint
  {http://arxiv.org/abs/1905.12618} {arXiv:1905.12618 [astro-ph.CO]}
  \BibitemShut {NoStop}%
\bibitem [{\citenamefont {Li}\ and\ \citenamefont
  {Shafieloo}(2019)}]{Li:2019yem}%
  \BibitemOpen
  \bibfield  {author} {\bibinfo {author} {\bibfnamefont {X.}~\bibnamefont
  {Li}}\ and\ \bibinfo {author} {\bibfnamefont {A.}~\bibnamefont {Shafieloo}},\
  }\href {\doibase 10.3847/2041-8213/ab3e09} {\bibfield  {journal} {\bibinfo
  {journal} {Astrophys. J. Lett.}\ }\textbf {\bibinfo {volume} {883}},\
  \bibinfo {pages} {L3} (\bibinfo {year} {2019})},\ \Eprint
  {http://arxiv.org/abs/1906.08275} {arXiv:1906.08275 [astro-ph.CO]}
  \BibitemShut {NoStop}%
\bibitem [{\citenamefont {Di~Valentino}\ \emph
  {et~al.}(2019{\natexlab{a}})\citenamefont {Di~Valentino}, \citenamefont
  {Ferreira}, \citenamefont {Visinelli},\ and\ \citenamefont
  {Danielsson}}]{DiValentino:2019exe}%
  \BibitemOpen
  \bibfield  {author} {\bibinfo {author} {\bibfnamefont {E.}~\bibnamefont
  {Di~Valentino}}, \bibinfo {author} {\bibfnamefont {R.~Z.}\ \bibnamefont
  {Ferreira}}, \bibinfo {author} {\bibfnamefont {L.}~\bibnamefont {Visinelli}},
  \ and\ \bibinfo {author} {\bibfnamefont {U.}~\bibnamefont {Danielsson}},\
  }\href {\doibase 10.1016/j.dark.2019.100385} {\bibfield  {journal} {\bibinfo
  {journal} {Phys. Dark Univ.}\ }\textbf {\bibinfo {volume} {26}},\ \bibinfo
  {pages} {100385} (\bibinfo {year} {2019}{\natexlab{a}})},\ \Eprint
  {http://arxiv.org/abs/1906.11255} {arXiv:1906.11255 [astro-ph.CO]}
  \BibitemShut {NoStop}%
\bibitem [{\citenamefont {Yang}\ \emph {et~al.}(2019)\citenamefont {Yang},
  \citenamefont {Pan}, \citenamefont {Vagnozzi}, \citenamefont {Di~Valentino},
  \citenamefont {Mota},\ and\ \citenamefont {Capozziello}}]{Yang:2019nhz}%
  \BibitemOpen
  \bibfield  {author} {\bibinfo {author} {\bibfnamefont {W.}~\bibnamefont
  {Yang}}, \bibinfo {author} {\bibfnamefont {S.}~\bibnamefont {Pan}}, \bibinfo
  {author} {\bibfnamefont {S.}~\bibnamefont {Vagnozzi}}, \bibinfo {author}
  {\bibfnamefont {E.}~\bibnamefont {Di~Valentino}}, \bibinfo {author}
  {\bibfnamefont {D.~F.}\ \bibnamefont {Mota}}, \ and\ \bibinfo {author}
  {\bibfnamefont {S.}~\bibnamefont {Capozziello}},\ }\href {\doibase
  10.1088/1475-7516/2019/11/044} {\bibfield  {journal} {\bibinfo  {journal}
  {JCAP}\ }\textbf {\bibinfo {volume} {11}},\ \bibinfo {pages} {044} (\bibinfo
  {year} {2019})},\ \Eprint {http://arxiv.org/abs/1907.05344} {arXiv:1907.05344
  [astro-ph.CO]} \BibitemShut {NoStop}%
\bibitem [{\citenamefont {Pan}\ \emph {et~al.}(2019)\citenamefont {Pan},
  \citenamefont {Yang}, \citenamefont {Di~Valentino}, \citenamefont
  {Saridakis},\ and\ \citenamefont {Chakraborty}}]{Pan:2019gop}%
  \BibitemOpen
  \bibfield  {author} {\bibinfo {author} {\bibfnamefont {S.}~\bibnamefont
  {Pan}}, \bibinfo {author} {\bibfnamefont {W.}~\bibnamefont {Yang}}, \bibinfo
  {author} {\bibfnamefont {E.}~\bibnamefont {Di~Valentino}}, \bibinfo {author}
  {\bibfnamefont {E.~N.}\ \bibnamefont {Saridakis}}, \ and\ \bibinfo {author}
  {\bibfnamefont {S.}~\bibnamefont {Chakraborty}},\ }\href {\doibase
  10.1103/PhysRevD.100.103520} {\bibfield  {journal} {\bibinfo  {journal}
  {Phys. Rev. D}\ }\textbf {\bibinfo {volume} {100}},\ \bibinfo {pages}
  {103520} (\bibinfo {year} {2019})},\ \Eprint
  {http://arxiv.org/abs/1907.07540} {arXiv:1907.07540 [astro-ph.CO]}
  \BibitemShut {NoStop}%
\bibitem [{\citenamefont {Vagnozzi}(2020)}]{Vagnozzi:2019ezj}%
  \BibitemOpen
  \bibfield  {author} {\bibinfo {author} {\bibfnamefont {S.}~\bibnamefont
  {Vagnozzi}},\ }\href {\doibase 10.1103/PhysRevD.102.023518} {\bibfield
  {journal} {\bibinfo  {journal} {Phys. Rev. D}\ }\textbf {\bibinfo {volume}
  {102}},\ \bibinfo {pages} {023518} (\bibinfo {year} {2020})},\ \Eprint
  {http://arxiv.org/abs/1907.07569} {arXiv:1907.07569 [astro-ph.CO]}
  \BibitemShut {NoStop}%
\bibitem [{\citenamefont {Visinelli}\ \emph {et~al.}(2019)\citenamefont
  {Visinelli}, \citenamefont {Vagnozzi},\ and\ \citenamefont
  {Danielsson}}]{Visinelli:2019qqu}%
  \BibitemOpen
  \bibfield  {author} {\bibinfo {author} {\bibfnamefont {L.}~\bibnamefont
  {Visinelli}}, \bibinfo {author} {\bibfnamefont {S.}~\bibnamefont {Vagnozzi}},
  \ and\ \bibinfo {author} {\bibfnamefont {U.}~\bibnamefont {Danielsson}},\
  }\href {\doibase 10.3390/sym11081035} {\bibfield  {journal} {\bibinfo
  {journal} {Symmetry}\ }\textbf {\bibinfo {volume} {11}},\ \bibinfo {pages}
  {1035} (\bibinfo {year} {2019})},\ \Eprint {http://arxiv.org/abs/1907.07953}
  {arXiv:1907.07953 [astro-ph.CO]} \BibitemShut {NoStop}%
\bibitem [{\citenamefont {Cai}\ \emph {et~al.}(2020)\citenamefont {Cai},
  \citenamefont {Khurshudyan},\ and\ \citenamefont {Saridakis}}]{Cai:2019bdh}%
  \BibitemOpen
  \bibfield  {author} {\bibinfo {author} {\bibfnamefont {Y.-F.}\ \bibnamefont
  {Cai}}, \bibinfo {author} {\bibfnamefont {M.}~\bibnamefont {Khurshudyan}}, \
  and\ \bibinfo {author} {\bibfnamefont {E.~N.}\ \bibnamefont {Saridakis}},\
  }\href {\doibase 10.3847/1538-4357/ab5a7f} {\bibfield  {journal} {\bibinfo
  {journal} {Astrophys. J.}\ }\textbf {\bibinfo {volume} {888}},\ \bibinfo
  {pages} {62} (\bibinfo {year} {2020})},\ \Eprint
  {http://arxiv.org/abs/1907.10813} {arXiv:1907.10813 [astro-ph.CO]}
  \BibitemShut {NoStop}%
\bibitem [{\citenamefont {Sch\"oneberg}\ \emph {et~al.}(2019)\citenamefont
  {Sch\"oneberg}, \citenamefont {Lesgourgues},\ and\ \citenamefont
  {Hooper}}]{Schoneberg:2019wmt}%
  \BibitemOpen
  \bibfield  {author} {\bibinfo {author} {\bibfnamefont {N.}~\bibnamefont
  {Sch\"oneberg}}, \bibinfo {author} {\bibfnamefont {J.}~\bibnamefont
  {Lesgourgues}}, \ and\ \bibinfo {author} {\bibfnamefont {D.~C.}\ \bibnamefont
  {Hooper}},\ }\href {\doibase 10.1088/1475-7516/2019/10/029} {\bibfield
  {journal} {\bibinfo  {journal} {JCAP}\ }\textbf {\bibinfo {volume} {10}},\
  \bibinfo {pages} {029} (\bibinfo {year} {2019})},\ \Eprint
  {http://arxiv.org/abs/1907.11594} {arXiv:1907.11594 [astro-ph.CO]}
  \BibitemShut {NoStop}%
\bibitem [{\citenamefont {Pan}\ \emph {et~al.}(2020)\citenamefont {Pan},
  \citenamefont {Yang}, \citenamefont {Di~Valentino}, \citenamefont
  {Shafieloo},\ and\ \citenamefont {Chakraborty}}]{Pan:2019hac}%
  \BibitemOpen
  \bibfield  {author} {\bibinfo {author} {\bibfnamefont {S.}~\bibnamefont
  {Pan}}, \bibinfo {author} {\bibfnamefont {W.}~\bibnamefont {Yang}}, \bibinfo
  {author} {\bibfnamefont {E.}~\bibnamefont {Di~Valentino}}, \bibinfo {author}
  {\bibfnamefont {A.}~\bibnamefont {Shafieloo}}, \ and\ \bibinfo {author}
  {\bibfnamefont {S.}~\bibnamefont {Chakraborty}},\ }\href {\doibase
  10.1088/1475-7516/2020/06/062} {\bibfield  {journal} {\bibinfo  {journal}
  {JCAP}\ }\textbf {\bibinfo {volume} {06}},\ \bibinfo {pages} {062} (\bibinfo
  {year} {2020})},\ \Eprint {http://arxiv.org/abs/1907.12551} {arXiv:1907.12551
  [astro-ph.CO]} \BibitemShut {NoStop}%
\bibitem [{\citenamefont {Di~Valentino}\ \emph
  {et~al.}(2020{\natexlab{a}})\citenamefont {Di~Valentino}, \citenamefont
  {Melchiorri}, \citenamefont {Mena},\ and\ \citenamefont
  {Vagnozzi}}]{DiValentino:2019ffd}%
  \BibitemOpen
  \bibfield  {author} {\bibinfo {author} {\bibfnamefont {E.}~\bibnamefont
  {Di~Valentino}}, \bibinfo {author} {\bibfnamefont {A.}~\bibnamefont
  {Melchiorri}}, \bibinfo {author} {\bibfnamefont {O.}~\bibnamefont {Mena}}, \
  and\ \bibinfo {author} {\bibfnamefont {S.}~\bibnamefont {Vagnozzi}},\ }\href
  {\doibase 10.1016/j.dark.2020.100666} {\bibfield  {journal} {\bibinfo
  {journal} {Phys. Dark Univ.}\ }\textbf {\bibinfo {volume} {30}},\ \bibinfo
  {pages} {100666} (\bibinfo {year} {2020}{\natexlab{a}})},\ \Eprint
  {http://arxiv.org/abs/1908.04281} {arXiv:1908.04281 [astro-ph.CO]}
  \BibitemShut {NoStop}%
\bibitem [{\citenamefont {Sol\`a~Peracaula}\ \emph {et~al.}(2019)\citenamefont
  {Sol\`a~Peracaula}, \citenamefont {Gomez-Valent}, \citenamefont
  {de~Cruz~P\'erez},\ and\ \citenamefont {Moreno-Pulido}}]{Sola:2019jek}%
  \BibitemOpen
  \bibfield  {author} {\bibinfo {author} {\bibfnamefont {J.}~\bibnamefont
  {Sol\`a~Peracaula}}, \bibinfo {author} {\bibfnamefont {A.}~\bibnamefont
  {Gomez-Valent}}, \bibinfo {author} {\bibfnamefont {J.}~\bibnamefont
  {de~Cruz~P\'erez}}, \ and\ \bibinfo {author} {\bibfnamefont {C.}~\bibnamefont
  {Moreno-Pulido}},\ }\href {\doibase 10.3847/2041-8213/ab53e9} {\bibfield
  {journal} {\bibinfo  {journal} {Astrophys. J. Lett.}\ }\textbf {\bibinfo
  {volume} {886}},\ \bibinfo {pages} {L6} (\bibinfo {year} {2019})},\ \Eprint
  {http://arxiv.org/abs/1909.02554} {arXiv:1909.02554 [astro-ph.CO]}
  \BibitemShut {NoStop}%
\bibitem [{\citenamefont {Escudero}\ and\ \citenamefont
  {Witte}(2020)}]{Escudero:2019gvw}%
  \BibitemOpen
  \bibfield  {author} {\bibinfo {author} {\bibfnamefont {M.}~\bibnamefont
  {Escudero}}\ and\ \bibinfo {author} {\bibfnamefont {S.~J.}\ \bibnamefont
  {Witte}},\ }\href {\doibase 10.1140/epjc/s10052-020-7854-5} {\bibfield
  {journal} {\bibinfo  {journal} {Eur. Phys. J. C}\ }\textbf {\bibinfo {volume}
  {80}},\ \bibinfo {pages} {294} (\bibinfo {year} {2020})},\ \Eprint
  {http://arxiv.org/abs/1909.04044} {arXiv:1909.04044 [astro-ph.CO]}
  \BibitemShut {NoStop}%
\bibitem [{\citenamefont {Di~Valentino}\ \emph
  {et~al.}(2020{\natexlab{b}})\citenamefont {Di~Valentino}, \citenamefont
  {Melchiorri}, \citenamefont {Mena},\ and\ \citenamefont
  {Vagnozzi}}]{DiValentino:2019jae}%
  \BibitemOpen
  \bibfield  {author} {\bibinfo {author} {\bibfnamefont {E.}~\bibnamefont
  {Di~Valentino}}, \bibinfo {author} {\bibfnamefont {A.}~\bibnamefont
  {Melchiorri}}, \bibinfo {author} {\bibfnamefont {O.}~\bibnamefont {Mena}}, \
  and\ \bibinfo {author} {\bibfnamefont {S.}~\bibnamefont {Vagnozzi}},\ }\href
  {\doibase 10.1103/PhysRevD.101.063502} {\bibfield  {journal} {\bibinfo
  {journal} {Phys. Rev. D}\ }\textbf {\bibinfo {volume} {101}},\ \bibinfo
  {pages} {063502} (\bibinfo {year} {2020}{\natexlab{b}})},\ \Eprint
  {http://arxiv.org/abs/1910.09853} {arXiv:1910.09853 [astro-ph.CO]}
  \BibitemShut {NoStop}%
\bibitem [{\citenamefont {Liu}\ \emph {et~al.}(2020)\citenamefont {Liu},
  \citenamefont {Huang}, \citenamefont {Luo}, \citenamefont {Miao},
  \citenamefont {Singh},\ and\ \citenamefont {Huang}}]{Liu:2019awo}%
  \BibitemOpen
  \bibfield  {author} {\bibinfo {author} {\bibfnamefont {M.}~\bibnamefont
  {Liu}}, \bibinfo {author} {\bibfnamefont {Z.}~\bibnamefont {Huang}}, \bibinfo
  {author} {\bibfnamefont {X.}~\bibnamefont {Luo}}, \bibinfo {author}
  {\bibfnamefont {H.}~\bibnamefont {Miao}}, \bibinfo {author} {\bibfnamefont
  {N.~K.}\ \bibnamefont {Singh}}, \ and\ \bibinfo {author} {\bibfnamefont
  {L.}~\bibnamefont {Huang}},\ }\href {\doibase 10.1007/s11433-019-1509-5}
  {\bibfield  {journal} {\bibinfo  {journal} {Sci. China Phys. Mech. Astron.}\
  }\textbf {\bibinfo {volume} {63}},\ \bibinfo {pages} {290405} (\bibinfo
  {year} {2020})},\ \Eprint {http://arxiv.org/abs/1912.00190} {arXiv:1912.00190
  [astro-ph.CO]} \BibitemShut {NoStop}%
\bibitem [{\citenamefont {Hart}\ and\ \citenamefont
  {Chluba}(2020)}]{Hart:2019dxi}%
  \BibitemOpen
  \bibfield  {author} {\bibinfo {author} {\bibfnamefont {L.}~\bibnamefont
  {Hart}}\ and\ \bibinfo {author} {\bibfnamefont {J.}~\bibnamefont {Chluba}},\
  }\href {\doibase 10.1093/mnras/staa412} {\bibfield  {journal} {\bibinfo
  {journal} {Mon. Not. Roy. Astron. Soc.}\ }\textbf {\bibinfo {volume} {493}},\
  \bibinfo {pages} {3255} (\bibinfo {year} {2020})},\ \Eprint
  {http://arxiv.org/abs/1912.03986} {arXiv:1912.03986 [astro-ph.CO]}
  \BibitemShut {NoStop}%
\bibitem [{\citenamefont {Akarsu}\ \emph
  {et~al.}(2020{\natexlab{a}})\citenamefont {Akarsu}, \citenamefont {Barrow},
  \citenamefont {Escamilla},\ and\ \citenamefont {Vazquez}}]{Akarsu:2019hmw}%
  \BibitemOpen
  \bibfield  {author} {\bibinfo {author} {\bibfnamefont {O.}~\bibnamefont
  {Akarsu}}, \bibinfo {author} {\bibfnamefont {J.~D.}\ \bibnamefont {Barrow}},
  \bibinfo {author} {\bibfnamefont {L.~A.}\ \bibnamefont {Escamilla}}, \ and\
  \bibinfo {author} {\bibfnamefont {J.~A.}\ \bibnamefont {Vazquez}},\ }\href
  {\doibase 10.1103/PhysRevD.101.063528} {\bibfield  {journal} {\bibinfo
  {journal} {Phys. Rev. D}\ }\textbf {\bibinfo {volume} {101}},\ \bibinfo
  {pages} {063528} (\bibinfo {year} {2020}{\natexlab{a}})},\ \Eprint
  {http://arxiv.org/abs/1912.08751} {arXiv:1912.08751 [astro-ph.CO]}
  \BibitemShut {NoStop}%
\bibitem [{\citenamefont {Benisty}(2019)}]{Benisty:2019pxb}%
  \BibitemOpen
  \bibfield  {author} {\bibinfo {author} {\bibfnamefont {D.}~\bibnamefont
  {Benisty}},\ }\href@noop {} {\  (\bibinfo {year} {2019})},\ \Eprint
  {http://arxiv.org/abs/1912.11124} {arXiv:1912.11124 [gr-qc]} \BibitemShut
  {NoStop}%
\bibitem [{\citenamefont {Yang}\ \emph {et~al.}(2020)\citenamefont {Yang},
  \citenamefont {Di~Valentino}, \citenamefont {Pan}, \citenamefont
  {Basilakos},\ and\ \citenamefont {Paliathanasis}}]{Yang:2020zuk}%
  \BibitemOpen
  \bibfield  {author} {\bibinfo {author} {\bibfnamefont {W.}~\bibnamefont
  {Yang}}, \bibinfo {author} {\bibfnamefont {E.}~\bibnamefont {Di~Valentino}},
  \bibinfo {author} {\bibfnamefont {S.}~\bibnamefont {Pan}}, \bibinfo {author}
  {\bibfnamefont {S.}~\bibnamefont {Basilakos}}, \ and\ \bibinfo {author}
  {\bibfnamefont {A.}~\bibnamefont {Paliathanasis}},\ }\href {\doibase
  10.1103/PhysRevD.102.063503} {\bibfield  {journal} {\bibinfo  {journal}
  {Phys. Rev. D}\ }\textbf {\bibinfo {volume} {102}},\ \bibinfo {pages}
  {063503} (\bibinfo {year} {2020})},\ \Eprint
  {http://arxiv.org/abs/2001.04307} {arXiv:2001.04307 [astro-ph.CO]}
  \BibitemShut {NoStop}%
\bibitem [{\citenamefont {Choi}\ \emph {et~al.}(2020)\citenamefont {Choi},
  \citenamefont {Suzuki},\ and\ \citenamefont {Yanagida}}]{Choi:2020tqp}%
  \BibitemOpen
  \bibfield  {author} {\bibinfo {author} {\bibfnamefont {G.}~\bibnamefont
  {Choi}}, \bibinfo {author} {\bibfnamefont {M.}~\bibnamefont {Suzuki}}, \ and\
  \bibinfo {author} {\bibfnamefont {T.~T.}\ \bibnamefont {Yanagida}},\ }\href
  {\doibase 10.1103/PhysRevD.101.075031} {\bibfield  {journal} {\bibinfo
  {journal} {Phys. Rev. D}\ }\textbf {\bibinfo {volume} {101}},\ \bibinfo
  {pages} {075031} (\bibinfo {year} {2020})},\ \Eprint
  {http://arxiv.org/abs/2002.00036} {arXiv:2002.00036 [hep-ph]} \BibitemShut
  {NoStop}%
\bibitem [{\citenamefont {Lucca}\ and\ \citenamefont
  {Hooper}(2020)}]{Lucca:2020zjb}%
  \BibitemOpen
  \bibfield  {author} {\bibinfo {author} {\bibfnamefont {M.}~\bibnamefont
  {Lucca}}\ and\ \bibinfo {author} {\bibfnamefont {D.~C.}\ \bibnamefont
  {Hooper}},\ }\href {\doibase 10.1103/PhysRevD.102.123502} {\bibfield
  {journal} {\bibinfo  {journal} {Phys. Rev. D}\ }\textbf {\bibinfo {volume}
  {102}},\ \bibinfo {pages} {123502} (\bibinfo {year} {2020})},\ \Eprint
  {http://arxiv.org/abs/2002.06127} {arXiv:2002.06127 [astro-ph.CO]}
  \BibitemShut {NoStop}%
\bibitem [{\citenamefont {Hogg}\ \emph {et~al.}(2020)\citenamefont {Hogg},
  \citenamefont {Bruni}, \citenamefont {Crittenden}, \citenamefont
  {Martinelli},\ and\ \citenamefont {Peirone}}]{Hogg:2020rdp}%
  \BibitemOpen
  \bibfield  {author} {\bibinfo {author} {\bibfnamefont {N.~B.}\ \bibnamefont
  {Hogg}}, \bibinfo {author} {\bibfnamefont {M.}~\bibnamefont {Bruni}},
  \bibinfo {author} {\bibfnamefont {R.}~\bibnamefont {Crittenden}}, \bibinfo
  {author} {\bibfnamefont {M.}~\bibnamefont {Martinelli}}, \ and\ \bibinfo
  {author} {\bibfnamefont {S.}~\bibnamefont {Peirone}},\ }\href {\doibase
  10.1016/j.dark.2020.100583} {\bibfield  {journal} {\bibinfo  {journal} {Phys.
  Dark Univ.}\ }\textbf {\bibinfo {volume} {29}},\ \bibinfo {pages} {100583}
  (\bibinfo {year} {2020})},\ \Eprint {http://arxiv.org/abs/2002.10449}
  {arXiv:2002.10449 [astro-ph.CO]} \BibitemShut {NoStop}%
\bibitem [{\citenamefont {Benevento}\ \emph {et~al.}(2020)\citenamefont
  {Benevento}, \citenamefont {Hu},\ and\ \citenamefont
  {Raveri}}]{Benevento:2020fev}%
  \BibitemOpen
  \bibfield  {author} {\bibinfo {author} {\bibfnamefont {G.}~\bibnamefont
  {Benevento}}, \bibinfo {author} {\bibfnamefont {W.}~\bibnamefont {Hu}}, \
  and\ \bibinfo {author} {\bibfnamefont {M.}~\bibnamefont {Raveri}},\ }\href
  {\doibase 10.1103/PhysRevD.101.103517} {\bibfield  {journal} {\bibinfo
  {journal} {Phys. Rev. D}\ }\textbf {\bibinfo {volume} {101}},\ \bibinfo
  {pages} {103517} (\bibinfo {year} {2020})},\ \Eprint
  {http://arxiv.org/abs/2002.11707} {arXiv:2002.11707 [astro-ph.CO]}
  \BibitemShut {NoStop}%
\bibitem [{\citenamefont {Barker}\ \emph {et~al.}(2020)\citenamefont {Barker},
  \citenamefont {Lasenby}, \citenamefont {Hobson},\ and\ \citenamefont
  {Handley}}]{Barker:2020gcp}%
  \BibitemOpen
  \bibfield  {author} {\bibinfo {author} {\bibfnamefont {W.~E.~V.}\
  \bibnamefont {Barker}}, \bibinfo {author} {\bibfnamefont {A.~N.}\
  \bibnamefont {Lasenby}}, \bibinfo {author} {\bibfnamefont {M.~P.}\
  \bibnamefont {Hobson}}, \ and\ \bibinfo {author} {\bibfnamefont {W.~J.}\
  \bibnamefont {Handley}},\ }\href {\doibase 10.1103/PhysRevD.102.024048}
  {\bibfield  {journal} {\bibinfo  {journal} {Phys. Rev. D}\ }\textbf {\bibinfo
  {volume} {102}},\ \bibinfo {pages} {024048} (\bibinfo {year} {2020})},\
  \Eprint {http://arxiv.org/abs/2003.02690} {arXiv:2003.02690 [gr-qc]}
  \BibitemShut {NoStop}%
\bibitem [{\citenamefont {G\'omez-Valent}\ \emph {et~al.}(2020)\citenamefont
  {G\'omez-Valent}, \citenamefont {Pettorino},\ and\ \citenamefont
  {Amendola}}]{Gomez-Valent:2020mqn}%
  \BibitemOpen
  \bibfield  {author} {\bibinfo {author} {\bibfnamefont {A.}~\bibnamefont
  {G\'omez-Valent}}, \bibinfo {author} {\bibfnamefont {V.}~\bibnamefont
  {Pettorino}}, \ and\ \bibinfo {author} {\bibfnamefont {L.}~\bibnamefont
  {Amendola}},\ }\href {\doibase 10.1103/PhysRevD.101.123513} {\bibfield
  {journal} {\bibinfo  {journal} {Phys. Rev. D}\ }\textbf {\bibinfo {volume}
  {101}},\ \bibinfo {pages} {123513} (\bibinfo {year} {2020})},\ \Eprint
  {http://arxiv.org/abs/2004.00610} {arXiv:2004.00610 [astro-ph.CO]}
  \BibitemShut {NoStop}%
\bibitem [{\citenamefont {Akarsu}\ \emph
  {et~al.}(2020{\natexlab{b}})\citenamefont {Akarsu}, \citenamefont
  {Kat\i{}rc\i{}}, \citenamefont {Kumar}, \citenamefont {Nunes}, \citenamefont
  {\"Ozt\"urk},\ and\ \citenamefont {Sharma}}]{Akarsu:2020yqa}%
  \BibitemOpen
  \bibfield  {author} {\bibinfo {author} {\bibfnamefont {O.}~\bibnamefont
  {Akarsu}}, \bibinfo {author} {\bibfnamefont {N.}~\bibnamefont
  {Kat\i{}rc\i{}}}, \bibinfo {author} {\bibfnamefont {S.}~\bibnamefont
  {Kumar}}, \bibinfo {author} {\bibfnamefont {R.~C.}\ \bibnamefont {Nunes}},
  \bibinfo {author} {\bibfnamefont {B.}~\bibnamefont {\"Ozt\"urk}}, \ and\
  \bibinfo {author} {\bibfnamefont {S.}~\bibnamefont {Sharma}},\ }\href
  {\doibase 10.1140/epjc/s10052-020-08586-4} {\bibfield  {journal} {\bibinfo
  {journal} {Eur. Phys. J. C}\ }\textbf {\bibinfo {volume} {80}},\ \bibinfo
  {pages} {1050} (\bibinfo {year} {2020}{\natexlab{b}})},\ \Eprint
  {http://arxiv.org/abs/2004.04074} {arXiv:2004.04074 [astro-ph.CO]}
  \BibitemShut {NoStop}%
\bibitem [{\citenamefont {Ballesteros}\ \emph {et~al.}(2020)\citenamefont
  {Ballesteros}, \citenamefont {Notari},\ and\ \citenamefont
  {Rompineve}}]{Ballesteros:2020sik}%
  \BibitemOpen
  \bibfield  {author} {\bibinfo {author} {\bibfnamefont {G.}~\bibnamefont
  {Ballesteros}}, \bibinfo {author} {\bibfnamefont {A.}~\bibnamefont {Notari}},
  \ and\ \bibinfo {author} {\bibfnamefont {F.}~\bibnamefont {Rompineve}},\
  }\href {\doibase 10.1088/1475-7516/2020/11/024} {\bibfield  {journal}
  {\bibinfo  {journal} {JCAP}\ }\textbf {\bibinfo {volume} {11}},\ \bibinfo
  {pages} {024} (\bibinfo {year} {2020})},\ \Eprint
  {http://arxiv.org/abs/2004.05049} {arXiv:2004.05049 [astro-ph.CO]}
  \BibitemShut {NoStop}%
\bibitem [{\citenamefont {Haridasu}\ and\ \citenamefont
  {Viel}(2020)}]{Haridasu:2020xaa}%
  \BibitemOpen
  \bibfield  {author} {\bibinfo {author} {\bibfnamefont {B.~S.}\ \bibnamefont
  {Haridasu}}\ and\ \bibinfo {author} {\bibfnamefont {M.}~\bibnamefont
  {Viel}},\ }\href {\doibase 10.1093/mnras/staa1991} {\bibfield  {journal}
  {\bibinfo  {journal} {Mon. Not. Roy. Astron. Soc.}\ }\textbf {\bibinfo
  {volume} {497}},\ \bibinfo {pages} {1757} (\bibinfo {year} {2020})},\ \Eprint
  {http://arxiv.org/abs/2004.07709} {arXiv:2004.07709 [astro-ph.CO]}
  \BibitemShut {NoStop}%
\bibitem [{\citenamefont {Alestas}\ \emph {et~al.}(2020)\citenamefont
  {Alestas}, \citenamefont {Kazantzidis},\ and\ \citenamefont
  {Perivolaropoulos}}]{Alestas:2020mvb}%
  \BibitemOpen
  \bibfield  {author} {\bibinfo {author} {\bibfnamefont {G.}~\bibnamefont
  {Alestas}}, \bibinfo {author} {\bibfnamefont {L.}~\bibnamefont
  {Kazantzidis}}, \ and\ \bibinfo {author} {\bibfnamefont {L.}~\bibnamefont
  {Perivolaropoulos}},\ }\href {\doibase 10.1103/PhysRevD.101.123516}
  {\bibfield  {journal} {\bibinfo  {journal} {Phys. Rev. D}\ }\textbf {\bibinfo
  {volume} {101}},\ \bibinfo {pages} {123516} (\bibinfo {year} {2020})},\
  \Eprint {http://arxiv.org/abs/2004.08363} {arXiv:2004.08363 [astro-ph.CO]}
  \BibitemShut {NoStop}%
\bibitem [{\citenamefont {Jedamzik}\ and\ \citenamefont
  {Pogosian}(2020)}]{Jedamzik:2020krr}%
  \BibitemOpen
  \bibfield  {author} {\bibinfo {author} {\bibfnamefont {K.}~\bibnamefont
  {Jedamzik}}\ and\ \bibinfo {author} {\bibfnamefont {L.}~\bibnamefont
  {Pogosian}},\ }\href {\doibase 10.1103/PhysRevLett.125.181302} {\bibfield
  {journal} {\bibinfo  {journal} {Phys. Rev. Lett.}\ }\textbf {\bibinfo
  {volume} {125}},\ \bibinfo {pages} {181302} (\bibinfo {year} {2020})},\
  \Eprint {http://arxiv.org/abs/2004.09487} {arXiv:2004.09487 [astro-ph.CO]}
  \BibitemShut {NoStop}%
\bibitem [{\citenamefont {Ballardini}\ \emph {et~al.}(2020)\citenamefont
  {Ballardini}, \citenamefont {Braglia}, \citenamefont {Finelli}, \citenamefont
  {Paoletti}, \citenamefont {Starobinsky},\ and\ \citenamefont
  {Umilt\`a}}]{Ballardini:2020iws}%
  \BibitemOpen
  \bibfield  {author} {\bibinfo {author} {\bibfnamefont {M.}~\bibnamefont
  {Ballardini}}, \bibinfo {author} {\bibfnamefont {M.}~\bibnamefont {Braglia}},
  \bibinfo {author} {\bibfnamefont {F.}~\bibnamefont {Finelli}}, \bibinfo
  {author} {\bibfnamefont {D.}~\bibnamefont {Paoletti}}, \bibinfo {author}
  {\bibfnamefont {A.~A.}\ \bibnamefont {Starobinsky}}, \ and\ \bibinfo {author}
  {\bibfnamefont {C.}~\bibnamefont {Umilt\`a}},\ }\href {\doibase
  10.1088/1475-7516/2020/10/044} {\bibfield  {journal} {\bibinfo  {journal}
  {JCAP}\ }\textbf {\bibinfo {volume} {10}},\ \bibinfo {pages} {044} (\bibinfo
  {year} {2020})},\ \Eprint {http://arxiv.org/abs/2004.14349} {arXiv:2004.14349
  [astro-ph.CO]} \BibitemShut {NoStop}%
\bibitem [{\citenamefont {Banerjee}\ \emph {et~al.}(2021)\citenamefont
  {Banerjee}, \citenamefont {Cai}, \citenamefont {Heisenberg}, \citenamefont
  {Colg\'ain}, \citenamefont {Sheikh-Jabbari},\ and\ \citenamefont
  {Yang}}]{Banerjee:2020xcn}%
  \BibitemOpen
  \bibfield  {author} {\bibinfo {author} {\bibfnamefont {A.}~\bibnamefont
  {Banerjee}}, \bibinfo {author} {\bibfnamefont {H.}~\bibnamefont {Cai}},
  \bibinfo {author} {\bibfnamefont {L.}~\bibnamefont {Heisenberg}}, \bibinfo
  {author} {\bibfnamefont {E.~O.}\ \bibnamefont {Colg\'ain}}, \bibinfo {author}
  {\bibfnamefont {M.~M.}\ \bibnamefont {Sheikh-Jabbari}}, \ and\ \bibinfo
  {author} {\bibfnamefont {T.}~\bibnamefont {Yang}},\ }\href {\doibase
  10.1103/PhysRevD.103.L081305} {\bibfield  {journal} {\bibinfo  {journal}
  {Phys. Rev. D}\ }\textbf {\bibinfo {volume} {103}},\ \bibinfo {pages}
  {L081305} (\bibinfo {year} {2021})},\ \Eprint
  {http://arxiv.org/abs/2006.00244} {arXiv:2006.00244 [astro-ph.CO]}
  \BibitemShut {NoStop}%
\bibitem [{\citenamefont {Elizalde}\ \emph {et~al.}(2020)\citenamefont
  {Elizalde}, \citenamefont {Khurshudyan}, \citenamefont {Odintsov},\ and\
  \citenamefont {Myrzakulov}}]{Elizalde:2020mfs}%
  \BibitemOpen
  \bibfield  {author} {\bibinfo {author} {\bibfnamefont {E.}~\bibnamefont
  {Elizalde}}, \bibinfo {author} {\bibfnamefont {M.}~\bibnamefont
  {Khurshudyan}}, \bibinfo {author} {\bibfnamefont {S.~D.}\ \bibnamefont
  {Odintsov}}, \ and\ \bibinfo {author} {\bibfnamefont {R.}~\bibnamefont
  {Myrzakulov}},\ }\href {\doibase 10.1103/PhysRevD.102.123501} {\bibfield
  {journal} {\bibinfo  {journal} {Phys. Rev. D}\ }\textbf {\bibinfo {volume}
  {102}},\ \bibinfo {pages} {123501} (\bibinfo {year} {2020})},\ \Eprint
  {http://arxiv.org/abs/2006.01879} {arXiv:2006.01879 [gr-qc]} \BibitemShut
  {NoStop}%
\bibitem [{\citenamefont {Gonzalez}\ \emph {et~al.}(2020)\citenamefont
  {Gonzalez}, \citenamefont {Hertzberg},\ and\ \citenamefont
  {Rompineve}}]{Gonzalez:2020fdy}%
  \BibitemOpen
  \bibfield  {author} {\bibinfo {author} {\bibfnamefont {M.}~\bibnamefont
  {Gonzalez}}, \bibinfo {author} {\bibfnamefont {M.~P.}\ \bibnamefont
  {Hertzberg}}, \ and\ \bibinfo {author} {\bibfnamefont {F.}~\bibnamefont
  {Rompineve}},\ }\href {\doibase 10.1088/1475-7516/2020/10/028} {\bibfield
  {journal} {\bibinfo  {journal} {JCAP}\ }\textbf {\bibinfo {volume} {10}},\
  \bibinfo {pages} {028} (\bibinfo {year} {2020})},\ \Eprint
  {http://arxiv.org/abs/2006.13959} {arXiv:2006.13959 [astro-ph.CO]}
  \BibitemShut {NoStop}%
\bibitem [{\citenamefont {Capozziello}\ \emph {et~al.}(2020)\citenamefont
  {Capozziello}, \citenamefont {Benetti},\ and\ \citenamefont
  {Spallicci}}]{Capozziello:2020nyq}%
  \BibitemOpen
  \bibfield  {author} {\bibinfo {author} {\bibfnamefont {S.}~\bibnamefont
  {Capozziello}}, \bibinfo {author} {\bibfnamefont {M.}~\bibnamefont
  {Benetti}}, \ and\ \bibinfo {author} {\bibfnamefont {A.~D. A.~M.}\
  \bibnamefont {Spallicci}},\ }\href {\doibase 10.1007/s10701-020-00356-2}
  {\bibfield  {journal} {\bibinfo  {journal} {Found. Phys.}\ }\textbf {\bibinfo
  {volume} {50}},\ \bibinfo {pages} {893} (\bibinfo {year} {2020})},\ \Eprint
  {http://arxiv.org/abs/2007.00462} {arXiv:2007.00462 [gr-qc]} \BibitemShut
  {NoStop}%
\bibitem [{\citenamefont {Das}\ and\ \citenamefont
  {Ghosh}(2020)}]{Das:2020xke}%
  \BibitemOpen
  \bibfield  {author} {\bibinfo {author} {\bibfnamefont {A.}~\bibnamefont
  {Das}}\ and\ \bibinfo {author} {\bibfnamefont {S.}~\bibnamefont {Ghosh}},\
  }\href@noop {} {\  (\bibinfo {year} {2020})},\ \Eprint
  {http://arxiv.org/abs/2011.12315} {arXiv:2011.12315 [astro-ph.CO]}
  \BibitemShut {NoStop}%
\bibitem [{\citenamefont {Banihashemi}\ \emph {et~al.}(2021)\citenamefont
  {Banihashemi}, \citenamefont {Khosravi},\ and\ \citenamefont
  {Shafieloo}}]{Banihashemi:2020wtb}%
  \BibitemOpen
  \bibfield  {author} {\bibinfo {author} {\bibfnamefont {A.}~\bibnamefont
  {Banihashemi}}, \bibinfo {author} {\bibfnamefont {N.}~\bibnamefont
  {Khosravi}}, \ and\ \bibinfo {author} {\bibfnamefont {A.}~\bibnamefont
  {Shafieloo}},\ }\href {\doibase 10.1088/1475-7516/2021/06/003} {\bibfield
  {journal} {\bibinfo  {journal} {JCAP}\ }\textbf {\bibinfo {volume} {06}},\
  \bibinfo {pages} {003} (\bibinfo {year} {2021})},\ \Eprint
  {http://arxiv.org/abs/2012.01407} {arXiv:2012.01407 [astro-ph.CO]}
  \BibitemShut {NoStop}%
\bibitem [{\citenamefont {Roy~Choudhury}\ \emph {et~al.}(2021)\citenamefont
  {Roy~Choudhury}, \citenamefont {Hannestad},\ and\ \citenamefont
  {Tram}}]{Choudhury:2020tka}%
  \BibitemOpen
  \bibfield  {author} {\bibinfo {author} {\bibfnamefont {S.}~\bibnamefont
  {Roy~Choudhury}}, \bibinfo {author} {\bibfnamefont {S.}~\bibnamefont
  {Hannestad}}, \ and\ \bibinfo {author} {\bibfnamefont {T.}~\bibnamefont
  {Tram}},\ }\href {\doibase 10.1088/1475-7516/2021/03/084} {\bibfield
  {journal} {\bibinfo  {journal} {JCAP}\ }\textbf {\bibinfo {volume} {03}},\
  \bibinfo {pages} {084} (\bibinfo {year} {2021})},\ \Eprint
  {http://arxiv.org/abs/2012.07519} {arXiv:2012.07519 [astro-ph.CO]}
  \BibitemShut {NoStop}%
\bibitem [{\citenamefont {Brinckmann}\ \emph {et~al.}(2020)\citenamefont
  {Brinckmann}, \citenamefont {Chang},\ and\ \citenamefont
  {LoVerde}}]{Brinckmann:2020bcn}%
  \BibitemOpen
  \bibfield  {author} {\bibinfo {author} {\bibfnamefont {T.}~\bibnamefont
  {Brinckmann}}, \bibinfo {author} {\bibfnamefont {J.~H.}\ \bibnamefont
  {Chang}}, \ and\ \bibinfo {author} {\bibfnamefont {M.}~\bibnamefont
  {LoVerde}},\ }\href@noop {} {\  (\bibinfo {year} {2020})},\ \Eprint
  {http://arxiv.org/abs/2012.11830} {arXiv:2012.11830 [astro-ph.CO]}
  \BibitemShut {NoStop}%
\bibitem [{\citenamefont {Moshafi}\ \emph {et~al.}(2020)\citenamefont
  {Moshafi}, \citenamefont {Baghram},\ and\ \citenamefont
  {Khosravi}}]{Moshafi:2020rkq}%
  \BibitemOpen
  \bibfield  {author} {\bibinfo {author} {\bibfnamefont {H.}~\bibnamefont
  {Moshafi}}, \bibinfo {author} {\bibfnamefont {S.}~\bibnamefont {Baghram}}, \
  and\ \bibinfo {author} {\bibfnamefont {N.}~\bibnamefont {Khosravi}},\
  }\href@noop {} {\  (\bibinfo {year} {2020})},\ \Eprint
  {http://arxiv.org/abs/2012.14377} {arXiv:2012.14377 [astro-ph.CO]}
  \BibitemShut {NoStop}%
\bibitem [{\citenamefont {Alestas}\ and\ \citenamefont
  {Perivolaropoulos}(2021)}]{Alestas:2021xes}%
  \BibitemOpen
  \bibfield  {author} {\bibinfo {author} {\bibfnamefont {G.}~\bibnamefont
  {Alestas}}\ and\ \bibinfo {author} {\bibfnamefont {L.}~\bibnamefont
  {Perivolaropoulos}},\ }\href {\doibase 10.1093/mnras/stab1070} {\bibfield
  {journal} {\bibinfo  {journal} {Mon. Not. Roy. Astron. Soc.}\ }\textbf
  {\bibinfo {volume} {504}},\ \bibinfo {pages} {3956} (\bibinfo {year}
  {2021})},\ \Eprint {http://arxiv.org/abs/2103.04045} {arXiv:2103.04045
  [astro-ph.CO]} \BibitemShut {NoStop}%
\bibitem [{\citenamefont {Beutler}\ \emph {et~al.}(2011)\citenamefont
  {Beutler}, \citenamefont {Blake}, \citenamefont {Colless}, \citenamefont
  {Jones}, \citenamefont {Staveley-Smith}, \citenamefont {Campbell},
  \citenamefont {Parker}, \citenamefont {Saunders},\ and\ \citenamefont
  {Watson}}]{Beutler:2011hx}%
  \BibitemOpen
  \bibfield  {author} {\bibinfo {author} {\bibfnamefont {F.}~\bibnamefont
  {Beutler}}, \bibinfo {author} {\bibfnamefont {C.}~\bibnamefont {Blake}},
  \bibinfo {author} {\bibfnamefont {M.}~\bibnamefont {Colless}}, \bibinfo
  {author} {\bibfnamefont {D.~H.}\ \bibnamefont {Jones}}, \bibinfo {author}
  {\bibfnamefont {L.}~\bibnamefont {Staveley-Smith}}, \bibinfo {author}
  {\bibfnamefont {L.}~\bibnamefont {Campbell}}, \bibinfo {author}
  {\bibfnamefont {Q.}~\bibnamefont {Parker}}, \bibinfo {author} {\bibfnamefont
  {W.}~\bibnamefont {Saunders}}, \ and\ \bibinfo {author} {\bibfnamefont
  {F.}~\bibnamefont {Watson}},\ }\href {\doibase
  10.1111/j.1365-2966.2011.19250.x} {\bibfield  {journal} {\bibinfo  {journal}
  {Mon. Not. Roy. Astron. Soc.}\ }\textbf {\bibinfo {volume} {416}},\ \bibinfo
  {pages} {3017} (\bibinfo {year} {2011})},\ \Eprint
  {http://arxiv.org/abs/1106.3366} {arXiv:1106.3366 [astro-ph.CO]} \BibitemShut
  {NoStop}%
\bibitem [{\citenamefont {Ross}\ \emph {et~al.}(2015)\citenamefont {Ross},
  \citenamefont {Samushia}, \citenamefont {Howlett}, \citenamefont {Percival},
  \citenamefont {Burden},\ and\ \citenamefont {Manera}}]{Ross:2014qpa}%
  \BibitemOpen
  \bibfield  {author} {\bibinfo {author} {\bibfnamefont {A.~J.}\ \bibnamefont
  {Ross}}, \bibinfo {author} {\bibfnamefont {L.}~\bibnamefont {Samushia}},
  \bibinfo {author} {\bibfnamefont {C.}~\bibnamefont {Howlett}}, \bibinfo
  {author} {\bibfnamefont {W.~J.}\ \bibnamefont {Percival}}, \bibinfo {author}
  {\bibfnamefont {A.}~\bibnamefont {Burden}}, \ and\ \bibinfo {author}
  {\bibfnamefont {M.}~\bibnamefont {Manera}},\ }\href {\doibase
  10.1093/mnras/stv154} {\bibfield  {journal} {\bibinfo  {journal} {Mon. Not.
  Roy. Astron. Soc.}\ }\textbf {\bibinfo {volume} {449}},\ \bibinfo {pages}
  {835} (\bibinfo {year} {2015})},\ \Eprint {http://arxiv.org/abs/1409.3242}
  {arXiv:1409.3242 [astro-ph.CO]} \BibitemShut {NoStop}%
\bibitem [{\citenamefont {Bernal}\ \emph {et~al.}(2016)\citenamefont {Bernal},
  \citenamefont {Verde},\ and\ \citenamefont {Riess}}]{Bernal:2016gxb}%
  \BibitemOpen
  \bibfield  {author} {\bibinfo {author} {\bibfnamefont {J.~L.}\ \bibnamefont
  {Bernal}}, \bibinfo {author} {\bibfnamefont {L.}~\bibnamefont {Verde}}, \
  and\ \bibinfo {author} {\bibfnamefont {A.~G.}\ \bibnamefont {Riess}},\ }\href
  {\doibase 10.1088/1475-7516/2016/10/019} {\bibfield  {journal} {\bibinfo
  {journal} {JCAP}\ }\textbf {\bibinfo {volume} {10}},\ \bibinfo {pages} {019}
  (\bibinfo {year} {2016})},\ \Eprint {http://arxiv.org/abs/1607.05617}
  {arXiv:1607.05617 [astro-ph.CO]} \BibitemShut {NoStop}%
\bibitem [{\citenamefont {Addison}\ \emph {et~al.}(2018)\citenamefont
  {Addison}, \citenamefont {Watts}, \citenamefont {Bennett}, \citenamefont
  {Halpern}, \citenamefont {Hinshaw},\ and\ \citenamefont
  {Weiland}}]{Addison:2017fdm}%
  \BibitemOpen
  \bibfield  {author} {\bibinfo {author} {\bibfnamefont {G.~E.}\ \bibnamefont
  {Addison}}, \bibinfo {author} {\bibfnamefont {D.~J.}\ \bibnamefont {Watts}},
  \bibinfo {author} {\bibfnamefont {C.~L.}\ \bibnamefont {Bennett}}, \bibinfo
  {author} {\bibfnamefont {M.}~\bibnamefont {Halpern}}, \bibinfo {author}
  {\bibfnamefont {G.}~\bibnamefont {Hinshaw}}, \ and\ \bibinfo {author}
  {\bibfnamefont {J.~L.}\ \bibnamefont {Weiland}},\ }\href {\doibase
  10.3847/1538-4357/aaa1ed} {\bibfield  {journal} {\bibinfo  {journal}
  {Astrophys. J.}\ }\textbf {\bibinfo {volume} {853}},\ \bibinfo {pages} {119}
  (\bibinfo {year} {2018})},\ \Eprint {http://arxiv.org/abs/1707.06547}
  {arXiv:1707.06547 [astro-ph.CO]} \BibitemShut {NoStop}%
\bibitem [{\citenamefont {Lemos}\ \emph {et~al.}(2019)\citenamefont {Lemos},
  \citenamefont {Lee}, \citenamefont {Efstathiou},\ and\ \citenamefont
  {Gratton}}]{Lemos:2018smw}%
  \BibitemOpen
  \bibfield  {author} {\bibinfo {author} {\bibfnamefont {P.}~\bibnamefont
  {Lemos}}, \bibinfo {author} {\bibfnamefont {E.}~\bibnamefont {Lee}}, \bibinfo
  {author} {\bibfnamefont {G.}~\bibnamefont {Efstathiou}}, \ and\ \bibinfo
  {author} {\bibfnamefont {S.}~\bibnamefont {Gratton}},\ }\href {\doibase
  10.1093/mnras/sty3082} {\bibfield  {journal} {\bibinfo  {journal} {Mon. Not.
  Roy. Astron. Soc.}\ }\textbf {\bibinfo {volume} {483}},\ \bibinfo {pages}
  {4803} (\bibinfo {year} {2019})},\ \Eprint {http://arxiv.org/abs/1806.06781}
  {arXiv:1806.06781 [astro-ph.CO]} \BibitemShut {NoStop}%
\bibitem [{\citenamefont {Aylor}\ \emph {et~al.}(2019)\citenamefont {Aylor},
  \citenamefont {Joy}, \citenamefont {Knox}, \citenamefont {Millea},
  \citenamefont {Raghunathan},\ and\ \citenamefont {Wu}}]{Aylor:2018drw}%
  \BibitemOpen
  \bibfield  {author} {\bibinfo {author} {\bibfnamefont {K.}~\bibnamefont
  {Aylor}}, \bibinfo {author} {\bibfnamefont {M.}~\bibnamefont {Joy}}, \bibinfo
  {author} {\bibfnamefont {L.}~\bibnamefont {Knox}}, \bibinfo {author}
  {\bibfnamefont {M.}~\bibnamefont {Millea}}, \bibinfo {author} {\bibfnamefont
  {S.}~\bibnamefont {Raghunathan}}, \ and\ \bibinfo {author} {\bibfnamefont
  {W.~L.~K.}\ \bibnamefont {Wu}},\ }\href {\doibase 10.3847/1538-4357/ab0898}
  {\bibfield  {journal} {\bibinfo  {journal} {Astrophys. J.}\ }\textbf
  {\bibinfo {volume} {874}},\ \bibinfo {pages} {4} (\bibinfo {year} {2019})},\
  \Eprint {http://arxiv.org/abs/1811.00537} {arXiv:1811.00537 [astro-ph.CO]}
  \BibitemShut {NoStop}%
\bibitem [{\citenamefont {Knox}\ and\ \citenamefont
  {Millea}(2020)}]{Knox:2019rjx}%
  \BibitemOpen
  \bibfield  {author} {\bibinfo {author} {\bibfnamefont {L.}~\bibnamefont
  {Knox}}\ and\ \bibinfo {author} {\bibfnamefont {M.}~\bibnamefont {Millea}},\
  }\href {\doibase 10.1103/PhysRevD.101.043533} {\bibfield  {journal} {\bibinfo
   {journal} {Phys. Rev. D}\ }\textbf {\bibinfo {volume} {101}},\ \bibinfo
  {pages} {043533} (\bibinfo {year} {2020})},\ \Eprint
  {http://arxiv.org/abs/1908.03663} {arXiv:1908.03663 [astro-ph.CO]}
  \BibitemShut {NoStop}%
\bibitem [{\citenamefont {Arendse}\ \emph {et~al.}(2020)\citenamefont {Arendse}
  \emph {et~al.}}]{Arendse:2019hev}%
  \BibitemOpen
  \bibfield  {author} {\bibinfo {author} {\bibfnamefont {N.}~\bibnamefont
  {Arendse}} \emph {et~al.},\ }\href {\doibase 10.1051/0004-6361/201936720}
  {\bibfield  {journal} {\bibinfo  {journal} {Astron. Astrophys.}\ }\textbf
  {\bibinfo {volume} {639}},\ \bibinfo {pages} {A57} (\bibinfo {year}
  {2020})},\ \Eprint {http://arxiv.org/abs/1909.07986} {arXiv:1909.07986
  [astro-ph.CO]} \BibitemShut {NoStop}%
\bibitem [{\citenamefont {Efstathiou}(2021)}]{Efstathiou:2021ocp}%
  \BibitemOpen
  \bibfield  {author} {\bibinfo {author} {\bibfnamefont {G.}~\bibnamefont
  {Efstathiou}},\ }\href@noop {} {\  (\bibinfo {year} {2021})},\ \Eprint
  {http://arxiv.org/abs/2103.08723} {arXiv:2103.08723 [astro-ph.CO]}
  \BibitemShut {NoStop}%
\bibitem [{\citenamefont {Motloch}(2020)}]{Motloch:2020lhu}%
  \BibitemOpen
  \bibfield  {author} {\bibinfo {author} {\bibfnamefont {P.}~\bibnamefont
  {Motloch}},\ }\href {\doibase 10.1103/PhysRevD.101.123509} {\bibfield
  {journal} {\bibinfo  {journal} {Phys. Rev. D}\ }\textbf {\bibinfo {volume}
  {101}},\ \bibinfo {pages} {123509} (\bibinfo {year} {2020})},\ \Eprint
  {http://arxiv.org/abs/2004.11351} {arXiv:2004.11351 [astro-ph.CO]}
  \BibitemShut {NoStop}%
\bibitem [{\citenamefont {Chu}\ and\ \citenamefont {Knox}(2005)}]{Chu:2004qx}%
  \BibitemOpen
  \bibfield  {author} {\bibinfo {author} {\bibfnamefont {M.}~\bibnamefont
  {Chu}}\ and\ \bibinfo {author} {\bibfnamefont {L.}~\bibnamefont {Knox}},\
  }\href {\doibase 10.1086/427064} {\bibfield  {journal} {\bibinfo  {journal}
  {Astrophys. J.}\ }\textbf {\bibinfo {volume} {620}},\ \bibinfo {pages} {1}
  (\bibinfo {year} {2005})},\ \Eprint {http://arxiv.org/abs/astro-ph/0407198}
  {arXiv:astro-ph/0407198} \BibitemShut {NoStop}%
\bibitem [{\citenamefont {Sachs}\ and\ \citenamefont
  {Wolfe}(1967)}]{Sachs:1967er}%
  \BibitemOpen
  \bibfield  {author} {\bibinfo {author} {\bibfnamefont {R.~K.}\ \bibnamefont
  {Sachs}}\ and\ \bibinfo {author} {\bibfnamefont {A.~M.}\ \bibnamefont
  {Wolfe}},\ }\href {\doibase 10.1007/s10714-007-0448-9} {\bibfield  {journal}
  {\bibinfo  {journal} {Astrophys. J.}\ }\textbf {\bibinfo {volume} {147}},\
  \bibinfo {pages} {73} (\bibinfo {year} {1967})}\BibitemShut {NoStop}%
\bibitem [{\citenamefont {Rees}\ and\ \citenamefont
  {Sciama}(1968)}]{Rees:1968zza}%
  \BibitemOpen
  \bibfield  {author} {\bibinfo {author} {\bibfnamefont {M.~J.}\ \bibnamefont
  {Rees}}\ and\ \bibinfo {author} {\bibfnamefont {D.~W.}\ \bibnamefont
  {Sciama}},\ }\href {\doibase 10.1038/217511a0} {\bibfield  {journal}
  {\bibinfo  {journal} {Nature}\ }\textbf {\bibinfo {volume} {217}},\ \bibinfo
  {pages} {511} (\bibinfo {year} {1968})}\BibitemShut {NoStop}%
\bibitem [{\citenamefont {Calabrese}\ \emph {et~al.}(2008)\citenamefont
  {Calabrese}, \citenamefont {Slosar}, \citenamefont {Melchiorri},
  \citenamefont {Smoot},\ and\ \citenamefont {Zahn}}]{Calabrese:2008rt}%
  \BibitemOpen
  \bibfield  {author} {\bibinfo {author} {\bibfnamefont {E.}~\bibnamefont
  {Calabrese}}, \bibinfo {author} {\bibfnamefont {A.}~\bibnamefont {Slosar}},
  \bibinfo {author} {\bibfnamefont {A.}~\bibnamefont {Melchiorri}}, \bibinfo
  {author} {\bibfnamefont {G.~F.}\ \bibnamefont {Smoot}}, \ and\ \bibinfo
  {author} {\bibfnamefont {O.}~\bibnamefont {Zahn}},\ }\href {\doibase
  10.1103/PhysRevD.77.123531} {\bibfield  {journal} {\bibinfo  {journal} {Phys.
  Rev. D}\ }\textbf {\bibinfo {volume} {77}},\ \bibinfo {pages} {123531}
  (\bibinfo {year} {2008})},\ \Eprint {http://arxiv.org/abs/0803.2309}
  {arXiv:0803.2309 [astro-ph]} \BibitemShut {NoStop}%
\bibitem [{\citenamefont {Krishnan}\ \emph {et~al.}(2020)\citenamefont
  {Krishnan}, \citenamefont {Colg\'ain}, \citenamefont {Ruchika}, \citenamefont
  {Sen}, \citenamefont {Sheikh-Jabbari},\ and\ \citenamefont
  {Yang}}]{Krishnan:2020obg}%
  \BibitemOpen
  \bibfield  {author} {\bibinfo {author} {\bibfnamefont {C.}~\bibnamefont
  {Krishnan}}, \bibinfo {author} {\bibfnamefont {E.~O.}\ \bibnamefont
  {Colg\'ain}}, \bibinfo {author} {\bibnamefont {Ruchika}}, \bibinfo {author}
  {\bibfnamefont {A.~A.}\ \bibnamefont {Sen}}, \bibinfo {author} {\bibfnamefont
  {M.~M.}\ \bibnamefont {Sheikh-Jabbari}}, \ and\ \bibinfo {author}
  {\bibfnamefont {T.}~\bibnamefont {Yang}},\ }\href {\doibase
  10.1103/PhysRevD.102.103525} {\bibfield  {journal} {\bibinfo  {journal}
  {Phys. Rev. D}\ }\textbf {\bibinfo {volume} {102}},\ \bibinfo {pages}
  {103525} (\bibinfo {year} {2020})},\ \Eprint
  {http://arxiv.org/abs/2002.06044} {arXiv:2002.06044 [astro-ph.CO]}
  \BibitemShut {NoStop}%
\bibitem [{\citenamefont {Jedamzik}\ \emph {et~al.}(2021)\citenamefont
  {Jedamzik}, \citenamefont {Pogosian},\ and\ \citenamefont
  {Zhao}}]{Jedamzik:2020zmd}%
  \BibitemOpen
  \bibfield  {author} {\bibinfo {author} {\bibfnamefont {K.}~\bibnamefont
  {Jedamzik}}, \bibinfo {author} {\bibfnamefont {L.}~\bibnamefont {Pogosian}},
  \ and\ \bibinfo {author} {\bibfnamefont {G.-B.}\ \bibnamefont {Zhao}},\
  }\href {\doibase 10.1038/s42005-021-00628-x} {\bibfield  {journal} {\bibinfo
  {journal} {Commun. Phys.}\ }\textbf {\bibinfo {volume} {4}},\ \bibinfo
  {pages} {123} (\bibinfo {year} {2021})},\ \Eprint
  {http://arxiv.org/abs/2010.04158} {arXiv:2010.04158 [astro-ph.CO]}
  \BibitemShut {NoStop}%
\bibitem [{\citenamefont {Lin}\ \emph {et~al.}(2021)\citenamefont {Lin},
  \citenamefont {Chen},\ and\ \citenamefont {Mack}}]{Lin:2021sfs}%
  \BibitemOpen
  \bibfield  {author} {\bibinfo {author} {\bibfnamefont {W.}~\bibnamefont
  {Lin}}, \bibinfo {author} {\bibfnamefont {X.}~\bibnamefont {Chen}}, \ and\
  \bibinfo {author} {\bibfnamefont {K.~J.}\ \bibnamefont {Mack}},\ }\href@noop
  {} {\  (\bibinfo {year} {2021})},\ \Eprint {http://arxiv.org/abs/2102.05701}
  {arXiv:2102.05701 [astro-ph.CO]} \BibitemShut {NoStop}%
\bibitem [{\citenamefont {Dainotti}\ \emph {et~al.}(2021)\citenamefont
  {Dainotti}, \citenamefont {De~Simone}, \citenamefont {Schiavone},
  \citenamefont {Montani}, \citenamefont {Rinaldi},\ and\ \citenamefont
  {Lambiase}}]{Dainotti:2021pqg}%
  \BibitemOpen
  \bibfield  {author} {\bibinfo {author} {\bibfnamefont {M.~G.}\ \bibnamefont
  {Dainotti}}, \bibinfo {author} {\bibfnamefont {B.}~\bibnamefont {De~Simone}},
  \bibinfo {author} {\bibfnamefont {T.}~\bibnamefont {Schiavone}}, \bibinfo
  {author} {\bibfnamefont {G.}~\bibnamefont {Montani}}, \bibinfo {author}
  {\bibfnamefont {E.}~\bibnamefont {Rinaldi}}, \ and\ \bibinfo {author}
  {\bibfnamefont {G.}~\bibnamefont {Lambiase}},\ }\href {\doibase
  10.3847/1538-4357/abeb73} {\bibfield  {journal} {\bibinfo  {journal}
  {Astrophys. J.}\ }\textbf {\bibinfo {volume} {912}},\ \bibinfo {pages} {150}
  (\bibinfo {year} {2021})},\ \Eprint {http://arxiv.org/abs/2103.02117}
  {arXiv:2103.02117 [astro-ph.CO]} \BibitemShut {NoStop}%
\bibitem [{\citenamefont {Krishnan}\ \emph
  {et~al.}(2021{\natexlab{a}})\citenamefont {Krishnan}, \citenamefont
  {Mohayaee}, \citenamefont {Colg\'ain}, \citenamefont {Sheikh-Jabbari},\ and\
  \citenamefont {Yin}}]{Krishnan:2021dyb}%
  \BibitemOpen
  \bibfield  {author} {\bibinfo {author} {\bibfnamefont {C.}~\bibnamefont
  {Krishnan}}, \bibinfo {author} {\bibfnamefont {R.}~\bibnamefont {Mohayaee}},
  \bibinfo {author} {\bibfnamefont {E.~O.}\ \bibnamefont {Colg\'ain}}, \bibinfo
  {author} {\bibfnamefont {M.~M.}\ \bibnamefont {Sheikh-Jabbari}}, \ and\
  \bibinfo {author} {\bibfnamefont {L.}~\bibnamefont {Yin}},\ }\href@noop {} {\
   (\bibinfo {year} {2021}{\natexlab{a}})},\ \Eprint
  {http://arxiv.org/abs/2105.09790} {arXiv:2105.09790 [astro-ph.CO]}
  \BibitemShut {NoStop}%
\bibitem [{\citenamefont {Vagnozzi}\ \emph
  {et~al.}(2021{\natexlab{a}})\citenamefont {Vagnozzi}, \citenamefont
  {Pacucci},\ and\ \citenamefont {Loeb}}]{Vagnozzi:2021tjv}%
  \BibitemOpen
  \bibfield  {author} {\bibinfo {author} {\bibfnamefont {S.}~\bibnamefont
  {Vagnozzi}}, \bibinfo {author} {\bibfnamefont {F.}~\bibnamefont {Pacucci}}, \
  and\ \bibinfo {author} {\bibfnamefont {A.}~\bibnamefont {Loeb}},\ }\href@noop
  {} {\  (\bibinfo {year} {2021}{\natexlab{a}})},\ \Eprint
  {http://arxiv.org/abs/2105.10421} {arXiv:2105.10421 [astro-ph.CO]}
  \BibitemShut {NoStop}%
\bibitem [{\citenamefont {Krishnan}\ \emph
  {et~al.}(2021{\natexlab{b}})\citenamefont {Krishnan}, \citenamefont
  {Mohayaee}, \citenamefont {Colg\'ain}, \citenamefont {Sheikh-Jabbari},\ and\
  \citenamefont {Yin}}]{Krishnan:2021jmh}%
  \BibitemOpen
  \bibfield  {author} {\bibinfo {author} {\bibfnamefont {C.}~\bibnamefont
  {Krishnan}}, \bibinfo {author} {\bibfnamefont {R.}~\bibnamefont {Mohayaee}},
  \bibinfo {author} {\bibfnamefont {E.~O.}\ \bibnamefont {Colg\'ain}}, \bibinfo
  {author} {\bibfnamefont {M.~M.}\ \bibnamefont {Sheikh-Jabbari}}, \ and\
  \bibinfo {author} {\bibfnamefont {L.}~\bibnamefont {Yin}},\ }\href@noop {} {\
   (\bibinfo {year} {2021}{\natexlab{b}})},\ \Eprint
  {http://arxiv.org/abs/2106.02532} {arXiv:2106.02532 [astro-ph.CO]}
  \BibitemShut {NoStop}%
\bibitem [{\citenamefont {Ma}\ and\ \citenamefont
  {Bertschinger}(1995)}]{Ma:1995ey}%
  \BibitemOpen
  \bibfield  {author} {\bibinfo {author} {\bibfnamefont {C.-P.}\ \bibnamefont
  {Ma}}\ and\ \bibinfo {author} {\bibfnamefont {E.}~\bibnamefont
  {Bertschinger}},\ }\href {\doibase 10.1086/176550} {\bibfield  {journal}
  {\bibinfo  {journal} {Astrophys. J.}\ }\textbf {\bibinfo {volume} {455}},\
  \bibinfo {pages} {7} (\bibinfo {year} {1995})},\ \Eprint
  {http://arxiv.org/abs/astro-ph/9506072} {arXiv:astro-ph/9506072} \BibitemShut
  {NoStop}%
\bibitem [{\citenamefont {Hou}\ \emph {et~al.}(2013)\citenamefont {Hou},
  \citenamefont {Keisler}, \citenamefont {Knox}, \citenamefont {Millea},\ and\
  \citenamefont {Reichardt}}]{Hou:2011ec}%
  \BibitemOpen
  \bibfield  {author} {\bibinfo {author} {\bibfnamefont {Z.}~\bibnamefont
  {Hou}}, \bibinfo {author} {\bibfnamefont {R.}~\bibnamefont {Keisler}},
  \bibinfo {author} {\bibfnamefont {L.}~\bibnamefont {Knox}}, \bibinfo {author}
  {\bibfnamefont {M.}~\bibnamefont {Millea}}, \ and\ \bibinfo {author}
  {\bibfnamefont {C.}~\bibnamefont {Reichardt}},\ }\href {\doibase
  10.1103/PhysRevD.87.083008} {\bibfield  {journal} {\bibinfo  {journal} {Phys.
  Rev. D}\ }\textbf {\bibinfo {volume} {87}},\ \bibinfo {pages} {083008}
  (\bibinfo {year} {2013})},\ \Eprint {http://arxiv.org/abs/1104.2333}
  {arXiv:1104.2333 [astro-ph.CO]} \BibitemShut {NoStop}%
\bibitem [{\citenamefont {Cabass}\ \emph {et~al.}(2015)\citenamefont {Cabass},
  \citenamefont {Gerbino}, \citenamefont {Giusarma}, \citenamefont
  {Melchiorri}, \citenamefont {Pagano},\ and\ \citenamefont
  {Salvati}}]{Cabass:2015xfa}%
  \BibitemOpen
  \bibfield  {author} {\bibinfo {author} {\bibfnamefont {G.}~\bibnamefont
  {Cabass}}, \bibinfo {author} {\bibfnamefont {M.}~\bibnamefont {Gerbino}},
  \bibinfo {author} {\bibfnamefont {E.}~\bibnamefont {Giusarma}}, \bibinfo
  {author} {\bibfnamefont {A.}~\bibnamefont {Melchiorri}}, \bibinfo {author}
  {\bibfnamefont {L.}~\bibnamefont {Pagano}}, \ and\ \bibinfo {author}
  {\bibfnamefont {L.}~\bibnamefont {Salvati}},\ }\href {\doibase
  10.1103/PhysRevD.92.063534} {\bibfield  {journal} {\bibinfo  {journal} {Phys.
  Rev. D}\ }\textbf {\bibinfo {volume} {92}},\ \bibinfo {pages} {063534}
  (\bibinfo {year} {2015})},\ \Eprint {http://arxiv.org/abs/1507.07586}
  {arXiv:1507.07586 [astro-ph.CO]} \BibitemShut {NoStop}%
\bibitem [{\citenamefont {Kable}\ \emph {et~al.}(2020)\citenamefont {Kable},
  \citenamefont {Addison},\ and\ \citenamefont {Bennett}}]{Kable:2020hcw}%
  \BibitemOpen
  \bibfield  {author} {\bibinfo {author} {\bibfnamefont {J.~A.}\ \bibnamefont
  {Kable}}, \bibinfo {author} {\bibfnamefont {G.~E.}\ \bibnamefont {Addison}},
  \ and\ \bibinfo {author} {\bibfnamefont {C.~L.}\ \bibnamefont {Bennett}},\
  }\href {\doibase 10.3847/1538-4357/abc4e7} {\bibfield  {journal} {\bibinfo
  {journal} {Astrophys. J.}\ }\textbf {\bibinfo {volume} {905}},\ \bibinfo
  {pages} {164} (\bibinfo {year} {2020})},\ \Eprint
  {http://arxiv.org/abs/2008.01785} {arXiv:2008.01785 [astro-ph.CO]}
  \BibitemShut {NoStop}%
\bibitem [{\citenamefont {Efstathiou}\ and\ \citenamefont
  {Gratton}(2019)}]{Efstathiou:2019mdh}%
  \BibitemOpen
  \bibfield  {author} {\bibinfo {author} {\bibfnamefont {G.}~\bibnamefont
  {Efstathiou}}\ and\ \bibinfo {author} {\bibfnamefont {S.}~\bibnamefont
  {Gratton}},\ }\href@noop {} {\  (\bibinfo {year} {2019})},\ \Eprint
  {http://arxiv.org/abs/1910.00483} {arXiv:1910.00483 [astro-ph.CO]}
  \BibitemShut {NoStop}%
\bibitem [{\citenamefont {Aghanim}\ \emph
  {et~al.}(2020{\natexlab{b}})\citenamefont {Aghanim} \emph
  {et~al.}}]{Aghanim:2018oex}%
  \BibitemOpen
  \bibfield  {author} {\bibinfo {author} {\bibfnamefont {N.}~\bibnamefont
  {Aghanim}} \emph {et~al.} (\bibinfo {collaboration} {Planck}),\ }\href
  {\doibase 10.1051/0004-6361/201833886} {\bibfield  {journal} {\bibinfo
  {journal} {Astron. Astrophys.}\ }\textbf {\bibinfo {volume} {641}},\ \bibinfo
  {pages} {A8} (\bibinfo {year} {2020}{\natexlab{b}})},\ \Eprint
  {http://arxiv.org/abs/1807.06210} {arXiv:1807.06210 [astro-ph.CO]}
  \BibitemShut {NoStop}%
\bibitem [{\citenamefont {Aghanim}\ \emph
  {et~al.}(2020{\natexlab{c}})\citenamefont {Aghanim} \emph
  {et~al.}}]{Aghanim:2019ame}%
  \BibitemOpen
  \bibfield  {author} {\bibinfo {author} {\bibfnamefont {N.}~\bibnamefont
  {Aghanim}} \emph {et~al.} (\bibinfo {collaboration} {Planck}),\ }\href
  {\doibase 10.1051/0004-6361/201936386} {\bibfield  {journal} {\bibinfo
  {journal} {Astron. Astrophys.}\ }\textbf {\bibinfo {volume} {641}},\ \bibinfo
  {pages} {A5} (\bibinfo {year} {2020}{\natexlab{c}})},\ \Eprint
  {http://arxiv.org/abs/1907.12875} {arXiv:1907.12875 [astro-ph.CO]}
  \BibitemShut {NoStop}%
\bibitem [{\citenamefont {Alam}\ \emph {et~al.}(2017)\citenamefont {Alam} \emph
  {et~al.}}]{Alam:2016hwk}%
  \BibitemOpen
  \bibfield  {author} {\bibinfo {author} {\bibfnamefont {S.}~\bibnamefont
  {Alam}} \emph {et~al.} (\bibinfo {collaboration} {BOSS}),\ }\href {\doibase
  10.1093/mnras/stx721} {\bibfield  {journal} {\bibinfo  {journal} {Mon. Not.
  Roy. Astron. Soc.}\ }\textbf {\bibinfo {volume} {470}},\ \bibinfo {pages}
  {2617} (\bibinfo {year} {2017})},\ \Eprint {http://arxiv.org/abs/1607.03155}
  {arXiv:1607.03155 [astro-ph.CO]} \BibitemShut {NoStop}%
\bibitem [{\citenamefont {Lewis}\ \emph {et~al.}(2000)\citenamefont {Lewis},
  \citenamefont {Challinor},\ and\ \citenamefont {Lasenby}}]{Lewis:1999bs}%
  \BibitemOpen
  \bibfield  {author} {\bibinfo {author} {\bibfnamefont {A.}~\bibnamefont
  {Lewis}}, \bibinfo {author} {\bibfnamefont {A.}~\bibnamefont {Challinor}}, \
  and\ \bibinfo {author} {\bibfnamefont {A.}~\bibnamefont {Lasenby}},\ }\href
  {\doibase 10.1086/309179} {\bibfield  {journal} {\bibinfo  {journal}
  {Astrophys. J.}\ }\textbf {\bibinfo {volume} {538}},\ \bibinfo {pages} {473}
  (\bibinfo {year} {2000})},\ \Eprint {http://arxiv.org/abs/astro-ph/9911177}
  {arXiv:astro-ph/9911177} \BibitemShut {NoStop}%
\bibitem [{\citenamefont {Lewis}\ and\ \citenamefont
  {Bridle}(2002)}]{Lewis:2002ah}%
  \BibitemOpen
  \bibfield  {author} {\bibinfo {author} {\bibfnamefont {A.}~\bibnamefont
  {Lewis}}\ and\ \bibinfo {author} {\bibfnamefont {S.}~\bibnamefont {Bridle}},\
  }\href {\doibase 10.1103/PhysRevD.66.103511} {\bibfield  {journal} {\bibinfo
  {journal} {Phys. Rev. D}\ }\textbf {\bibinfo {volume} {66}},\ \bibinfo
  {pages} {103511} (\bibinfo {year} {2002})},\ \Eprint
  {http://arxiv.org/abs/astro-ph/0205436} {arXiv:astro-ph/0205436} \BibitemShut
  {NoStop}%
\bibitem [{\citenamefont {Gelman}\ and\ \citenamefont
  {Rubin}(1992)}]{Gelman:1992zz}%
  \BibitemOpen
  \bibfield  {author} {\bibinfo {author} {\bibfnamefont {A.}~\bibnamefont
  {Gelman}}\ and\ \bibinfo {author} {\bibfnamefont {D.~B.}\ \bibnamefont
  {Rubin}},\ }\href {\doibase 10.1214/ss/1177011136} {\bibfield  {journal}
  {\bibinfo  {journal} {Statist. Sci.}\ }\textbf {\bibinfo {volume} {7}},\
  \bibinfo {pages} {457} (\bibinfo {year} {1992})}\BibitemShut {NoStop}%
\bibitem [{\citenamefont {Vagnozzi}(2019)}]{Vagnozzi:2019utt}%
  \BibitemOpen
  \bibfield  {author} {\bibinfo {author} {\bibfnamefont {S.}~\bibnamefont
  {Vagnozzi}},\ }\href@noop {} {\  (\bibinfo {year} {2019})},\ \Eprint
  {http://arxiv.org/abs/1907.08010} {arXiv:1907.08010 [astro-ph.CO]}
  \BibitemShut {NoStop}%
\bibitem [{\citenamefont {Park}\ and\ \citenamefont
  {Ratra}(2019)}]{Park:2017xbl}%
  \BibitemOpen
  \bibfield  {author} {\bibinfo {author} {\bibfnamefont {C.-G.}\ \bibnamefont
  {Park}}\ and\ \bibinfo {author} {\bibfnamefont {B.}~\bibnamefont {Ratra}},\
  }\href {\doibase 10.3847/1538-4357/ab3641} {\bibfield  {journal} {\bibinfo
  {journal} {Astrophys. J.}\ }\textbf {\bibinfo {volume} {882}},\ \bibinfo
  {pages} {158} (\bibinfo {year} {2019})},\ \Eprint
  {http://arxiv.org/abs/1801.00213} {arXiv:1801.00213 [astro-ph.CO]}
  \BibitemShut {NoStop}%
\bibitem [{\citenamefont {Handley}(2021)}]{Handley:2019tkm}%
  \BibitemOpen
  \bibfield  {author} {\bibinfo {author} {\bibfnamefont {W.}~\bibnamefont
  {Handley}},\ }\href {\doibase 10.1103/PhysRevD.103.L041301} {\bibfield
  {journal} {\bibinfo  {journal} {Phys. Rev. D}\ }\textbf {\bibinfo {volume}
  {103}},\ \bibinfo {pages} {L041301} (\bibinfo {year} {2021})},\ \Eprint
  {http://arxiv.org/abs/1908.09139} {arXiv:1908.09139 [astro-ph.CO]}
  \BibitemShut {NoStop}%
\bibitem [{\citenamefont {Di~Valentino}\ \emph
  {et~al.}(2019{\natexlab{b}})\citenamefont {Di~Valentino}, \citenamefont
  {Melchiorri},\ and\ \citenamefont {Silk}}]{DiValentino:2019qzk}%
  \BibitemOpen
  \bibfield  {author} {\bibinfo {author} {\bibfnamefont {E.}~\bibnamefont
  {Di~Valentino}}, \bibinfo {author} {\bibfnamefont {A.}~\bibnamefont
  {Melchiorri}}, \ and\ \bibinfo {author} {\bibfnamefont {J.}~\bibnamefont
  {Silk}},\ }\href {\doibase 10.1038/s41550-019-0906-9} {\bibfield  {journal}
  {\bibinfo  {journal} {Nature Astron.}\ }\textbf {\bibinfo {volume} {4}},\
  \bibinfo {pages} {196} (\bibinfo {year} {2019}{\natexlab{b}})},\ \Eprint
  {http://arxiv.org/abs/1911.02087} {arXiv:1911.02087 [astro-ph.CO]}
  \BibitemShut {NoStop}%
\bibitem [{\citenamefont {Efstathiou}\ and\ \citenamefont
  {Gratton}(2020)}]{Efstathiou:2020wem}%
  \BibitemOpen
  \bibfield  {author} {\bibinfo {author} {\bibfnamefont {G.}~\bibnamefont
  {Efstathiou}}\ and\ \bibinfo {author} {\bibfnamefont {S.}~\bibnamefont
  {Gratton}},\ }\href {\doibase 10.1093/mnrasl/slaa093} {\bibfield  {journal}
  {\bibinfo  {journal} {Mon. Not. Roy. Astron. Soc.}\ }\textbf {\bibinfo
  {volume} {496}},\ \bibinfo {pages} {L91} (\bibinfo {year} {2020})},\ \Eprint
  {http://arxiv.org/abs/2002.06892} {arXiv:2002.06892 [astro-ph.CO]}
  \BibitemShut {NoStop}%
\bibitem [{\citenamefont {Di~Valentino}\ \emph
  {et~al.}(2021{\natexlab{b}})\citenamefont {Di~Valentino}, \citenamefont
  {Melchiorri},\ and\ \citenamefont {Silk}}]{DiValentino:2020hov}%
  \BibitemOpen
  \bibfield  {author} {\bibinfo {author} {\bibfnamefont {E.}~\bibnamefont
  {Di~Valentino}}, \bibinfo {author} {\bibfnamefont {A.}~\bibnamefont
  {Melchiorri}}, \ and\ \bibinfo {author} {\bibfnamefont {J.}~\bibnamefont
  {Silk}},\ }\href {\doibase 10.3847/2041-8213/abe1c4} {\bibfield  {journal}
  {\bibinfo  {journal} {Astrophys. J. Lett.}\ }\textbf {\bibinfo {volume}
  {908}},\ \bibinfo {pages} {L9} (\bibinfo {year} {2021}{\natexlab{b}})},\
  \Eprint {http://arxiv.org/abs/2003.04935} {arXiv:2003.04935 [astro-ph.CO]}
  \BibitemShut {NoStop}%
\bibitem [{\citenamefont {Di~Valentino}\ \emph
  {et~al.}(2020{\natexlab{c}})\citenamefont {Di~Valentino}, \citenamefont
  {Gariazzo}, \citenamefont {Mena},\ and\ \citenamefont
  {Vagnozzi}}]{DiValentino:2020evt}%
  \BibitemOpen
  \bibfield  {author} {\bibinfo {author} {\bibfnamefont {E.}~\bibnamefont
  {Di~Valentino}}, \bibinfo {author} {\bibfnamefont {S.}~\bibnamefont
  {Gariazzo}}, \bibinfo {author} {\bibfnamefont {O.}~\bibnamefont {Mena}}, \
  and\ \bibinfo {author} {\bibfnamefont {S.}~\bibnamefont {Vagnozzi}},\ }\href
  {\doibase 10.1088/1475-7516/2020/07/045} {\bibfield  {journal} {\bibinfo
  {journal} {JCAP}\ }\textbf {\bibinfo {volume} {07}},\ \bibinfo {pages} {045}
  (\bibinfo {year} {2020}{\natexlab{c}})},\ \Eprint
  {http://arxiv.org/abs/2005.02062} {arXiv:2005.02062 [astro-ph.CO]}
  \BibitemShut {NoStop}%
\bibitem [{\citenamefont {Di~Valentino}\ \emph
  {et~al.}(2021{\natexlab{c}})\citenamefont {Di~Valentino} \emph
  {et~al.}}]{DiValentino:2020srs}%
  \BibitemOpen
  \bibfield  {author} {\bibinfo {author} {\bibfnamefont {E.}~\bibnamefont
  {Di~Valentino}} \emph {et~al.},\ }\href {\doibase
  10.1016/j.astropartphys.2021.102607} {\bibfield  {journal} {\bibinfo
  {journal} {Astropart. Phys.}\ }\textbf {\bibinfo {volume} {131}},\ \bibinfo
  {pages} {102607} (\bibinfo {year} {2021}{\natexlab{c}})},\ \Eprint
  {http://arxiv.org/abs/2008.11286} {arXiv:2008.11286 [astro-ph.CO]}
  \BibitemShut {NoStop}%
\bibitem [{\citenamefont {Vagnozzi}\ \emph
  {et~al.}(2021{\natexlab{b}})\citenamefont {Vagnozzi}, \citenamefont
  {Di~Valentino}, \citenamefont {Gariazzo}, \citenamefont {Melchiorri},
  \citenamefont {Mena},\ and\ \citenamefont {Silk}}]{Vagnozzi:2020rcz}%
  \BibitemOpen
  \bibfield  {author} {\bibinfo {author} {\bibfnamefont {S.}~\bibnamefont
  {Vagnozzi}}, \bibinfo {author} {\bibfnamefont {E.}~\bibnamefont
  {Di~Valentino}}, \bibinfo {author} {\bibfnamefont {S.}~\bibnamefont
  {Gariazzo}}, \bibinfo {author} {\bibfnamefont {A.}~\bibnamefont
  {Melchiorri}}, \bibinfo {author} {\bibfnamefont {O.}~\bibnamefont {Mena}}, \
  and\ \bibinfo {author} {\bibfnamefont {J.}~\bibnamefont {Silk}},\ }\href
  {\doibase 10.1016/j.dark.2021.100851} {\bibfield  {journal} {\bibinfo
  {journal} {Phys. Dark Univ.}\ }\textbf {\bibinfo {volume} {33}},\ \bibinfo
  {pages} {100851} (\bibinfo {year} {2021}{\natexlab{b}})},\ \Eprint
  {http://arxiv.org/abs/2010.02230} {arXiv:2010.02230 [astro-ph.CO]}
  \BibitemShut {NoStop}%
\bibitem [{\citenamefont {Di~Valentino}\ \emph
  {et~al.}(2021{\natexlab{d}})\citenamefont {Di~Valentino}, \citenamefont
  {Melchiorri}, \citenamefont {Mena}, \citenamefont {Pan},\ and\ \citenamefont
  {Yang}}]{DiValentino:2020kpf}%
  \BibitemOpen
  \bibfield  {author} {\bibinfo {author} {\bibfnamefont {E.}~\bibnamefont
  {Di~Valentino}}, \bibinfo {author} {\bibfnamefont {A.}~\bibnamefont
  {Melchiorri}}, \bibinfo {author} {\bibfnamefont {O.}~\bibnamefont {Mena}},
  \bibinfo {author} {\bibfnamefont {S.}~\bibnamefont {Pan}}, \ and\ \bibinfo
  {author} {\bibfnamefont {W.}~\bibnamefont {Yang}},\ }\href {\doibase
  10.1093/mnrasl/slaa207} {\bibfield  {journal} {\bibinfo  {journal} {Mon. Not.
  Roy. Astron. Soc.}\ }\textbf {\bibinfo {volume} {502}},\ \bibinfo {pages}
  {L23} (\bibinfo {year} {2021}{\natexlab{d}})},\ \Eprint
  {http://arxiv.org/abs/2011.00283} {arXiv:2011.00283 [astro-ph.CO]}
  \BibitemShut {NoStop}%
\bibitem [{\citenamefont {Vagnozzi}\ \emph
  {et~al.}(2021{\natexlab{c}})\citenamefont {Vagnozzi}, \citenamefont {Loeb},\
  and\ \citenamefont {Moresco}}]{Vagnozzi:2020dfn}%
  \BibitemOpen
  \bibfield  {author} {\bibinfo {author} {\bibfnamefont {S.}~\bibnamefont
  {Vagnozzi}}, \bibinfo {author} {\bibfnamefont {A.}~\bibnamefont {Loeb}}, \
  and\ \bibinfo {author} {\bibfnamefont {M.}~\bibnamefont {Moresco}},\ }\href
  {\doibase 10.3847/1538-4357/abd4df} {\bibfield  {journal} {\bibinfo
  {journal} {Astrophys. J.}\ }\textbf {\bibinfo {volume} {908}},\ \bibinfo
  {pages} {84} (\bibinfo {year} {2021}{\natexlab{c}})},\ \Eprint
  {http://arxiv.org/abs/2011.11645} {arXiv:2011.11645 [astro-ph.CO]}
  \BibitemShut {NoStop}%
\bibitem [{\citenamefont {Cao}\ \emph {et~al.}(2021)\citenamefont {Cao},
  \citenamefont {Ryan},\ and\ \citenamefont {Ratra}}]{Cao:2021ldv}%
  \BibitemOpen
  \bibfield  {author} {\bibinfo {author} {\bibfnamefont {S.}~\bibnamefont
  {Cao}}, \bibinfo {author} {\bibfnamefont {J.}~\bibnamefont {Ryan}}, \ and\
  \bibinfo {author} {\bibfnamefont {B.}~\bibnamefont {Ratra}},\ }\href
  {\doibase 10.1093/mnras/stab942} {\bibfield  {journal} {\bibinfo  {journal}
  {Mon. Not. Roy. Astron. Soc.}\ }\textbf {\bibinfo {volume} {504}},\ \bibinfo
  {pages} {300} (\bibinfo {year} {2021})},\ \Eprint
  {http://arxiv.org/abs/2101.08817} {arXiv:2101.08817 [astro-ph.CO]}
  \BibitemShut {NoStop}%
\bibitem [{\citenamefont {Dhawan}\ \emph {et~al.}(2021)\citenamefont {Dhawan},
  \citenamefont {Alsing},\ and\ \citenamefont {Vagnozzi}}]{Dhawan:2021mel}%
  \BibitemOpen
  \bibfield  {author} {\bibinfo {author} {\bibfnamefont {S.}~\bibnamefont
  {Dhawan}}, \bibinfo {author} {\bibfnamefont {J.}~\bibnamefont {Alsing}}, \
  and\ \bibinfo {author} {\bibfnamefont {S.}~\bibnamefont {Vagnozzi}},\ }\href
  {\doibase 10.1093/mnrasl/slab058} {\bibfield  {journal} {\bibinfo  {journal}
  {Mon. Not. Roy. Astron. Soc.}\ }\textbf {\bibinfo {volume} {506}},\ \bibinfo
  {pages} {L1} (\bibinfo {year} {2021})},\ \Eprint
  {http://arxiv.org/abs/2104.02485} {arXiv:2104.02485 [astro-ph.CO]}
  \BibitemShut {NoStop}%
\bibitem [{\citenamefont {Gonzalez}\ \emph {et~al.}(2021)\citenamefont
  {Gonzalez}, \citenamefont {Benetti}, \citenamefont {von Marttens},\ and\
  \citenamefont {Alcaniz}}]{Gonzalez:2021ojp}%
  \BibitemOpen
  \bibfield  {author} {\bibinfo {author} {\bibfnamefont {J.~E.}\ \bibnamefont
  {Gonzalez}}, \bibinfo {author} {\bibfnamefont {M.}~\bibnamefont {Benetti}},
  \bibinfo {author} {\bibfnamefont {R.}~\bibnamefont {von Marttens}}, \ and\
  \bibinfo {author} {\bibfnamefont {J.}~\bibnamefont {Alcaniz}},\ }\href@noop
  {} {\  (\bibinfo {year} {2021})},\ \Eprint {http://arxiv.org/abs/2104.13455}
  {arXiv:2104.13455 [astro-ph.CO]} \BibitemShut {NoStop}%
\bibitem [{\citenamefont {Cabass}\ \emph {et~al.}(2016)\citenamefont {Cabass},
  \citenamefont {Di~Valentino}, \citenamefont {Melchiorri}, \citenamefont
  {Pajer},\ and\ \citenamefont {Silk}}]{Cabass:2016ldu}%
  \BibitemOpen
  \bibfield  {author} {\bibinfo {author} {\bibfnamefont {G.}~\bibnamefont
  {Cabass}}, \bibinfo {author} {\bibfnamefont {E.}~\bibnamefont
  {Di~Valentino}}, \bibinfo {author} {\bibfnamefont {A.}~\bibnamefont
  {Melchiorri}}, \bibinfo {author} {\bibfnamefont {E.}~\bibnamefont {Pajer}}, \
  and\ \bibinfo {author} {\bibfnamefont {J.}~\bibnamefont {Silk}},\ }\href
  {\doibase 10.1103/PhysRevD.94.023523} {\bibfield  {journal} {\bibinfo
  {journal} {Phys. Rev. D}\ }\textbf {\bibinfo {volume} {94}},\ \bibinfo
  {pages} {023523} (\bibinfo {year} {2016})},\ \Eprint
  {http://arxiv.org/abs/1605.00209} {arXiv:1605.00209 [astro-ph.CO]}
  \BibitemShut {NoStop}%
\bibitem [{\citenamefont {Addison}\ \emph {et~al.}(2016)\citenamefont
  {Addison}, \citenamefont {Huang}, \citenamefont {Watts}, \citenamefont
  {Bennett}, \citenamefont {Halpern}, \citenamefont {Hinshaw},\ and\
  \citenamefont {Weiland}}]{Addison:2015wyg}%
  \BibitemOpen
  \bibfield  {author} {\bibinfo {author} {\bibfnamefont {G.~E.}\ \bibnamefont
  {Addison}}, \bibinfo {author} {\bibfnamefont {Y.}~\bibnamefont {Huang}},
  \bibinfo {author} {\bibfnamefont {D.~J.}\ \bibnamefont {Watts}}, \bibinfo
  {author} {\bibfnamefont {C.~L.}\ \bibnamefont {Bennett}}, \bibinfo {author}
  {\bibfnamefont {M.}~\bibnamefont {Halpern}}, \bibinfo {author} {\bibfnamefont
  {G.}~\bibnamefont {Hinshaw}}, \ and\ \bibinfo {author} {\bibfnamefont
  {J.~L.}\ \bibnamefont {Weiland}},\ }\href {\doibase
  10.3847/0004-637X/818/2/132} {\bibfield  {journal} {\bibinfo  {journal}
  {Astrophys. J.}\ }\textbf {\bibinfo {volume} {818}},\ \bibinfo {pages} {132}
  (\bibinfo {year} {2016})},\ \Eprint {http://arxiv.org/abs/1511.00055}
  {arXiv:1511.00055 [astro-ph.CO]} \BibitemShut {NoStop}%
\bibitem [{\citenamefont {Galli}\ \emph {et~al.}(2014)\citenamefont {Galli},
  \citenamefont {Benabed}, \citenamefont {Bouchet}, \citenamefont {Cardoso},
  \citenamefont {Elsner}, \citenamefont {Hivon}, \citenamefont {Mangilli},
  \citenamefont {Prunet},\ and\ \citenamefont {Wandelt}}]{Galli:2014kla}%
  \BibitemOpen
  \bibfield  {author} {\bibinfo {author} {\bibfnamefont {S.}~\bibnamefont
  {Galli}}, \bibinfo {author} {\bibfnamefont {K.}~\bibnamefont {Benabed}},
  \bibinfo {author} {\bibfnamefont {F.}~\bibnamefont {Bouchet}}, \bibinfo
  {author} {\bibfnamefont {J.-F.}\ \bibnamefont {Cardoso}}, \bibinfo {author}
  {\bibfnamefont {F.}~\bibnamefont {Elsner}}, \bibinfo {author} {\bibfnamefont
  {E.}~\bibnamefont {Hivon}}, \bibinfo {author} {\bibfnamefont
  {A.}~\bibnamefont {Mangilli}}, \bibinfo {author} {\bibfnamefont
  {S.}~\bibnamefont {Prunet}}, \ and\ \bibinfo {author} {\bibfnamefont
  {B.}~\bibnamefont {Wandelt}},\ }\href {\doibase 10.1103/PhysRevD.90.063504}
  {\bibfield  {journal} {\bibinfo  {journal} {Phys. Rev. D}\ }\textbf {\bibinfo
  {volume} {90}},\ \bibinfo {pages} {063504} (\bibinfo {year} {2014})},\
  \Eprint {http://arxiv.org/abs/1403.5271} {arXiv:1403.5271 [astro-ph.CO]}
  \BibitemShut {NoStop}%
\bibitem [{\citenamefont {Ade}\ \emph {et~al.}(2019)\citenamefont {Ade} \emph
  {et~al.}}]{Ade:2018sbj}%
  \BibitemOpen
  \bibfield  {author} {\bibinfo {author} {\bibfnamefont {P.}~\bibnamefont
  {Ade}} \emph {et~al.} (\bibinfo {collaboration} {Simons Observatory}),\
  }\href {\doibase 10.1088/1475-7516/2019/02/056} {\bibfield  {journal}
  {\bibinfo  {journal} {JCAP}\ }\textbf {\bibinfo {volume} {02}},\ \bibinfo
  {pages} {056} (\bibinfo {year} {2019})},\ \Eprint
  {http://arxiv.org/abs/1808.07445} {arXiv:1808.07445 [astro-ph.CO]}
  \BibitemShut {NoStop}%
\bibitem [{\citenamefont {Abazajian}\ \emph {et~al.}(2016)\citenamefont
  {Abazajian} \emph {et~al.}}]{Abazajian:2016yjj}%
  \BibitemOpen
  \bibfield  {author} {\bibinfo {author} {\bibfnamefont {K.~N.}\ \bibnamefont
  {Abazajian}} \emph {et~al.} (\bibinfo {collaboration} {CMB-S4}),\ }\href@noop
  {} {\  (\bibinfo {year} {2016})},\ \Eprint {http://arxiv.org/abs/1610.02743}
  {arXiv:1610.02743 [astro-ph.CO]} \BibitemShut {NoStop}%
\bibitem [{\citenamefont {Svrcek}\ and\ \citenamefont
  {Witten}(2006)}]{Svrcek:2006yi}%
  \BibitemOpen
  \bibfield  {author} {\bibinfo {author} {\bibfnamefont {P.}~\bibnamefont
  {Svrcek}}\ and\ \bibinfo {author} {\bibfnamefont {E.}~\bibnamefont
  {Witten}},\ }\href {\doibase 10.1088/1126-6708/2006/06/051} {\bibfield
  {journal} {\bibinfo  {journal} {JHEP}\ }\textbf {\bibinfo {volume} {06}},\
  \bibinfo {pages} {051} (\bibinfo {year} {2006})},\ \Eprint
  {http://arxiv.org/abs/hep-th/0605206} {arXiv:hep-th/0605206} \BibitemShut
  {NoStop}%
\bibitem [{\citenamefont {Arvanitaki}\ \emph {et~al.}(2010)\citenamefont
  {Arvanitaki}, \citenamefont {Dimopoulos}, \citenamefont {Dubovsky},
  \citenamefont {Kaloper},\ and\ \citenamefont
  {March-Russell}}]{Arvanitaki:2009fg}%
  \BibitemOpen
  \bibfield  {author} {\bibinfo {author} {\bibfnamefont {A.}~\bibnamefont
  {Arvanitaki}}, \bibinfo {author} {\bibfnamefont {S.}~\bibnamefont
  {Dimopoulos}}, \bibinfo {author} {\bibfnamefont {S.}~\bibnamefont
  {Dubovsky}}, \bibinfo {author} {\bibfnamefont {N.}~\bibnamefont {Kaloper}}, \
  and\ \bibinfo {author} {\bibfnamefont {J.}~\bibnamefont {March-Russell}},\
  }\href {\doibase 10.1103/PhysRevD.81.123530} {\bibfield  {journal} {\bibinfo
  {journal} {Phys. Rev. D}\ }\textbf {\bibinfo {volume} {81}},\ \bibinfo
  {pages} {123530} (\bibinfo {year} {2010})},\ \Eprint
  {http://arxiv.org/abs/0905.4720} {arXiv:0905.4720 [hep-th]} \BibitemShut
  {NoStop}%
\bibitem [{\citenamefont {Kamionkowski}\ \emph {et~al.}(2014)\citenamefont
  {Kamionkowski}, \citenamefont {Pradler},\ and\ \citenamefont
  {Walker}}]{Kamionkowski:2014zda}%
  \BibitemOpen
  \bibfield  {author} {\bibinfo {author} {\bibfnamefont {M.}~\bibnamefont
  {Kamionkowski}}, \bibinfo {author} {\bibfnamefont {J.}~\bibnamefont
  {Pradler}}, \ and\ \bibinfo {author} {\bibfnamefont {D.~G.~E.}\ \bibnamefont
  {Walker}},\ }\href {\doibase 10.1103/PhysRevLett.113.251302} {\bibfield
  {journal} {\bibinfo  {journal} {Phys. Rev. Lett.}\ }\textbf {\bibinfo
  {volume} {113}},\ \bibinfo {pages} {251302} (\bibinfo {year} {2014})},\
  \Eprint {http://arxiv.org/abs/1409.0549} {arXiv:1409.0549 [hep-ph]}
  \BibitemShut {NoStop}%
\bibitem [{\citenamefont {Karwal}\ and\ \citenamefont
  {Kamionkowski}(2016)}]{Karwal:2016vyq}%
  \BibitemOpen
  \bibfield  {author} {\bibinfo {author} {\bibfnamefont {T.}~\bibnamefont
  {Karwal}}\ and\ \bibinfo {author} {\bibfnamefont {M.}~\bibnamefont
  {Kamionkowski}},\ }\href {\doibase 10.1103/PhysRevD.94.103523} {\bibfield
  {journal} {\bibinfo  {journal} {Phys. Rev. D}\ }\textbf {\bibinfo {volume}
  {94}},\ \bibinfo {pages} {103523} (\bibinfo {year} {2016})},\ \Eprint
  {http://arxiv.org/abs/1608.01309} {arXiv:1608.01309 [astro-ph.CO]}
  \BibitemShut {NoStop}%
\bibitem [{\citenamefont {Visinelli}\ and\ \citenamefont
  {Vagnozzi}(2019)}]{Visinelli:2018utg}%
  \BibitemOpen
  \bibfield  {author} {\bibinfo {author} {\bibfnamefont {L.}~\bibnamefont
  {Visinelli}}\ and\ \bibinfo {author} {\bibfnamefont {S.}~\bibnamefont
  {Vagnozzi}},\ }\href {\doibase 10.1103/PhysRevD.99.063517} {\bibfield
  {journal} {\bibinfo  {journal} {Phys. Rev. D}\ }\textbf {\bibinfo {volume}
  {99}},\ \bibinfo {pages} {063517} (\bibinfo {year} {2019})},\ \Eprint
  {http://arxiv.org/abs/1809.06382} {arXiv:1809.06382 [hep-ph]} \BibitemShut
  {NoStop}%
\bibitem [{\citenamefont {Rudelius}(2015)}]{Rudelius:2015xta}%
  \BibitemOpen
  \bibfield  {author} {\bibinfo {author} {\bibfnamefont {T.}~\bibnamefont
  {Rudelius}},\ }\href {\doibase 10.1088/1475-7516/2015/9/020} {\bibfield
  {journal} {\bibinfo  {journal} {JCAP}\ }\textbf {\bibinfo {volume} {09}},\
  \bibinfo {pages} {020} (\bibinfo {year} {2015})},\ \Eprint
  {http://arxiv.org/abs/1503.00795} {arXiv:1503.00795 [hep-th]} \BibitemShut
  {NoStop}%
\bibitem [{\citenamefont {Montero}\ \emph {et~al.}(2015)\citenamefont
  {Montero}, \citenamefont {Uranga},\ and\ \citenamefont
  {Valenzuela}}]{Montero:2015ofa}%
  \BibitemOpen
  \bibfield  {author} {\bibinfo {author} {\bibfnamefont {M.}~\bibnamefont
  {Montero}}, \bibinfo {author} {\bibfnamefont {A.~M.}\ \bibnamefont {Uranga}},
  \ and\ \bibinfo {author} {\bibfnamefont {I.}~\bibnamefont {Valenzuela}},\
  }\href {\doibase 10.1007/JHEP08(2015)032} {\bibfield  {journal} {\bibinfo
  {journal} {JHEP}\ }\textbf {\bibinfo {volume} {08}},\ \bibinfo {pages} {032}
  (\bibinfo {year} {2015})},\ \Eprint {http://arxiv.org/abs/1503.03886}
  {arXiv:1503.03886 [hep-th]} \BibitemShut {NoStop}%
\bibitem [{\citenamefont {Blas}\ \emph {et~al.}(2011)\citenamefont {Blas},
  \citenamefont {Lesgourgues},\ and\ \citenamefont {Tram}}]{Blas:2011rf}%
  \BibitemOpen
  \bibfield  {author} {\bibinfo {author} {\bibfnamefont {D.}~\bibnamefont
  {Blas}}, \bibinfo {author} {\bibfnamefont {J.}~\bibnamefont {Lesgourgues}}, \
  and\ \bibinfo {author} {\bibfnamefont {T.}~\bibnamefont {Tram}},\ }\href
  {\doibase 10.1088/1475-7516/2011/07/034} {\bibfield  {journal} {\bibinfo
  {journal} {JCAP}\ }\textbf {\bibinfo {volume} {07}},\ \bibinfo {pages} {034}
  (\bibinfo {year} {2011})},\ \Eprint {http://arxiv.org/abs/1104.2933}
  {arXiv:1104.2933 [astro-ph.CO]} \BibitemShut {NoStop}%
\bibitem [{\citenamefont {Ho}\ \emph {et~al.}(2008)\citenamefont {Ho},
  \citenamefont {Hirata}, \citenamefont {Padmanabhan}, \citenamefont {Seljak},\
  and\ \citenamefont {Bahcall}}]{Ho:2008bz}%
  \BibitemOpen
  \bibfield  {author} {\bibinfo {author} {\bibfnamefont {S.}~\bibnamefont
  {Ho}}, \bibinfo {author} {\bibfnamefont {C.}~\bibnamefont {Hirata}}, \bibinfo
  {author} {\bibfnamefont {N.}~\bibnamefont {Padmanabhan}}, \bibinfo {author}
  {\bibfnamefont {U.}~\bibnamefont {Seljak}}, \ and\ \bibinfo {author}
  {\bibfnamefont {N.}~\bibnamefont {Bahcall}},\ }\href {\doibase
  10.1103/PhysRevD.78.043519} {\bibfield  {journal} {\bibinfo  {journal} {Phys.
  Rev. D}\ }\textbf {\bibinfo {volume} {78}},\ \bibinfo {pages} {043519}
  (\bibinfo {year} {2008})},\ \Eprint {http://arxiv.org/abs/0801.0642}
  {arXiv:0801.0642 [astro-ph]} \BibitemShut {NoStop}%
\bibitem [{\citenamefont {Giannantonio}\ and\ \citenamefont
  {Percival}(2014)}]{Giannantonio:2013kqa}%
  \BibitemOpen
  \bibfield  {author} {\bibinfo {author} {\bibfnamefont {T.}~\bibnamefont
  {Giannantonio}}\ and\ \bibinfo {author} {\bibfnamefont {W.~J.}\ \bibnamefont
  {Percival}},\ }\href {\doibase 10.1093/mnrasl/slu036} {\bibfield  {journal}
  {\bibinfo  {journal} {Mon. Not. Roy. Astron. Soc.}\ }\textbf {\bibinfo
  {volume} {441}},\ \bibinfo {pages} {L16} (\bibinfo {year} {2014})},\ \Eprint
  {http://arxiv.org/abs/1312.5154} {arXiv:1312.5154 [astro-ph.CO]} \BibitemShut
  {NoStop}%
\bibitem [{\citenamefont {Renk}\ \emph {et~al.}(2016)\citenamefont {Renk},
  \citenamefont {Zumalacarregui},\ and\ \citenamefont
  {Montanari}}]{Renk:2016olm}%
  \BibitemOpen
  \bibfield  {author} {\bibinfo {author} {\bibfnamefont {J.}~\bibnamefont
  {Renk}}, \bibinfo {author} {\bibfnamefont {M.}~\bibnamefont
  {Zumalacarregui}}, \ and\ \bibinfo {author} {\bibfnamefont {F.}~\bibnamefont
  {Montanari}},\ }\href {\doibase 10.1088/1475-7516/2016/07/040} {\bibfield
  {journal} {\bibinfo  {journal} {JCAP}\ }\textbf {\bibinfo {volume} {07}},\
  \bibinfo {pages} {040} (\bibinfo {year} {2016})},\ \Eprint
  {http://arxiv.org/abs/1604.03487} {arXiv:1604.03487 [astro-ph.CO]}
  \BibitemShut {NoStop}%
\bibitem [{\citenamefont {Giusarma}\ \emph {et~al.}(2018)\citenamefont
  {Giusarma}, \citenamefont {Vagnozzi}, \citenamefont {Ho}, \citenamefont
  {Ferraro}, \citenamefont {Freese}, \citenamefont {Kamen-Rubio},\ and\
  \citenamefont {Luk}}]{Giusarma:2018jei}%
  \BibitemOpen
  \bibfield  {author} {\bibinfo {author} {\bibfnamefont {E.}~\bibnamefont
  {Giusarma}}, \bibinfo {author} {\bibfnamefont {S.}~\bibnamefont {Vagnozzi}},
  \bibinfo {author} {\bibfnamefont {S.}~\bibnamefont {Ho}}, \bibinfo {author}
  {\bibfnamefont {S.}~\bibnamefont {Ferraro}}, \bibinfo {author} {\bibfnamefont
  {K.}~\bibnamefont {Freese}}, \bibinfo {author} {\bibfnamefont
  {R.}~\bibnamefont {Kamen-Rubio}}, \ and\ \bibinfo {author} {\bibfnamefont
  {K.-B.}\ \bibnamefont {Luk}},\ }\href {\doibase 10.1103/PhysRevD.98.123526}
  {\bibfield  {journal} {\bibinfo  {journal} {Phys. Rev. D}\ }\textbf {\bibinfo
  {volume} {98}},\ \bibinfo {pages} {123526} (\bibinfo {year} {2018})},\
  \Eprint {http://arxiv.org/abs/1802.08694} {arXiv:1802.08694 [astro-ph.CO]}
  \BibitemShut {NoStop}%
\bibitem [{\citenamefont {Vagnozzi}\ \emph {et~al.}(2020)\citenamefont
  {Vagnozzi}, \citenamefont {Visinelli}, \citenamefont {Mena},\ and\
  \citenamefont {Mota}}]{Vagnozzi:2019kvw}%
  \BibitemOpen
  \bibfield  {author} {\bibinfo {author} {\bibfnamefont {S.}~\bibnamefont
  {Vagnozzi}}, \bibinfo {author} {\bibfnamefont {L.}~\bibnamefont {Visinelli}},
  \bibinfo {author} {\bibfnamefont {O.}~\bibnamefont {Mena}}, \ and\ \bibinfo
  {author} {\bibfnamefont {D.~F.}\ \bibnamefont {Mota}},\ }\href {\doibase
  10.1093/mnras/staa311} {\bibfield  {journal} {\bibinfo  {journal} {Mon. Not.
  Roy. Astron. Soc.}\ }\textbf {\bibinfo {volume} {493}},\ \bibinfo {pages}
  {1139} (\bibinfo {year} {2020})},\ \Eprint {http://arxiv.org/abs/1911.12374}
  {arXiv:1911.12374 [gr-qc]} \BibitemShut {NoStop}%
\bibitem [{\citenamefont {Dodelson}(2003)}]{Dodelson:2003ft}%
  \BibitemOpen
  \bibfield  {author} {\bibinfo {author} {\bibfnamefont {S.}~\bibnamefont
  {Dodelson}},\ }\href@noop {} {\emph {\bibinfo {title} {{Modern Cosmology}}}}\
  (\bibinfo  {publisher} {Academic Press},\ \bibinfo {address} {Amsterdam},\
  \bibinfo {year} {2003})\BibitemShut {NoStop}%
\bibitem [{\citenamefont {Nunes}\ and\ \citenamefont
  {Vagnozzi}(2021)}]{Nunes:2021ipq}%
  \BibitemOpen
  \bibfield  {author} {\bibinfo {author} {\bibfnamefont {R.~C.}\ \bibnamefont
  {Nunes}}\ and\ \bibinfo {author} {\bibfnamefont {S.}~\bibnamefont
  {Vagnozzi}},\ }\href {\doibase 10.1093/mnras/stab1613} {\  (\bibinfo {year}
  {2021}),\ 10.1093/mnras/stab1613},\ \Eprint {http://arxiv.org/abs/2106.01208}
  {arXiv:2106.01208 [astro-ph.CO]} \BibitemShut {NoStop}%
\bibitem [{\citenamefont {Escudero}\ and\ \citenamefont
  {Witte}(2021)}]{Escudero:2021rfi}%
  \BibitemOpen
  \bibfield  {author} {\bibinfo {author} {\bibfnamefont {M.}~\bibnamefont
  {Escudero}}\ and\ \bibinfo {author} {\bibfnamefont {S.~J.}\ \bibnamefont
  {Witte}},\ }\href@noop {} {\  (\bibinfo {year} {2021})},\ \Eprint
  {http://arxiv.org/abs/2103.03249} {arXiv:2103.03249 [hep-ph]} \BibitemShut
  {NoStop}%
\bibitem [{\citenamefont {Lombriser}(2020)}]{Lombriser:2019ahl}%
  \BibitemOpen
  \bibfield  {author} {\bibinfo {author} {\bibfnamefont {L.}~\bibnamefont
  {Lombriser}},\ }\href {\doibase 10.1016/j.physletb.2020.135303} {\bibfield
  {journal} {\bibinfo  {journal} {Phys. Lett. B}\ }\textbf {\bibinfo {volume}
  {803}},\ \bibinfo {pages} {135303} (\bibinfo {year} {2020})},\ \Eprint
  {http://arxiv.org/abs/1906.12347} {arXiv:1906.12347 [astro-ph.CO]}
  \BibitemShut {NoStop}%
\bibitem [{\citenamefont {Desmond}\ \emph {et~al.}(2019)\citenamefont
  {Desmond}, \citenamefont {Jain},\ and\ \citenamefont
  {Sakstein}}]{Desmond:2019ygn}%
  \BibitemOpen
  \bibfield  {author} {\bibinfo {author} {\bibfnamefont {H.}~\bibnamefont
  {Desmond}}, \bibinfo {author} {\bibfnamefont {B.}~\bibnamefont {Jain}}, \
  and\ \bibinfo {author} {\bibfnamefont {J.}~\bibnamefont {Sakstein}},\ }\href
  {\doibase 10.1103/PhysRevD.100.043537} {\bibfield  {journal} {\bibinfo
  {journal} {Phys. Rev. D}\ }\textbf {\bibinfo {volume} {100}},\ \bibinfo
  {pages} {043537} (\bibinfo {year} {2019})},\ \bibinfo {note} {[Erratum:
  Phys.Rev.D 101, 069904 (2020), Erratum: Phys.Rev.D 101, 129901 (2020)]},\
  \Eprint {http://arxiv.org/abs/1907.03778} {arXiv:1907.03778 [astro-ph.CO]}
  \BibitemShut {NoStop}%
\bibitem [{\citenamefont {Ding}\ \emph {et~al.}(2020)\citenamefont {Ding},
  \citenamefont {Nakama},\ and\ \citenamefont {Wang}}]{Ding:2019mmw}%
  \BibitemOpen
  \bibfield  {author} {\bibinfo {author} {\bibfnamefont {Q.}~\bibnamefont
  {Ding}}, \bibinfo {author} {\bibfnamefont {T.}~\bibnamefont {Nakama}}, \ and\
  \bibinfo {author} {\bibfnamefont {Y.}~\bibnamefont {Wang}},\ }\href {\doibase
  10.1007/s11433-020-1531-0} {\bibfield  {journal} {\bibinfo  {journal} {Sci.
  China Phys. Mech. Astron.}\ }\textbf {\bibinfo {volume} {63}},\ \bibinfo
  {pages} {290403} (\bibinfo {year} {2020})},\ \Eprint
  {http://arxiv.org/abs/1912.12600} {arXiv:1912.12600 [astro-ph.CO]}
  \BibitemShut {NoStop}%
\bibitem [{\citenamefont {Desmond}\ and\ \citenamefont
  {Sakstein}(2020)}]{Desmond:2020wep}%
  \BibitemOpen
  \bibfield  {author} {\bibinfo {author} {\bibfnamefont {H.}~\bibnamefont
  {Desmond}}\ and\ \bibinfo {author} {\bibfnamefont {J.}~\bibnamefont
  {Sakstein}},\ }\href {\doibase 10.1103/PhysRevD.102.023007} {\bibfield
  {journal} {\bibinfo  {journal} {Phys. Rev. D}\ }\textbf {\bibinfo {volume}
  {102}},\ \bibinfo {pages} {023007} (\bibinfo {year} {2020})},\ \Eprint
  {http://arxiv.org/abs/2003.12876} {arXiv:2003.12876 [astro-ph.CO]}
  \BibitemShut {NoStop}%
\bibitem [{\citenamefont {Alestas}\ \emph {et~al.}(2021)\citenamefont
  {Alestas}, \citenamefont {Kazantzidis},\ and\ \citenamefont
  {Perivolaropoulos}}]{Alestas:2020zol}%
  \BibitemOpen
  \bibfield  {author} {\bibinfo {author} {\bibfnamefont {G.}~\bibnamefont
  {Alestas}}, \bibinfo {author} {\bibfnamefont {L.}~\bibnamefont
  {Kazantzidis}}, \ and\ \bibinfo {author} {\bibfnamefont {L.}~\bibnamefont
  {Perivolaropoulos}},\ }\href {\doibase 10.1103/PhysRevD.103.083517}
  {\bibfield  {journal} {\bibinfo  {journal} {Phys. Rev. D}\ }\textbf {\bibinfo
  {volume} {103}},\ \bibinfo {pages} {083517} (\bibinfo {year} {2021})},\
  \Eprint {http://arxiv.org/abs/2012.13932} {arXiv:2012.13932 [astro-ph.CO]}
  \BibitemShut {NoStop}%
\bibitem [{\citenamefont {Cai}\ \emph {et~al.}(2021)\citenamefont {Cai},
  \citenamefont {Guo}, \citenamefont {Li}, \citenamefont {Wang},\ and\
  \citenamefont {Yu}}]{Cai:2021wgv}%
  \BibitemOpen
  \bibfield  {author} {\bibinfo {author} {\bibfnamefont {R.-G.}\ \bibnamefont
  {Cai}}, \bibinfo {author} {\bibfnamefont {Z.-K.}\ \bibnamefont {Guo}},
  \bibinfo {author} {\bibfnamefont {L.}~\bibnamefont {Li}}, \bibinfo {author}
  {\bibfnamefont {S.-J.}\ \bibnamefont {Wang}}, \ and\ \bibinfo {author}
  {\bibfnamefont {W.-W.}\ \bibnamefont {Yu}},\ }\href {\doibase
  10.1103/PhysRevD.103.L121302} {\bibfield  {journal} {\bibinfo  {journal}
  {Phys. Rev. D}\ }\textbf {\bibinfo {volume} {103}},\ \bibinfo {pages}
  {L121302} (\bibinfo {year} {2021})},\ \Eprint
  {http://arxiv.org/abs/2102.02020} {arXiv:2102.02020 [astro-ph.CO]}
  \BibitemShut {NoStop}%
\bibitem [{\citenamefont {Marra}\ and\ \citenamefont
  {Perivolaropoulos}(2021)}]{Marra:2021fvf}%
  \BibitemOpen
  \bibfield  {author} {\bibinfo {author} {\bibfnamefont {V.}~\bibnamefont
  {Marra}}\ and\ \bibinfo {author} {\bibfnamefont {L.}~\bibnamefont
  {Perivolaropoulos}},\ }\href@noop {} {\  (\bibinfo {year} {2021})},\ \Eprint
  {http://arxiv.org/abs/2102.06012} {arXiv:2102.06012 [astro-ph.CO]}
  \BibitemShut {NoStop}%
\bibitem [{\citenamefont {Efstathiou}(2020)}]{Efstathiou:2020wxn}%
  \BibitemOpen
  \bibfield  {author} {\bibinfo {author} {\bibfnamefont {G.}~\bibnamefont
  {Efstathiou}},\ }\href@noop {} {\  (\bibinfo {year} {2020})},\ \Eprint
  {http://arxiv.org/abs/2007.10716} {arXiv:2007.10716 [astro-ph.CO]}
  \BibitemShut {NoStop}%
\bibitem [{\citenamefont {Mortsell}\ \emph
  {et~al.}(2021{\natexlab{a}})\citenamefont {Mortsell}, \citenamefont {Goobar},
  \citenamefont {Johansson},\ and\ \citenamefont {Dhawan}}]{Mortsell:2021nzg}%
  \BibitemOpen
  \bibfield  {author} {\bibinfo {author} {\bibfnamefont {E.}~\bibnamefont
  {Mortsell}}, \bibinfo {author} {\bibfnamefont {A.}~\bibnamefont {Goobar}},
  \bibinfo {author} {\bibfnamefont {J.}~\bibnamefont {Johansson}}, \ and\
  \bibinfo {author} {\bibfnamefont {S.}~\bibnamefont {Dhawan}},\ }\href@noop {}
  {\  (\bibinfo {year} {2021}{\natexlab{a}})},\ \Eprint
  {http://arxiv.org/abs/2105.11461} {arXiv:2105.11461 [astro-ph.CO]}
  \BibitemShut {NoStop}%
\bibitem [{\citenamefont {Mortsell}\ \emph
  {et~al.}(2021{\natexlab{b}})\citenamefont {Mortsell}, \citenamefont {Goobar},
  \citenamefont {Johansson},\ and\ \citenamefont {Dhawan}}]{Mortsell:2021tcx}%
  \BibitemOpen
  \bibfield  {author} {\bibinfo {author} {\bibfnamefont {E.}~\bibnamefont
  {Mortsell}}, \bibinfo {author} {\bibfnamefont {A.}~\bibnamefont {Goobar}},
  \bibinfo {author} {\bibfnamefont {J.}~\bibnamefont {Johansson}}, \ and\
  \bibinfo {author} {\bibfnamefont {S.}~\bibnamefont {Dhawan}},\ }\href@noop {}
  {\  (\bibinfo {year} {2021}{\natexlab{b}})},\ \Eprint
  {http://arxiv.org/abs/2106.09400} {arXiv:2106.09400 [astro-ph.CO]}
  \BibitemShut {NoStop}%
\bibitem [{\citenamefont {Abitbol}\ \emph {et~al.}(2019)\citenamefont {Abitbol}
  \emph {et~al.}}]{Abitbol:2019nhf}%
  \BibitemOpen
  \bibfield  {author} {\bibinfo {author} {\bibfnamefont {M.~H.}\ \bibnamefont
  {Abitbol}} \emph {et~al.} (\bibinfo {collaboration} {Simons Observatory}),\
  }\href@noop {} {\bibfield  {journal} {\bibinfo  {journal} {Bull. Am. Astron.
  Soc.}\ }\textbf {\bibinfo {volume} {51}},\ \bibinfo {pages} {147} (\bibinfo
  {year} {2019})},\ \Eprint {http://arxiv.org/abs/1907.08284} {arXiv:1907.08284
  [astro-ph.IM]} \BibitemShut {NoStop}%
\end{thebibliography}%

\end{document}